\documentclass{article}
\usepackage{cite}
\usepackage{array}
\usepackage{epsfig}
\usepackage{hyperref}
\usepackage{amsmath}
\usepackage{multirow}
\usepackage{amssymb}
\usepackage{color}
\usepackage{indentfirst}
\usepackage{graphicx,floatflt,subfig}
\usepackage{axodraw4j}
\usepackage{pstricks}
\numberwithin{equation}{section}
\def\rmd{{\rm d}}
\def\bea{\begin{eqnarray}}
\def\eea{\end{eqnarray}}

\def\Log{\mbox{ Log} }

\begin{document}
%%%%%%%%%%%%%%%%%%%%SECTION1%%%%%%%%%%%%%%%%%%%%%%%%%%%%%%%%%%%%%%%%%%%%
\baselineskip18pt
%\tableofcontents

\begin{titlepage}
\rightline{SNUST-14-07}
\vspace*{0.7cm}

\centerline{\Large \bf Antenna Operator Product Expansion}
\vskip0.3cm
\centerline{\Large \bf of}
\vskip0.3cm
\centerline{\Large \bf  ABJ(M) Lightlike Polygon Wilson Loop}
\vskip0.5cm
%\centerline{\huge \bf in}
%\vskip0.5cm
%\centerline{\huge \bf ABJM Theory}
\vspace*{1.3cm}
\centerline{\Large Jinbeom Bae $^a$, \ \ \ \ Soo-Jong Rey $^{a,b}$}
\vspace*{1.3cm}
\centerline{\sl $^a$ School of Physics \& Astronomy and Center for Theoretical Physics}
\vspace*{0.15cm}
\centerline{\sl Seoul National University, Seoul 151-747 \rm KOREA}
\vspace*{0.30cm}
\centerline{\sl $^b$ Center for Theoretical Physics of the Universe}
\vspace*{0.15cm}
\centerline{\sl Institute for Basic Sciences, Daejeon 305-811 \rm KOREA}
\vspace{2cm}
\centerline{ABSTRACT}
\vskip0.75cm
\noindent
Expectation value of lightlike polygon Wilson loop is computed in the three-dimensional ABJM theory up to second-order in `t Hooft coupling in the limit of infinitely many colors and the result is critically compared with that in the four-dimensional ${\cal N}=4$ super Yang-Mills theory. We first obtain analytic result for hexagon Wilson loop by combining Mellin-Barnes transformation, high precision numerical computation and the PSLQ algorithm. We then derive a version of operator product expansion (OPE) that reduces  lightlike $n$-gon to a linear combination of $(n-2)$-gons in the soft-collinear limit of the polygon geometry. The Wilson coefficient of the OPE is the universal antenna function defined by a collapsed lightlike tetragon Wilson loop. Using this, we first construct a all order recursion relation among the lightlike Wilson loops and then solve it for arbitrary polygon with the hexagon Wilson loop as the initial condition. The functional form of the polygon Wilson loop takes the structure remarkably similar to the four-dimensional ${\cal N}=4$ super Yang-Mills theory. We also observe that Gram subdeterminant conditions for polygon moduli variables restricts that the Wilson loop contour should be restricted even-sided.  As a consistency check, we take thermodynamic limit of regular polygon and reproduce the known results for spacelike circular Wilson loop expectation value.

\end{titlepage}
\rightline{\sl "When you have eliminated the impossible, whatever remains, however improbable, must be the truth"}
\rightline{A. Conan Doyle, The Sign of Four}
\section{Introduction and Results}
\subsection{Introduction and Background}

In gauge theories, the Wilson loop $W_{\bf R} [C]$ constitutes an important class of defect operator localized on a contour $C$. Carrying representation ${\bf R}$ under gauge group $G$, it has played important roles in probing ground-state phases of gauge theories. It has also been used for studying excitation dynamics: Wilson loops on a light-like cusp or a polygon $C_n$ with $n$-cusps located at $x_1, \cdots, x_n$ probed deeply inelastic scattering of hadrons or high-energy parton-parton scattering with associated evolution equation for regge trajectory.
\cite{Korchemskaya:1992je}

In this paper, we undertake perturbtive study of light-like Wilson loop in the ABJM theory\cite{Maldacena:2008a} for the contour $C$ a lightlike polygon with $n$-cusps ($n \ge 4)$ located at $x_1, \cdots, x_n$. The ABJM theory is $(2+1)$-dimensional, parity-invariant Chern-Simons matter theory having ${\cal N}=6$ superconformal symmetry and $SU (N) \times SU (N)$ gauge symmetry with Chern-Simons levels $+k, -k $, respectively. We shall take planar limit $N \rightarrow \infty$ while holding `t Hooft coupling $\lambda = (N/k)$ fixed and study the Wilson loop up to second order in $\lambda$.

Our motivation comes from the equivalence of the lightlike polygon Wilson loops with parton scattering amplitudes on one hand and correlation functions on another in $(3+1)$-dimensional, planalr $\mathcal{N}=4$ Super Yang-Mills (SYM) theory, so we shall briefly review the results. A stepping stone was provided by the remarkable ansatz of Bern, Dixon and Smirnov (BDS) for the all-loop resummation formula for the four-point scattering amplitudes \cite{Bern:2005a}, built upon recursion relations at one-, two- and three-loops and upon universality of the splitting function for extending the ansatz to higher-point scattering amplitudes. The BDS ansatz prompted to investigate the scattering amplitudes in strongly coupled ${\cal N}=4$ SYM theory. Utilizing the AdS/CFT correspondence,  Alday and Maldacena mapped the problem to a problem in string theory on weakly curved, $(4+1)$-dimensional anti-de Sitter (AdS) spacetime background. By exploiting conformal invariance, they calculated the four-point scattering amplitude on a D-brane displaced to the Coulomb branch by fermionic T-dualizing it to a holographic Wilson loop on a tetragon enclosed by light-like edges. The infrared (IR) finite part of the Wilson loop matched with the ultraviolet (UV) finite part of the all-loop BDS conjecture for four-point scattering amplitudes. Replacing cusp anomalous dimension to weakly coupled counterpart, the IR divergent part also agreed with the all-loop BDS conjecture. At weak `t Hooft coupling regime, the scattering amplitude - Wilson loop duality relates the $n$-particle MHV scattering amplitudes $A_n^{\rm MHV} = A_n^{\rm tree} K_n$ and the lightlike $n$-gon Wilson loops $\langle W_n \rangle$ by \cite{Drummond:2007a}
\bea
\log K_n (p_1, \cdots, p_n) = \log \langle W_n \rangle (x_1, \cdots, x_n).
\label{duality-relation}
%\quad
%\mbox{where} \quad p_1 = x_1 - x_{n}, \cdots, p_n = x_n - x_{n-1}.
\eea
The equivalence was confirmed by explicit computation for $n=4, 5$ and up to two loops.

The situation changed for higher-point scattering amplitudes. A discrepancy was found between the BDS conjecture of six-point scattering amplitude and the hexagon Wilson loop \cite{Drummond:2007b}. This result suggested either breakdown of the fermionic T-duality between scattering amplitude and Wilson loop or breakdown of the BDS conjecture. The first possibility was excluded by direct computation of parity even part of six gauge boson amplitude \cite{Bern:2008a}. There also appeared non-trivial differences to the BDS conjecture. This quantity, the discrepancy from the BDS conjecture, constitutes the remainder function. Detailed investigation of the lightlike hexagon Wilson loop at two-loop order \cite{Heslop:2009a} concluded that the BDS conjecture indeed breaks down. Numerically, they found that this remainder function depends only on conformal cross-ratios. Moreover, absence of the remainder function at tetragon and pentagon Wilson loops was understood as a consequence of the dual conformal symmetry and conformal Ward identity \cite{Drummond:2007a, Drummond:2008b}. This conformal Ward identity fixed functional structure of the Wilson loop expectation value and limited the number of free variables in the remainder function.

The analytic structure of this remainder function of hexagon Wilson loop was obtained in \cite{Duhr:2010a}. There, central tool was to utilize the special limit of the quasi-multi Regge kinematics(QMRK). It was found that the remainder function could be written with the uniform transcendentality in terms of the Goncharov polylogarithms. The structure was very involved but was reduced to a combination of classical polylogarithms of uniform transcendentality of degree four \cite{Spradlin:2010a}.  These techniques were also utilized in subsequent works investigating analytic structure of the remainder function up to three loop order \cite{Heslop:2011a,Drummond:2011a}. It was further extended to analytic structure of the non-maximal helicity violating (nMHV) amplitudes at two loop order with help of the symbols \cite{Drummond:2011b}.

%Recently, the light-like polygon Wilson loop was studied extensively in $(3+1)$-dimensional ${\cal N}=4$ superconformal Yang-Mills (SYM) theory at planar and fixed `t Hooft coupling limit. There, the motivation stemmed from the Wilson loop / scattering amplitudes duality.

%It was noted that the scattering amplitudes are constrained by superconformal invariance, while the lightlike Wilson loops are constrained by dual superconformal invariance. At strong `t Hooft coupling, this equivalence was confirmed by the fermionic T-duality symmetry for open superstrings in AdS$_5 \times \mathbb{S}^5$. It was noted that the T-duality interchanges scattering amplitudes and lightlike Wilson loops and also interchanges superconformal isometries with dual superconformal isometries.

Moreover, there is a growing evidence that, in the collinear limit of adjacent edges of the polygon Wilson loop, the remainder function can be systematically expanded in powers of collinearity and that each term can be computed to all order in the `t Hooft coupling by utilizing the integrability of the four-dimensional planar ${\cal N}=4$ SYM theory  \cite{Basso:2013vsa}, \cite{Basso:2013aha},   \cite{Basso:2014koa},   \cite{Basso:2014nra}. For the hexagon Wilson loop, the expansion takes schematically the form
\bea
\mbox{Rem}_6 = \sum_{\ell=1}^\infty \lambda^\ell \mbox{Rem}_6^{(\ell)}
\eea
where, in terms of suitable parametrization $(\tau, \sigma, \phi)$ of kinematical variables, 
\bea
\mbox{Rem}_6^{(\ell)} = \sum_{m=1}^\infty e^{- m \tau} \sum_{p=0}^{[m/2]} \cos (m- 2 p) \phi \sum_{n=0}^{\ell - 1} \tau^n F^{(\ell)}_{m,p,n}(\sigma).
\eea
The collinear limit corresponds $\tau \rightarrow \infty$ and $m$ is the number of particles excited on a color-electric flux tube stretching between two edges that become collinear. Most interestingly, the dynamics of these excitations on the flux tube is described by an integrable spin chain.    

We now turn to the ABJM theory and compare known results about scattering amplitudes and lightlike Wilson loops with those of the ${\cal N}=4 $ SYM theory. Given that both theories are superconformal, one might anticipate from the AdS/CFT correspondence that the Wilson loops / scattering amplitudes duality would hold equally and result in identical functional structure, at least at strong coupling regime. This is not the case. In the ABJM theory, the $n$-parton scattering amplitudes is definable only for even integers $n = 4, 6, 8, \cdots$ (since the dynamical matter fields are in bifundamental representations),  while the $n$-gon Wilson loops exist for all integers $n = 4, 5, 6, \cdots$. So, a possibility is that the duality Eq.(\ref{duality-relation}) holds only for even integers $n$, while leaving Wilson loops of odd-sided polygons a class of its own. Secondly, the open superstring in AdS$_4 \times \mathbb{CP}^3$ does not possess the fermionic T-duality. Absence of fermionic T-duality symmetry implies no relation between scattering amplitudes and light-like Wilson loops. This then rules out the duality relation Eq.(\ref{duality-relation}) in the ABJM theory.

The four-point scattering amplitude was computed in two different approaches : the unitarity-cut method \cite{Huang:2011a} and direct superspace diagram calculation \cite{Bianchi:2011a}. Parallel to the $\mathcal{N}=4$ SYM theory, the four-point scattering amplitudes turned out related to the tetragon Wilson loop expectation value. It was observed that the scattering amplitudes in ABJM and $\mathcal{N}=4$ SYM theories are similar: two-loop four-point amplitude in the ABJM theory is equivalent to that of $\mathcal{N}=4$ SYM theory to all orders in $\epsilon$ \cite{Bianchi:2011b}. The sixe-point scattering amplitudes were computed at one-loop \cite{Beisert:2012a,Leoni:2012a,Brandhuber:2012a} and  at two-loop \cite{Huang:2012a}, respectively. The one-loop six-point scattering amplitudes turned out non-vanishing, being proportional to shifted tree-level six-point amplitude.

The Wilson loops come in two types, one for $SU(N)_k$ and another for $SU(N)_{-k}$. Firstly, from the strong coupling side, one expects that the $AdS_4$ structure dictates that the minimal surface associated with the ABJM Wilson loop ought to behave the same as that for the ${\cal N}=4$ SYM Wilson loop.

there are two Wilson loop operators the expectation value at weak `t Hooft coupling regime starts  perturbative corrections start at two loops. The Wilson loop operator of each gauge group preserves $1/6$ of supersymmetry and its expectation value  is a function of $\lambda$, starting at the second order. On the other hand, average of the two Wilson loop operators preserves $1/2$ of the supersymmetry and its expectation value is a function of $\lambda^2$, starting at the second order.

Secondly, the perturbative corrections depend crucially on the radiatively-induced, parity-even gauge boson self-energy. Feynman rules for this gauge boson self-energy propagation in ABJM theory resembles propagators of gauge boson in $\mathcal{N}=4$ SYM\cite{Rey:2008a, Plefka:2008a}. This suggests that matter contributed part of two-loop Wilson loop in ABJM theory could be considered as one-loop Wilson loop in $\mathcal{N}=4$ SYM. Moreover, it was found that pure Chern-Simons part could be absorbed to matter part by replacing regularization energy scale. Indeed, anomalous conformal Ward identity again works for the ABJM theory. Indeed, it was conjectured that Wilson loop in the ABJM theory almost have same structure with Wilson loop in $\mathcal{N}=4$ SYM. Up to 4-point, this was numerically confirmed \cite{Plefka:2010a}. This suggests that bosonic Wilson loop expectation value is not equivalent object to scattering amplitude since odd loop Wilson loop expectation values are 0 in ABJM theory. 2-loop 6 point result seems more complex, appeared by linear combination of tree level amplitude and shifted tree level amplitude.

Despite of all these, explicit computation poses interesting puzzle. The light-like tetragon Wilson loop at two loops displays striking similarity to the Wilson loop of SYM theory at one loop. Moreover, the Wilson loop result was dual to the four-particle scattering amplitudes, thus satisfying the duality relation Eq.(\ref{duality-relation}). This might be a coincidence because, in ABJM theory, scattering amplitudes and tetragon Wilson loops are MHV-like observables for $n=4$ but not for higher $n$.

%The light-like polygon Wilson loop was studied extensively in $(3+1)$-dimensional, parity invariant super Yang-Mills theory having ${\cal N}=4$ superconformal symmetry and $SU(N)$ gauge symmetry with gauge coupling $g$. Taking the planar limit $N \rightarrow \infty$ while holding `t Hooft coupling $\lambda = 4 \pi g^2 N$ fixed, the Wilson loop was studied for $n \le 6$ and up to second order in $\lambda$.

%from various  duality between MHV amplitude and Wilson loop expectation value on light-like polygon at weak coupling region\cite{Drummond:2007a}. In $\mathcal{N}=4$ Super Yang-Mills theory, this duality examined at weak coupling side by explicit perturbative calculation. In both sides, conformal Ward identity is another way to construct scattering amplitude or Wilson loop expectation value. Although they did not gives full information about both objects, part of them are given by solving conformal Ward identity\cite{Drummond:2007a1}. Naturally we can expect such kind of properties are hold in another conformal field theory, ABJM theory. By explicit Feynman diagram calculation, we investigated various properties of Wilson loop in ABJM theory.

Another our motivation concerns comparative study between the lightlike polygon Wilson loop in $(2+1)$-dimensional ABJM theory and that in  $(3+1)$-dimensional $\mathcal{N}=4$ Super Yang-Mills theory. In \cite{Plefka:2010a}, they computed Wilson loop expectation value for two loop tetragon case. The result was quiet remarkable, its structure is almost equivalent to one-loop Wilson loop in $\mathcal{N}=4$ Super Yang-Mills theory. There is no reason that existence of such similarity for arbitrary n($n>4$)-gon case. Our goal is constructing n-gon Wilson loop expectation value and comparing with  $\mathcal{N}=4$ Super Yang-Mills theory's one.

Yet another motivation is to investigate whether and how the color electric flux-tube picture advanced in \cite{Basso:2013vsa}, \cite{Basso:2013aha}, \cite{Basso:2014koa}, \cite{Basso:2014nra} can be extended to the ABJM theory. While such an extension is much anticipated from the geometric similarity in the strong coupling regime, it is far from being clear in the weak coupling regime. The flux tube configuration in the ABJM theory is not transparent and also particle excitations on the flux tube worldsheet appear quite different. Nevertheless, results in this work indicates that a variety of geometric structures are quite similar for both theories but for rather different reasons. We expect that the soft-collinear operator product expansion we develop in this paper and the universal antenna function we derived would provide a starting point bridging conceptual and caculational gaps between the two theories. 

%%%%%%%%%%%%%%%%%%%%%%%%%%%%%%%%%%%%%%
\subsection{Main Results}
Firstly,  we obtained  analytic result of the light-like hexagon Wilson loop at second-order in `t Hooft coupling constant $\lambda = (N/k)$. Regularizing the UV divergence by supersymmetric dimensional reduction scheme with $d = 3 -2 \epsilon$, the result reads
\bea
\big< W_\Box [C_6] \big>^{(2)}_{\rm ABJM}= \lambda^2  \left[-{1 \over 2} \sum_{i=1}^6 \frac{(x_{i,i+2}^2 \tilde{\mu}^2)^{2\epsilon}}{(2\epsilon)^2}+\mbox{BDS}^{(2)}_{6}(x) + \left( \frac{9}{2} \mbox{Log}^2(2) + \frac{\pi^2}{3} \right)  \right].
\label{hexagon}
\eea
Here, $x_i \ (i=1, \cdots, 6)$ are hexagon vertex positions, and $\tilde{\mu}^2 := 8 \pi e^{\gamma_E} \mu^2$. The second term is the UV-finite "BDS" function \footnote{Hereafter, we shall adopt the terminology of corresponding quantities in the ${\cal N}=4$ SYM theory} :
\bea
\mbox{BDS}^{(2)}_{6}(x) \! &=& \! {1 \over 2} \! \sum_{i=1}^6 \left[\frac{1}{4} \Log^2\left(\frac{x_{i,i+3}^2}{x_{i+1,i+4}^2}\right)-  \! \Log\left(\frac{x_{i,i+2}^2}{x_{i,i+3}^2}\right) \! \! \Log\left(\frac{x_{i+1,i+3}^2}{x_{i,i+3}^2}\right) -\frac{1}{2} \! \mbox{Li}_2 \left(1-\frac{x_{i,i+2}^2 x_{i+3,i+5}^2}{x_{i,i+3}^2 x_{i+2,i+5}^2} \right)\right] \nonumber \\
&+& {\pi^2 \over 2}. \label{BDS6}
\eea
This function depends on the vertex positions the same way as the UV-finite, leading order BDS function of the $\mathcal{N}=4$ SYM theory. All the finite parts in Eqs.(\ref{hexagon}, \ref{BDS6}) exhibit the property of the uniform transcendentality.

Secondly, we obtained the ABJM antenna function relevant for lightlike polygon Wilson loops. Recall that, in the ${\cal N}=4$ SYM theory, the splitting function in the scattering amplitudes is defined by the IR factorization associated with multiple collinear limit of massless particles. Here, in ABJM theory, we dwell on special kinematic configuration and focus on the limit of the Wilson loop contour. For reasons that will become clear later, we focus on a sort of operator product expansion involving two collinear edges of fraction $h_1, h_3$ and one soft edge of fraction $h_2$ in between. This limit defines the triple antenna function: lightlike $n$-gon Wilson loop is decomposable into a product of $(n-2)$-gon light-like Wilson loop and this antenna function. At second order in `t Hooft coupling $\lambda$, the triple antenna function consists of two parts: the pure Chern-Simons part and matter-dependent part. Our result is
\bea
\left[ {\mbox{Ant}^{(2)}[C_n]
\over \mbox{Ant}^{(0)}[C_n]}
\right]_{\rm CS} &=& \frac{\Log(2)}{2\epsilon} \nonumber \\
&+&\frac{1}{2}\log(2)   \mbox{Log} (h_1) +\frac{1}{2}\Log(2)  \Log(h_3) +\frac{1}{2}\Log(2) \Log (x_{24}^2) +\frac{1}{2}\Log(2) \Log (x_{35}^2)
\nonumber \\
&-&\frac{7\pi^2}{24}+\Log^2(2)
\eea
for the pure Chern-Simons part and
\bea
\left[ {\mbox{Ant}^{(2)}[C_n] \over \mbox{Ant}^{(0)}[C_n] } \right]_{\rm matter} &=& \frac{1}{4\epsilon^2} + \frac{1}{4\epsilon}(\Log (h_1) + \Log(h_3) + \Log(x_{24}^2) + \Log(x_{35}^2)) \nonumber \\
&+&\frac{1}{2} \Log( h_1) \Log(x_{24}^2) +\frac{1}{2} \Log (h_3) \Log(x_{35}^2) +\frac{1}{2} \Log (x_{35}^2) \Log(x_{24}^2) \nonumber \\
&-&\frac{1}{2} \Log (h_1) \Log (h_3) - \frac{\pi^2}{6}
\eea
for the matter-dependent part. The result demonstrates that splitting function also displays the property of maximal transcendentality. Moreover, the result is independent of $n$, suggesting that the triple antenna function holds universally for all $n$. The total triple antenna function is strikingly similar to triple splitting function in the limit middle parton becomes soft. We found, however, they are still subtly different.

Thirdly, combining the two results above, we obtained the simplest functional form of lightlike polygon ABJM Wilson loop to the second-order in `t Hooft coupling. The requisite shape of lightlike polygon must obey the positivity condition and satisfy vanishing Gram determinant condition.
%In doing so, we note that there is a significant difference between the three-dimensional ABJM theory and the four-dimensional SYM theory.
%In both theories, the conformal Ward identity fixes the functional form of the Wilson loop as
%
%\bea
%\Big< W_\Box [C_n] \Big> = {(s)^{2 \epsilon} \over \epsilon^2} + {\cal F}_n (s) + G_n (u(s))
%\eea
%
%where $s$ is the set of Mandelstam variables and $u(s)$ is the set of conformal cross-ratios. %In general, there can be a sequence of $s, s', s", \cdots$ satisfying $u(s) = u(s') = u(s")= \cdots$. In the ${\cal N]=4$ SYM theory, for reasons not well understood yet, the remainder function $G_n(u(s))$ can be computed for arbitrary $s$ exhibited curious property that $G_n(u(s)) = G_n(u(s')) = G_n(u(s")) = \cdots$. In particular, one can adopt Mandelstam variables that does not satisfy the Gram determinant condition (which amounts to total momentum conservation) in so far as it gives the same conformal cross-ratio. Here in ABJM theory, explicit numerical computations show that this property no longer holds. Moreover,
It turns out such kinematics requirements limit the lightlike polygon only to the one with even number of cusps, marking a stark difference from the ${\cal N}=4$ SYM theory. Demanding IR factorization of the Wilson loop with the universal antenna function, we obtained a version of operator product expansion, leading to a linear recursion relation among the light-like Wilson loops:
\bea
\Big< W_\Box [C_n ] \Big> \quad \longrightarrow \quad \mbox{Ant}[C_n] \cdot
\Big< W_\Box[C_{n-2}] \Big> .
\eea
%
%This means that the operator product expansion (OPE) of three cusps into one cusp is summarized by the universal splitting function.
Solving this recursion with the hexagon Wilson loop (\ref{hexagon}) as an input, we finally find that
\bea
\boxed{
\Big< W_\Box [C_n] \Big>^{(2)}_{\rm ABJM}=\lambda^2  \left[ -{1 \over 2} \sum_{i=1}^n \frac{(x_{i,i+2}^2 \tilde{\mu}^2)^{2\epsilon}}{(2\epsilon)^2} +\mbox{BDS}^{(2)}_n(x) + \mbox{Rem}^{(2)}_n(u) \right],
}
\eea
where $\mbox{Rem}_n(u)$ is the remainder function that depends on the Mandelstam invariants only through conformal cross-ratios $u$\rq{}s. At two loops, the remainder function is independent of $u$\rq{}s and reads
\bea
\boxed{
\mbox{Rem}^{(2)}_n(u) = \left[ n\left(\frac{\pi^2}{12} +\frac{3}{4} \log^2(2) \right) - \frac{\pi^2}{6}  \right].
}
\eea
Here, we extracted this analytic result by utilizing the PSLQ algorithm to the high precision numerical integrations. As a nontrivial check, we derived the spacelike circular Wilson loop expectation value from the $n \rightarrow \infty$ continuum limit and found perfect agreement with the previous results.

%%%%%%%%%%%%%%%%%%%%%%%%%%%%%%%%%%%%%%%%%%%%%%%%%%%%%%%%%%%%%
This paper is organized as follows. In section 2, we set our notations for ABJM Wilson loops and summarize known results for lightlike tetragon Wilson loop. In section 3, we study lightlike hexagon Wilson loop at two loop order. Planar Feynman diagrams contributing to this order include ladder diagrams, triple-vertex diagrams and matter-dependent diagrams. %The first two originate purely from Chern-Simons interactions, while the last originates from matter coupled to the Chern-Simons gauge fields.
Details of integrand are relegated to the Appendix. In section 4, we explain computational details. We evaluate the matter-dependent part analytically. The ladder and the triple-vertex diagrams are more involved. We compute them numerically. To achieve high precision,  we study so-called two-parameter configurations and utilize Mellin-Barnes transformation. In section 5, we construct the antenna function for the ABJM theory. Requiring the OPE-like factorization conditions, we show that the antenna function relates $n$-gon Wilson loop to $(n-2)$-gon Wilson loop. Solving this relation recursively general expression of the lightlike $n$-gon Wilson loop can be obtained. In section 6, we carry this out and obtain analytic expression for arbitrary $n$-gon. As a consistency check, we examine the $n \rightarrow \infty$ limit and reproduce the known exact result of circular Wilson loop. In section 7, we compare the structure of the lightlike polygon Wilson loop for the ABJM theory with that for the ${\cal N}=4$ SYM theory, and discuss physical implications.

%%%%%%%%%%%%%%%%%%%%%%%%%%%%%%%%%%%%%%%SECTION2%%%%%%%%%%%%%%%%%%%%%%%%%%%%%%%%%%%%%%%%%%
\section{Light-like Polygon Wilson loop in ABJM Theory}
\subsection{ABJM Theory}
The ABJM theory describes $(2+1)$-dimensional supersymmetric matter interacting with Chern-Simons gauge system. It has ${\cal N}=6$ superconformal symmetry (having 24 conserved supercharges) and $U(N) \times \overline{U(N)}$ gauge group with Chern-Simons levels $+k, -k$, respectively. The gauge fields are denoted as $A_m(x) \in u(N)$ and $\overline{A}_m(x) \in \overline{u(N)}$. For our notations and conventions of the field contents, Lagrangian and Feynman rules, see the Appendix A. For foregoing considerations, it suffices to note that the action includes the pure Chern-Simons density \cite{Maldacena:2008a}
\bea
S_{\rm CS}&=& + \frac{k}{4\pi} \int \rmd^dx \epsilon^{mnp} \mbox{Tr} \Big( A_m \partial_{n} A_p + \frac{2i}{3} A_{m} A_{n} A_{p} \Big) \label{CS_action_a1}\\
\overline{S}_{\rm CS}&=& -\frac{k}{4\pi} \int \rmd^dx \epsilon^{mnp} \overline{\mbox{Tr}} \Big( \overline{A}_m \partial_{n} \overline{A}_p + \frac{2i}{3} \overline{A}_{m} \overline{A}_{n} \overline{A}_{p} \Big). \label{CS_action_b1}
\eea
Here, the Chern-Simons density has levels $+k$ and $-k$, respectively. Invariance of the action under large gauge transformation puts $k$ integer-valued. The action is invariant under the generalized parity that simultaneously reverts one spatial coordinates and exchanges the two gauge fields. In this theory, the ABJM Wilson loop operator in the fundamental representation $(\Box, 1) \oplus (1, \Box)$ of the gauge group $U(N) \times \overline{U(N)}$ is defined by \cite{Rey:2008a}:
\bea
\mathcal{W}_\Box [C]:=\frac{1}{2}\Big(W_\Box [C]+\overline{W}_\Box [C] \Big), \label{def:Wilson}
\eea
where $W_\Box [C]$ and $\overline{W}_\Box [C]$ refer to the Wilson loop of the fundamental representation of $U(N)$ and $\overline{U(N)}$ gauge groups, respectively.

%This definition of the Wilson loop operator also holds for the ABJ theory provided that their tree-level expectation values are normalized to 1 for $W[C], \overline{W}]C]$ and hence for ${\cal W}[C]$.

The close contour $C$ is a geometric datum of the Wilson loop operator. Hereafter, we shall exclusively deal with Lorentzian contour $C_n$ connecting $n$ vertices $x_1, x_2, \cdots, x_n$ whose adjacent points are lightlike-separated. The total set $C_n$ with $n = 4, 5, 6, \cdots$ form lightlike $n$-gons. Denote the distance vectors between a pair of vertices by
\bea
x_{i,j} \equiv [x_i - x_j] \qquad \quad i,j = 1, \cdots, n.
\eea
Among them are the lightlike-separated edges $x_{i+1, i}$. Denote a point on $i$-th edge by $z_i$. In parametrized form, it is
\bea
z_i (\tau) = x_i + y_i \tau \qquad \mbox{where} \qquad y_i \equiv x_{i+1, i}, \qquad  0 \le \tau \le 1.
\eea
We relegate notations for various Lorentz invariants of $x_i$'s to Appendix A.

The lightlike $n$-gon Wilson loop operators for $SU(N)$ and $\overline{SU(N)}$ gauge groups take the form
\bea
W_\Box [C_n]&=&\frac{1}{N} \mbox{Tr}\mathcal{P} \mbox{exp} \left[i \oint_{C_n} \rmd\tau A_m(x(\tau)) \dot{x}^m(\tau)  \right] \label{def:Wilson_a}\\
\overline{W}_\Box [C_n]&=& \frac{1}{N} \overline{\mbox{Tr}}\mathcal{P} \mbox{exp} \left[i \oint_{C_n} \rmd\tau \overline{A}_m(x(\tau)) \dot{x}^m(\tau) \right]. \label{def:Wilson_b}
\eea
Both are $1/6$-BPS operators preserving 4 supercharges. Under the generalized parity, the two Wilson operators are interchanged each other. On the other hand, the ABJM Wilson loop $\mathcal{W}_\Box [C]$ is $1/2$-BPS operator preserving 12 supercharges. By construction, it is invariant under the generalized parity. The $n$ vertices of $C_n$ break all supersymmetries. This implies that the expectation values of these Wilson loops receive quantum corrections. Analyzing these corrections in the regime of infinite number of color $N \rightarrow \infty$ and weak `t Hooft coupling $\lambda = (N/k) \ll 1$ is the main focus of this paper.

\subsection{Previous Results}
Our goal is to compute the vacuum expectation value of the lightlike polygon Wilson loop. In the planar limit, we evaluate it in perturbation theory of the `t Hooft coupling $\lambda$:
\bea
\big< {\cal W}_\Box[C] \big> = \sum_{\ell = 0}^\infty \lambda^\ell \ \big< {\cal W}_\Box[C] \big>^{(\ell)}
\eea
and similarly for $\big< W_\Box[C] \big>$ and $\big< \overline{W}_\Box [C] \big>$. The Wilson loops $\big< W_\Box[C] \big>$ and $\big< \overline{W}_\Box [C] \big>$ are 1/6-BPS configurations and in general receive perturbative corrections to all orders in $\lambda$. On the other hand, the ABJM Wilson loop $\big< {\cal W}_\Box[C] \big>$ is 1/2-BPS configuration and receive perturbative corrections only at even order of $\lambda$. This is an elementary consequence of the fact that the ABJM Wilson loop is invariant under the generalized parity. Since the net effect of the generalized parity is to flip $k$ to $-k$, equivalently, $\lambda$ to $-\lambda$, it follows immediately that
\begin{equation}
\big< W_\Box [C]\big>^{\ell=\mbox{odd}}=-\big< \overline{W}_\Box [C]\big>^{\ell = \mbox{odd}}
\end{equation}
Actually, the result is stronger at linear order in $\lambda$. At this order, kinematical considerations indicate that  $\big< W_\Box [C_n] \big>^{(1)}$ and  $\big< \overline{W}_\Box [C_n]\big>^{(1)}$ vanish separately.  By the generalized parity transformation, it also follows that
\begin{equation}
\big< W_\Box [C]\big>^{\ell = \mbox{even}}=\big< \overline{W}_N[C]\big>^{\ell = \mbox{even}}.
\end{equation}
We conclude that
\bea
\big< {\cal W}_\Box[C]\big> = \sum_{\ell = 0}^\infty \lambda^{2 \ell} \big< W_\Box[C]\big>^{(2 \ell)} = \sum_{\ell = 0}^\infty \lambda^{2 \ell} \big< \overline{W}_\Box[C]\big>^{(2\ell)}.
\eea

The leading-order correction arises at two-loop order $O(\lambda^2)$. The diagrams contributing to this order are categorized to three groups \cite{Plefka:2010a}: matter-dependent diagrams, gauge boson ladder diagrams, and gauge boson triple-vertex diagrams. The contribution of the matter diagrams is equivalent to one-loop contribution in the $\mathcal{N}=4$ SYM theory. This is because, in the ABJM theory,  the finite one-loop correction to the gauge boson propagator is precisely the same as the tree-level gauge boson propagator in the ${\cal N} = 4$ SYM theory \cite{Rey:2008a}. This means that differences between the ABJM theory and the $\mathcal{N}=4$ SYM originate from ladder diagrams and triple-vertex diagrams. Both diagrams originate from gauge boson interactions through the Chern-Simons parts. Computationally, these two contributions are the most complicated.

The general structure of the two-loop corrections to the light-like Wilson loop expectation value can be obtained by requiring the anomalous conformal Ward identities. For this consideration, we can split the contributions to two parts: the matter contribution and the Chern-Simons contribution.
%So, this already exemplifies that the scattering amplitudes - Wilson loop equivalence breaks down in three-dimensional conformal field theories.
%The matter contribution takes the same form as the one-loop result for the ${\cal N}=4$ SYM theory. In particular, it includes the BDS contribution. The Chern-Simons system has no propagating degrees of freedom.
As explained above, the matter contribution is structurally the same as the one-loop contribution to the lightlike Wilson loops in $\mathcal{N}=4$ SYM theory. Therefore, it is useful to recall how the anomalous conformal Ward identities determined the Wilson loop expectation value in the $(3+1)$-dimensional SYM theory. There, the dilatation generator $\mathbb{D}$ and the special conformal generator $\mathbb{K}$ were perturbatively modified by quantum corrections. The dilatation symmetry is broken by the UV regularization and its Ward identity gets anomalous. To ${\cal O}(\lambda_{\rm SYM})$,
%Derivation of this Ward identity requires diagram calculation, I did not repeated that. In Korchemsky's original paper, their results of dilaton anomalous equation up to $\lambda$ order is given by
the $(3+1)$-dimensional SYM theory exhibits
\begin{equation}
\mathbb{D}\big< W [C_n]\big> \Big\vert_{\rm SYM} = -\lambda_{\rm SYM} \left[ \sum \frac{(x_{i-1,i+1}^2 \mu^2)^\epsilon}{\epsilon} + \mathcal{O}(\epsilon^0) \right].
\end{equation}
The $\mathcal{O}(\epsilon^0)$ term refers that this Ward identity is verified up to $\epsilon^0$-order.
Using the elementary relation
\begin{equation}
\mathbb{D} \Big( (x_{i,j}^2)^\epsilon \Big) = 2\epsilon (x_{i,j}^2)^{\epsilon},
\end{equation}
we can find particular solution to the dilatational Ward identity as
\begin{equation}
\big< W [C_n]\big> \Big\vert_{\rm SYM} = \lambda_{\rm SYM} \left[ -\frac{1}{2} \sum \frac{(x_{i-1,i+1}^2 \mu^2)^\epsilon}{\epsilon^2} + \mathcal{O}(\epsilon^0) \right].
\end{equation}
Consideration of the special conformal generator $\mathbb{K}$ confirmes the result and further provides information for the ${\cal O}(\epsilon^0)$ part, so-called the BDS function, $\mbox{BDS}_n$. Homogeneous solution to the conformal Ward identities is referred as the remainder function $\mbox{Rem}_n$. It depends only on the conformal cross-ratios $u$ of the $n$-sided polygon. Putting together and replacing $\lambda_{\rm SYM}$ by $\lambda^2$, we deduce that the matter contribution in the ABJM theory takes the form
\bea
\big< W_\Box [C_n] \big>^{(2)}\Big|_{\rm matter} = \left[ - {1 \over 2} \sum_{i=1}^n {( - x^2_{i, i +2} \mu^2)^{2 \epsilon} \over (2 \epsilon)^2} + \mbox{BDS}^{(2)}_n (x) + \mbox{Rem}^{(2)}_{n, \rm SYM} (u) + {\cal O}(\epsilon) \right].
\eea
The subscript in the remainder function refers to the fact that it was deduced from the one-loop counterpart in the ${\cal N}=4$ SYM theory.

The pure Chern-Simons contribution is subject to the UV divergence. To regulate the divergence while preserving the supersymmetry, we use the dimensional reduction scheme, $d = (3 - 2 \epsilon)$. The scheme also contributes anomalies to the conformal and special conformal Ward identities. The resulting anomalous Ward identities are \cite{Plefka:2010a}
\bea
\mathbb{D}\big< W [C_n]\big> \Big\vert_{\rm CS} &=& -%\lambda^2
\Log(2) \Big(\sum_{i=1}^n 1 \Big) + {\cal O}(\epsilon) \nonumber \\
\mathbb{K}^m \big< W [C_n]\big> \Big\vert_{\rm CS} &=& -2%\lambda^2
\Log(2) \Big(\sum_{i=1}^n x_i^m \Big) + {\cal O}(\epsilon)
\eea
The full solution to these equations takes the form
\begin{equation}
\big< W_\Box [C_n]\big>^{(2)} \Big\vert_{\rm CS} = - {\Log(2) \over 2} \sum_{i=1}^n \frac{(-x_{i,i+2}^2 \mu^2)^{2\epsilon}}{2\epsilon}+\mbox{Rem}^{(2)}_{n, \rm CS}(u) +\mathcal{O}(\epsilon). \label{Wilson-anomalous}
\end{equation}
\color{black}
%\color{red} IS THIS CORRECT? DON'T WE HAVE -N/2* LOG(2) IN THE LAST EXPRESSION? \color{blue} This is correct answer. This is not obtained by explicit calculation. It is, rather than, expectations from anomalous Ward identity. Hence, up to here we CANNOT anticipate appearance of $R_{CS}-\frac{n}{2}$Log(2), such things.\color{black}
%where $u$'s denote the conformal cross-ratios of the $n$-sided polygon.

For the tetragon Wilson loop, $n=4$, the two-loop result was computed in \cite{Plefka:2010a}. The Chern-Simons contribution in (\ref{Wilson-anomalous}) is absorbable to the matter-dependent part by redefining the UV regularization scale $\mu$. Remarkably, the final result coincides with the one loop result in $\mathcal{N}=4$ SYM theory. Explicitly, the matter-dependent contribution and the ladder plus triple-vertex contribution take the form \cite{Bianchi:2013a}
\bea
\big< W_\Box [C_4]\big>^{(2)}\Big\vert_{\rm matter}&=&
- \frac{(-x_{13}^2 4\pi e^{\gamma_E}\mu^2)^{2\epsilon}}{(2\epsilon)^2} - \frac{(-x_{24}^2 4\pi e^{\gamma_E}\mu^2)^{2\epsilon}}{(2\epsilon)^2}
+ {1 \over 2} \Log^2 \Big(\frac{x_{13}^2}{x_{24}^2} \Big) + \mbox{Rem}_{4}^{(2)} (u) \Big\vert_{\rm matter}
%+\mathcal{O}(\epsilon) \qquad
\label{4pt_matter}\\
\big< W_\Box [C_4]\big>^{(2)}\Big\vert_{\rm CS}\ \ \ &=&
- {\Log(2) \over 2} \sum_{i=1}^4 \frac{(-x_{i,i+2} \pi e^{\gamma_E} \mu^2)^{2\epsilon}}{2\epsilon}
+ \mbox{Rem}_{4}^{(2)} (u) \Big\vert_{\rm CS}
\label{4pt_CS}
\eea
Hereafter, we denote $\mbox{Rem}_{n,\rm{matter}}^{(2)} (u)$ for the IR finite part of $\big< W_\Box [C_n]\big>^{(2)}_{\rm matter}$ modulo the BDS finite part. Also, $\mbox{Rem}_{n,\rm{CS}}^{(2)} (u)$ is the IR finite part of $\big< W_\Box [C_n]\big>^{(2)}_{\rm CS}$. For the tetragon Wilson loop, $n=4$, these numerical constants are given by
\bea
&& \mbox{Rem}_{4}^{(2)} (u) \Big\vert_{\rm matter} =  {\pi^2 \over 4} \nonumber \\
&& \mbox{Rem}_{4}^{(2)} (u) \Big\vert_{\rm CS} = \frac{5 \pi^2}{12}  -2 \mbox{Log}^2 (2).
\label{constants}
\eea
%
%Here,  $\alpha_4$ is a calculable numerical constant. \color{red} CITE. WHY DO WE DECOMPOSE THE CS FINITE PART INTO ALPHA4 AND OTHER NUMBERS?? WHY IS IT NATURAL? \color{black}
%A careful application of the dimensional reduction scheme revealed that \cite{Bianchi:2013a}
%
%\bea
%\alpha_4 = - 2 \log(2) -  2\log^2(2) +\frac{1}{6}\pi^2.
%\eea
%
Finally, the two contributions,  (\ref{4pt_matter}) and (\ref{4pt_CS}), can be combined to the following compact form
for the ABJM theory
\bea
\big< W_\Box [C_4]\big>^{(2)}_{\rm ABJM}=
-\frac{(-x_{13}^2 \tilde\mu^{2})^{2\epsilon}}{(2\epsilon)^2}
-\frac{(-x_{24}^2 \tilde\mu^{2})^{2\epsilon}}{(2\epsilon)^2}
+{1 \over 2}\mbox{Log}^2 \left(\frac{x_{13}^2}{x_{24}^2} \right)
+\mbox{Rem}^{(2)}_4 (u)
+\mathcal{O}(\epsilon). \label{4pt_ABJM}
\eea
Here, $\tilde{\mu}$ is the uniformizing UV regulator scale related to $\mu$ by
\bea
\widetilde\mu^{2}=8\pi e^{\gamma_E} \mu^2.
\label{regulator}
\eea
The remainder function $\mbox{Rem}^{(2)}_4(u)$ is
\bea
\mbox{Rem}_4^{(2)} (u) &=& \mbox{Rem}_4(u) \Big\vert_{\rm matter} + \mbox{Rem}_4 (u) \Big\vert_{\rm CS} + 5 \Log^2(2) \nonumber \\
%5 \Log^2(2)+ {1 \over 2}\pi^2 +  \alpha_4 \\
%
&=&  + 3 \Log^2(2) +\frac{2\pi^2}{3}.
\label{res_tet}
\eea
The last term in the first line is from the uniformization (\ref{regulator}) of the regulator scale.
The remainder function is independent of the conformal cross-ratios $u$'s, much the same way as the one-loop result in the ${\cal N}=4$ SYM theory. Moreover, it displays the uniform transcendentality property. \color{black}

\section{Hexagon Wilson Loops at Two Loops}
Our goal in this paper is to obtain the remainder function $\mbox{Rem}^{(2)}_n(u)$ in (\ref{Wilson-anomalous}) for general $n \ge 6$. For later convenience, we decompose the second-order corrections to the Wilson loop expectation value as
\bea
\big< W_\Box [C_n]\big>^{(2)}_{\rm ABJM} &=&
\left[ \big< W_\Box [C_n]\big>^{(2)}_{\rm matter}+\big< W_\Box [C_n]\big>^{(2)}_{\rm ladder}+\big< W_\Box [C_n]\big>^{(2)}_{\rm vertex}\right]_{\widetilde{\mu}}
\nonumber \\
&=& \left[ \big< W_\Box [C_n]\big>^{(1)}_{\mathcal{N} = 4 \ \rm SYM} \right]_{\rm BDS} +\mbox{Rem}^{(2)}_n \ .
\label{wilson-dec}
\eea
In the second line, we related the functional form of the ABJM Wilson loop expectation value to that of the ${\cal N}=4$ SYM Wilson loop expectation value. The BDS part is abelian, so it must be that both are the same. The remainder function is theory specific. In ABJM theory,
$\mbox{Rem}_n$ is related by%it contains two sources:
\begin{equation}
\mbox{Rem}^{(2)}_n := \mbox{Rem}^{(2)}_{n}\Big\vert_{\rm{matter}}+\mbox{Rem}^{(2)}_{n}\Big\vert_{\rm CS} + \frac{5}{4}n \Log^2(2).
\end{equation}
The last term constant originated from uniformizing the UV regulator scale as in (\ref{regulator}).  The contribution ${\rm Rem}^{(2)}_{n, \rm CS}$ is computationally most complicated.
%
%it contains two sources:
%\begin{equation} \mbox{Rem}_n(u) := \mbox{Rem}_{n, \mu} +\mbox{Rem}_{n, \rm CS}(u).\end{equation}
%$R_{n, \mu}$ is the constant that came from adjusting the UV regularization energy scale $\mu$.  Calculationally, ${\rm Rem}_{n, \rm CS}(u)$ is the most complicated.
%

Our first task is to compute ${\rm Rem}^{(2)}_{n}(u)$ for $n=6$ analytically. For $n > 6$, we will determine ${\rm Rem}^{(2)}_n(u)$ using recursion relations that we will derive in section 7 from soft-collinear factorization of the light-like Wilson loop and analytic result for $n=6$ as an input.
\vskip0.5cm
\begin{center}
\bigskip
\fcolorbox{white}{white}{
  \begin{picture}(88,76) (76,-58)
    \SetWidth{2.0}
    \SetColor{Black}
    \Line(96,16)(144,16)
    \Line(96,16)(78,-20)
    \Line(78,-20)(96,-56)
    \Line(96,-56)(144,-56)
    \Line(144,-56)(162,-20)
    \Line(162,-20)(144,16)
    \SetWidth{1.0}
    \PhotonArc[clock](82.875,-45.688)(50.196,81.84,-11.856){4.5}{6.5}
    \PhotonArc(69,7)(55.154,-67.62,-22.38){4.5}{3.5}
  \end{picture}
}
\fcolorbox{white}{white}{
  \begin{picture}(88,76) (76,-58)
    \SetWidth{2.0}
    \SetColor{Black}
    \Line(96,16)(144,16)
    \Line(96,16)(78,-20)
    \Line(78,-20)(96,-56)
    \Line(96,-56)(144,-56)
    \Line(144,-56)(162,-20)
    \Line(162,-20)(144,16)
    \SetWidth{1.0}
    \PhotonArc(208.5,142.75)(182.466,-130.499,-106.722){4.5}{6.5}
    \PhotonArc(45,43)(97.949,-62.65,-40.03){4.5}{2.5}
  \end{picture}
}
\fcolorbox{white}{white}{
  \begin{picture}(88,76) (76,-58)
    \SetWidth{2.0}
    \SetColor{Black}
    \Line(96,16)(144,16)
    \Line(96,16)(78,-20)
    \Line(78,-20)(96,-56)
    \Line(96,-56)(144,-56)
    \Line(144,-56)(162,-20)
    \Line(162,-20)(144,16)
    \SetWidth{1.0}
    \PhotonArc(208.5,142.75)(182.466,-130.499,-106.722){4.5}{6.5}
    \PhotonArc[clock](132,-44)(26.833,-153.435,-243.435){4.5}{3.5}
  \end{picture}
}
\fcolorbox{white}{white}{
  \begin{picture}(88,76) (76,-58)
    \SetWidth{2.0}
    \SetColor{Black}
    \Line(96,16)(144,16)
    \Line(96,16)(78,-20)
    \Line(78,-20)(96,-56)
    \Line(96,-56)(144,-56)
    \Line(144,-56)(162,-20)
    \Line(162,-20)(144,16)
    \SetWidth{1.0}
    \PhotonArc(120,10)(30.594,-168.69,-11.31){4.5}{6.5}
    \Photon(120,-26)(120,-56){4.5}{3}
  \end{picture}
}
\\
(a) \quad \quad \quad \quad \quad \quad \quad \quad \quad (b) \quad \quad \quad \quad \quad \quad \quad \quad \quad (c) \quad \quad \quad \quad \quad \quad \quad \quad \quad (d)
\\
\bigskip
\fcolorbox{white}{white}{
  \begin{picture}(88,76) (76,-58)
    \SetWidth{2.0}
    \SetColor{Black}
    \Line(96,16)(144,16)
    \Line(96,16)(78,-20)
    \Line(78,-20)(96,-56)
    \Line(96,-56)(144,-56)
    \Line(144,-56)(162,-20)
    \Line(162,-20)(144,16)
    \SetWidth{1.0}
    \Photon(84,-8)(150,-44){4.5}{6}
    \Photon(90,4)(156,-32){4.5}{6}
  \end{picture}
}
\fcolorbox{white}{white}{
  \begin{picture}(88,76) (76,-58)
    \SetWidth{2.0}
    \SetColor{Black}
    \Line(96,16)(144,16)
    \Line(96,16)(78,-20)
    \Line(78,-20)(96,-56)
    \Line(96,-56)(144,-56)
    \Line(144,-56)(162,-20)
    \Line(162,-20)(144,16)
    \SetWidth{1.0}
    \Photon(84,-8)(150,-44){4.5}{6}
    \Photon(90,4)(150,4){4.5}{5}
  \end{picture}
}
\fcolorbox{white}{white}{
  \begin{picture}(88,76) (76,-58)
    \SetWidth{2.0}
    \SetColor{Black}
    \Line(96,16)(144,16)
    \Line(96,16)(78,-20)
    \Line(78,-20)(96,-56)
    \Line(96,-56)(144,-56)
    \Line(144,-56)(162,-20)
    \Line(162,-20)(144,16)
    \SetWidth{1.0}
    \Photon(84,-8)(126,-56){4.5}{5}
    \Photon(90,4)(156,-32){4.5}{6}
  \end{picture}
}
\\
(e) \quad \quad \quad \quad \quad \quad \quad \quad \quad (f) \quad \quad \quad \quad \quad \quad \quad \quad \quad (g)
\\
\bigskip
\fcolorbox{white}{white}{
  \begin{picture}(88,76) (76,-58)
    \SetWidth{2.0}
    \SetColor{Black}
    \Line(96,16)(144,16)
    \Line(96,16)(78,-20)
    \Line(78,-20)(96,-56)
    \Line(96,-56)(144,-56)
    \Line(144,-56)(162,-20)
    \Line(162,-20)(144,16)
    \SetWidth{1.0}
    \Photon(84,-8)(126,-56){4.5}{5}
    \Photon(90,4)(150,4){4.5}{5}
  \end{picture}
}
\fcolorbox{white}{white}{
  \begin{picture}(88,76) (76,-58)
    \SetWidth{2.0}
    \SetColor{Black}
    \Line(96,16)(144,16)
    \Line(96,16)(78,-20)
    \Line(78,-20)(96,-56)
    \Line(96,-56)(144,-56)
    \Line(144,-56)(162,-20)
    \Line(162,-20)(144,16)
    \SetWidth{1.0}
    \Photon(84,-8)(120,-56){4.5}{5}
    \Photon(90,4)(138,-56){4.5}{6}
  \end{picture}
}
\fcolorbox{white}{white}{
  \begin{picture}(88,76) (76,-58)
    \SetWidth{2.0}
    \SetColor{Black}
    \Line(96,16)(144,16)
    \Line(96,16)(78,-20)
    \Line(78,-20)(96,-56)
    \Line(96,-56)(144,-56)
    \Line(144,-56)(162,-20)
    \Line(162,-20)(144,16)
    \SetWidth{1.0}
    \Photon(90,-44)(150,-44){4.5}{5}
    \Photon(90,4)(150,4){4.5}{5}
  \end{picture}
}
\\
(h) \quad \quad \quad \quad \quad \quad \quad \quad \quad (i) \quad \quad \quad \quad \quad \quad \quad \quad \quad (j)
\\
\bigskip
\fcolorbox{white}{white}{
  \begin{picture}(88,76) (76,-58)
    \SetWidth{2.0}
    \SetColor{Black}
    \Line(96,16)(144,16)
    \Line(96,16)(78,-20)
    \Line(78,-20)(96,-56)
    \Line(96,-56)(144,-56)
    \Line(144,-56)(162,-20)
    \Line(162,-20)(144,16)
    \SetWidth{1.0}
    \PhotonArc[clock](85,-21)(23.537,77.735,-77.735){4.5}{6.5}
    \Vertex(108,-20){6}
  \end{picture}
}
\fcolorbox{white}{white}{
  \begin{picture}(88,76) (76,-58)
    \SetWidth{2.0}
    \SetColor{Black}
    \Line(96,16)(144,16)
    \Line(96,16)(78,-20)
    \Line(78,-20)(96,-56)
    \Line(96,-56)(144,-56)
    \Line(144,-56)(162,-20)
    \Line(162,-20)(144,16)
    \SetWidth{1.0}
    \PhotonArc[clock](-400.333,-331)(593.845,34.341,27.586){4.5}{7.5}
    \Vertex(108,-26){6}
  \end{picture}
}
\fcolorbox{white}{white}{
  \begin{picture}(88,76) (76,-58)
    \SetWidth{2.0}
    \SetColor{Black}
    \Line(96,16)(144,16)
    \Line(96,16)(78,-20)
    \Line(78,-20)(96,-56)
    \Line(96,-56)(144,-56)
    \Line(144,-56)(162,-20)
    \Line(162,-20)(144,16)
    \SetWidth{1.0}
    \PhotonArc(171.857,75.571)(108.734,-138.835,-98.386){4.5}{7.5}
    \Vertex(120,-20){6}
  \end{picture}
}
\\
(k) \quad \quad \quad \quad \quad \quad \quad \quad \quad (l) \quad \quad \quad \quad \quad \quad \quad \quad \quad (m)
\\
\end{center}
\noindent
Figure 1 :
{\sl
The Feynman diagrams contributing to the lightlike hexagon Wilson loop consist of (a) $\sim$ (m) and cyclic permutations of the six edges. We classify them by (a)  $\sim$  (d) as triple-vertex contributions, (e)  $\sim$ (j) as ladder contributions, and (k)  $\sim$ (m) as matter contributions.}
\rm
\vskip0.5cm

It turns out the anomalous conformal Ward identities demand that the Wilson loop expectation value must take the form Eq.(\ref{wilson-dec}). Here, we want to determine the remainder function $\mbox{Rem}_n$ in Eq.(\ref{wilson-dec}). To this end, we evaluate all contributing Feynman diagrams to two loop orders. We shall regularize the UV divergences in the dimensional regularization $d = (3 - 2 \epsilon)$ and adopt the dimensional reduction scheme DRED that treats the Levi-Civita symbol $\epsilon_{mnp}$ as 3-dimensional tensor while all others as $d$-dimensional tensors.

In Figure 1, we display the relevant diagrams. The complete list of the contributing diagrams include them and their cyclic permutations with respect to the hexagon edges. For foregoing discussions, we classify the diagrams in Figure 1 into three groups: triple-vertex diagrams for (a)-(d), ladder diagrams for (e)-(j), and matter-dependent diagrams (k)-(m). Computationally, we found that the triple-vertex diagrams the most complex. All of them involve the gauge field propagator ${(\Delta_{mn})}(x,y)$. We take the Landau gauge. In this gauge, the tree-level gauge field propagator is parity-odd and is given in position space by
\bea
{(\Delta_{mn})}^{(0)} (x,y)= {\lambda \over N} \ \mathbb{I} \otimes \mathbb{I} \ Z_{\rm o} \frac{\epsilon_{mnp} (x-y)^p}{[(x-y)^2]^{\frac{d}{2}}} \qquad
\mbox{where} \qquad Z_{\rm o} =
\pi^{(2-d)/2} \Gamma(d/2). \label{gluon_CS}
\eea
For derivation, see Appendix B. In the rest of this section, we present integral expressions of each group.

%%%%%%%%%%%%%%%%%%%%%%%%%%%%%%%%%%%%%
\subsection{Matter Contribution}
For the diagrams (k)-(m) in Figure 1, it suffices to first consider the self-energy of the gauge fields. At one-loop, the gauge field propagators receive corrections from vacuum polarization of matter fields. The one-loop corrected self energy is equal to the tree-level gauge field propagator in $\mathcal{N}=4$ SYM\cite{Rey:2008a, Plefka:2008a} In position space, the one-loop corrected gauge field propagator ${\Delta}^{(1)}_{mn}(x,y)$ is parity-even and takes the form
\bea
{\Delta}^{(1)}_{mn}(x,y)=- \frac{\lambda^2}{N} \ \mathbb{I} \otimes \mathbb{I} \ Z_{\rm e} \frac{g_{mn}}{((x-y)^2)^{d-2}} \qquad
\mbox{where} \qquad
Z_{\rm e} = \pi^{2-d} \Gamma^2 \Big(d/2 -1 \Big).
\label{gluonSYM}
\eea
See Appendix C for derivation and physical interpretation.

The matter contribution is computable parallel to the leading-order in the $\mathcal{N}=4$ SYM theory, except replacing the propagator with ${\Delta}^{(1)}_{mn}(x,y)$ in (\ref{gluonSYM}):
\bea
\big< W_\Box [C_n]\big>^{(2)}_{\rm matter}&=& {1 \over \lambda^2} {1 \over N} \mbox{Tr} {\cal P} \oint \rmd x_i^m \oint \rmd x_j^n \big( i^2 {\Delta}^{(1)}_{mn}(z_i,z_j)\big) \nonumber \\
&=&  \Big((4\pi e^{\gamma_E})^{2\epsilon}+\frac{\pi^2}{2}\epsilon^2+\mathcal{O}(\epsilon^3) \Big) \sum_{i>j = 1}^n I_{ij} \label{matter_expression}
\eea
Here, $I_{ij}$ is the integral of one gauge boson exchange between edges $i, j$ along the contour $C_n$:
\bea
I_{ij} (x) =\int_0^1 \rmd \tau_i \int_0^1 \rmd \tau_j \frac{y_i \cdot y_j}{[(z_i-z_j)^2]^{d-2}} \qquad (i,j=1,2,\cdots,n). \label{one_loop}
\eea
It is straightforward to evaluate these integrals (\ref{one_loop}), as was done in \cite{Heslop:2007a}. Singular loci of the denominator are where the UV divergences arise and they occur precisely at the cusps, viz. when the gauge propagator connects two points on adjacent edges and approach toward the cusp in between.

For the adjacent diagrams, the integration is straightforward. The leading UV divergence is readily obtained as
\bea
I_{i+1,i} (x) &=& \int \rmd\tau_{i} \int \rmd\tau_{i+1} \ {y_i \cdot y_{i+1} \over [(z_{i+1}-z_{i})^2]^{d-2}}
%&=&   \frac{1}{2} x_{i,i+2}^2 \int \rmd \tau_{i} \int \rmd\tau_{i+1} \frac{1}{[\bar{\tau_i}\tau_{i+1} x_{i,i+2}^2]^{d-2}} \nonumber \\
=   -\frac{1}{2} {(x_{i,i+2}^2)^{2\epsilon} \over (2 \epsilon)^2}.
% + {\cal O}(\epsilon^{-1}).
\eea
%
%Note that the divergent terms in Eq.(\ref{4pt_matter_a}) originates from this adjacent connected part.
%
Non-adjacent diagrams are UV finite. Summing them over all possible distinct permutations, we obtain the so-called the BDS function $\mbox{BDS}_n^{(2)}$:
\bea
\boxed{
\mbox{BDS}_n^{(2)}(x) \equiv \sum_{i>j+1}^n I_{ij} (x).
}
\eea
These integrals can be evaluated analytically, as was done in \cite{Heslop:2007a}:
\bea
I_{ij}(x) = {1 \over 2} \left[-\mbox{Li}_2(1-as)-\mbox{Li}_2(1-at)+\mbox{Li}_2(1-aP^2)+\mbox{Li}_2(1-aQ^2) \right]_{ij}.
 \label{mat_a}
\eea
%
%\color{red} IS THIS ONLY FOR TETRAGON? IF NOT, THERE SHOULD BE SUM OVER i, j INDICES ETC. \color{blue} I agree. LHS should be exchanged to $I_{ij}(x)$ (It was $\mbox{BDS}^{(2)}_n (x)$). \color{black}
%In the right-hand side, the parameter $a$ defined by
%
%\bea
%{(1-as)(1-at) \over [(1-aP^2)(1-aQ^2)]} =1.
%\eea
%
%\color{red} WHERE IS THIS RELATION FROM? \color{blue} It came from explicit diagram calculation.
Here, the parameter $a$  is given by \cite{Heslop:2007a}
%satisfies the relation
%
%\bea
%{(1-as)(1-at) \over [(1-aP^2)(1-aQ^2)]} =1.
%\eea
%
%\color{red} WHERE IS THIS RELATION FROM? \color{black}
%The solution reads
%
\bea
a=\frac{s+t-P^2-Q^2}{st-P^2 Q^2}, \quad \mbox{where} \quad P^2=x_{i,j+1}^2, \quad Q^2=x_{i+1,j}^2, \quad s=x_{i,j}^2, \quad t=x_{i+1,j+1}^2.
\label{mat_b}
\eea
%
%
%Extending the BDS conjecture to the $(2+1)$-dimensional ABJM theory,  we find that $\sum_{i>j} I_{ij}(x) $ is given by
%\begin{equation}
%\sum_{i>j} I_{ij}(x) = -{1 \over 2} \sum_{\substack{i=1}}^n \frac{1}{(2\epsilon)^2}(x_{i,i+2}^2)^{2\epsilon} +\mbox{BDS}_n^{(2)} (x).
%\end{equation}
Combining this with Eq.(\ref{matter_expression}), it follows that the matter contribution to the Wilson loop expectation value is given by
\bea
\boxed{
\big< W_\Box[C_n]\big>^{(2)}_{\rm matter}
%&=&  \Big((4\pi e^{\gamma_E})^{2\epsilon}+\frac{\pi^2}{2}\epsilon^2+\mathcal{O}(\epsilon^3) \Big)
%\Big( - \frac{1}{2}\sum_{i=1}^n  {(x_{i,i+2}^2 \mu^2)^{2\epsilon} \over (2\epsilon)^2} +\mbox{BDS}_n^{(2)} \Big) \nonumber \\
= - \frac{1}{2}\sum_{i=1}^n
{(x_{i,i+2}^2 4\pi e^{\gamma_E} \mu^2)^{2\epsilon} \over (2\epsilon)^2}
+\mbox{BDS}^{(2)}_n(x)
+ \mbox{Rem}^{(2)}_ n(u)\Big\vert_{\rm matter}
+{\cal O}(\epsilon).}
\label{matterWvev}
\eea
Here, the matter contribution to the remainder function is given by
\bea
\boxed{
\mbox{Rem}_n^{(2)}(u) \Big\vert_{\rm matter} = -\frac{1}{16}n \pi^2 .
}
\label{remainder-matter}
\eea
For the special case of $n=4$, this result reproduces (\ref{4pt_matter}) and the remainder function (\ref{constants}).
%%%%%%%%%%%%%%%%%%%%%%%%%%%%%%%%%%%%
\subsection{Gauge Boson Ladder diagram}
The pure Chern-Simons term generates ladder diagrams and triple-vetex diagrams. The ladder diagram contributes to the Wilson loop expectation value as
\begin{equation}
\big< W_\Box[C_n]\big>^{(2)}_{\rm ladder}= \left(\frac{\Gamma\big(\frac{d}{2}\big)}{\pi^{\frac{d-2}{2}}} \right)^2 \sum_{{\cal P}(i,j,k,l)}I_{\rm ladder}(x)
\end{equation}
Here, ${\cal P}(i,j,k,l)$ refers to sum over path-ordered, pairwise connections among the four segments $(i,j,k, l)$ and the $I_{\rm ladder}$  integral is given by
\begin{equation}
I_{\rm ladder}^{\{i,j,k,l\}}=\int \rmd \tau_i \cdots \int \rmd\tau_{l} \frac{\epsilon(y_i,y_l,z_i-z_l)}{[(z_i-z_l)^2]^{\frac{d}{2}}} \frac{\epsilon(y_j,y_k,z_j-z_k)}{[(z_j-z_k)^2]^{\frac{d}{2}}},  \label{Ladder}
\end{equation}
where the superscript $\{i,j,k,l\}$ labels the edges that the gauge field is attached. For instance, for the hexagon, the six configurations
\begin{equation}
\{i,j,k,l\} = \{4,4,1,1\} ,\{5,4,1,1\} ,\{4,3,1,1\} ,\{5,3,1,1\} ,\{3,3,1,1\} , \{5,4,2,1\}
\end{equation}
and their cyclic permutations should be summed over . Importantly, these ladder diagrams are all UV finite. Explicit form of the integrals are tabulated in Appendix E.
%%%%%%%%%%%%%%%%%%%%%%%%%%%%%%%%%%%%%
\subsection{Triple-Vertex Diagram}
 The triple-vertex diagrams are reduced to tensor integrals involving the Levi-Civita tensor $\varepsilon_{mnp}$. We deal with such tensor integrals by reducing them to scalar integrals via the relations
\begin{equation}
I^{mnp}(x,y,z)={\partial \over \partial y^n} {\partial \over \partial z^p} I^m(y-x,z-x), \label{trick}
\end{equation}
where $I^m(y-x,z-x)$ is given by
\begin{equation}
I^m(a,b)=\int \rmd^dw \frac{w^m}{|w|^d |w-(y-x)|^d |w-(z-x)|^d} \ .
\end{equation}
Contracting the Levi-Civita tensors with the segment vectors of the polygon, one obtains integrals in readily evaluatable forms.
%, for example, (\ref{321ex}).

Triple-vertex diagram contributes to the Wilson loop expectation value as
\bea
\big< W_\Box[C_n]\big>^{(2)}_{\rm vertex}= \frac{i}{2\pi} \Big(\frac{\Gamma\big(\frac{d}{2} \big)}{\pi^{\frac{d-2}{2}}} \Big)^3 \sum_{\tiny{\mbox{path-ordered}}} I^{\{i,j,k\}}_{\rm vertex}
\eea
Here again, the superscript $\{i,j,k\}$ labels the edges where the gauge field is attached. The path ordering restricts $i>j>k$ case only. In self-explaining notation, the integral takes the form
\bea
I_{\rm vertex}^{\{i,j,k\}} (x) =\int_{\mathbb{R}^{2,1}} \rmd^d w \left[ \int \cdots \int \rmd \tau_i \rmd \tau_j \rmd \tau_k \epsilon^{abc}  \frac{\epsilon(y_i, a, w-z_i) \epsilon(y_j, b, w-z_j) \epsilon(y_k, b, w-z_k)}{[(w-z_i)^2]^{\frac{d}{2}} [(w-z_j)^2]^{\frac{d}{2}}[(w-z_k)^2]^{\frac{d}{2}}} \right]. \label{Vertex}
\eea
In the case of hexagon, the four configurations
\bea
\{i,j,k\} = \{3,2,1\}, \{4,2,1\}, \{4,3,1\}, \{5,3,1\}
\eea
and their cyclic permutation generate all possible diagrams. Among them, divergence appears only through $\{3,2,1\}$-type configuration. For integral expression of the triple-vertex diagrams, see Appendix G.

%%%%%%%%%%%%%%%%%%%%%%
%\subsection{Ultraviolet Divergences}
Note that the triple-vertex diagrams are UV-divergent. These divergences arise from configurations whose three attached points of the gauge bosons  approach a single segment. The $\{3,2,1\}$ diagram is an example of such configuration. After the Mellin-Barnes transformation, the integral $I_{\rm vertex}^{\{3,2,1\}}$ can be brought to a form that can be evaluated in part analytically and in part numerically with high precision.
%
%\bea
%I_{\rm vertex}^{\{3,2,1\}} &=& \kappa \int \rmd \tau_1 \cdots \int \rmd \tau_{3} \int \int \rmd x \rmd y \ (x\bar{x}\bar{y})^{{d \over 2}-1} {(x_{13}^2 x_{24}^2) \over \Delta_y^d}\nonumber \\
%&& \hskip3.3cm \times \Big[(d-2) \Delta_y+2(d-1) x\bar{y} \bar{\tau_1} \tau_3 (x_{13}^2-x_{14}^2+x_{24}^2) \Big]. \label{I_321_a}
%\eea
%
%\color{red}
%WHAT ARE THE BAR NOTATIONS? \color{blue} $\bar{x}$ means $1-x$.
%\color{black}
%Here, $\kappa$ and $\Delta_y$ denote
%
%\bea
%\kappa =
%\frac{i\pi^{\frac{d}{2}}\Gamma(d-1)}{8\Gamma(\frac{d}{2})^3}, \quad \Delta_y= \bar{\tau_1}x (\bar{\tau_3}\bar{y}+\tau_2\bar{x}y)  x_{13}^2+\tau_3 \bar{y}  (\bar{\tau_2}\bar{x}+\tau_1x) x_{24}^2+\bar{\tau_1}\tau_3  x\bar{y} x_{14}^2
%\eea
%
%We calculated the first term in (\ref{I_321_a}) analytically,  while the second term %$I^{\{3,2,1 \}}_{\mbox{\tiny{Fin}}}$
%numerically.
The result reads
\bea
I_{\rm vertex}^{\{3,2,1\}}(x) =%\kappa
\frac{i\pi^{\frac{d}{2}}\Gamma(d-1)}{8\Gamma(\frac{d}{2})^3}
\Big(4\pi \Log(2) \frac{(x_{13}^2 \mu^2)^{2\epsilon}}{\epsilon} +4\pi \Log(2) \frac{(x_{24}^2 \mu^2)^{2\epsilon}}{\epsilon} + I^{\{3,2,1 \}}_{\rm finite}\Big).
\label{I_321_a}
\eea
Summing over all possible path-ordered triples $(i,j,k)$, we find that
\bea
\sum_{{\cal P}(i,j,k)} I^{\{i,j,k\}}_{\rm vertex}(x) = %\kappa
\frac{i\pi^{\frac{d}{2}}\Gamma(d-1)}{8\Gamma(\frac{d}{2})^3}
\left(8\pi \Log(2) \sum_{i=1}^n \frac{(x_{i,i+2}^2 \mu^2)^{2\epsilon}}{2\epsilon} + I_{\rm finite}\right),
\label{vertex_result_a}
\eea
%
%In the right-hand side, the UV divergent terms are entirely from the $\{3,2,1\}$ diagram and its cyclic permutations. The UV divergence arises whenever all of the three attached points approach a single segment.
The leading UV-divergence is  $\frac{1}{\epsilon}$, in contrast to $\frac{1}{\epsilon^2}$ leading UV-divergence in matter contribution.
%%%%%%%%%%%%%%%%%%%%%%%%%%%%%%%%%%%%%%%

\subsection{Wilson Loop of the Pure Chern-Simons Theory}
In pure Chern-Simons theory, the  contribution $\big< W_\Box[C_n]\big>^{(2)}_{\rm CS}$ is obtained by combining $\big< W_\Box[C_n]\big>^{(2)}_{\rm ladder}$ and $\big< W_\Box[C_n]\big>^{(2)}_{\rm vertex}$. To evaluate these expectation values, we carry out tensor integral $\sum_{i>j>k>l} I_{i,j,k,l}^{\rm ladder}$ and $\sum_{i>j>k} I_{i,j,k}^{\rm vertex}$.
%\color{red} WHY IS THE FIRST SUM ONLY OVER TWO EDGES? ISN'T IT OVER FOUR EDGES? \color{blue} Errata. I fixed it. \color{black}
%However, both objects themselves do not agree to $\big< W_\Box[C_n]\big>^{(2)}_{\rm ladder}$ and $\big< W_\Box[C_n]\big>^{(2)}_{\rm vertex}$, completely. To match with them, we should be dressed with prefactors $\Big(\frac{\Gamma\big(\frac{d}{2}\big)}{\pi^{\frac{d-2}{2}}} \Big)^2$ and $\frac{i}{2\pi} \Big(\frac{\Gamma\big(\frac{d}{2} \big)}{\pi^{\frac{d-2}{2}}} \Big)^3$ in front of the tensor integrals.
%
The result is
\bea
\big< W_\Box[C_n]\big>^{(2)}_{\rm CS} &=&
\big< W_\Box[C_n]\big>^{(2)}_{\rm ladder} + \big< W_\Box[C_n]\big>^{(2)}_{\rm vertex} \nonumber \\
&=& \Big(\frac{\Gamma\big(\frac{d}{2}\big)}{\pi^{\frac{d-2}{2}}} \Big)^2 \sum_{i>j=1}^n I^{i,j}_{\rm ladder} + \frac{i}{2\pi} \Big(\frac{\Gamma\big(\frac{d}{2} \big)}{\pi^{\frac{d-2}{2}}} \Big)^3 \sum_{i>j>k=1}^n I^{i,j,k}_{\rm vertex}.
\eea
Inserting (\ref{vertex_result_a}), we finally obtain
\bea
\big< W_\Box[C_n]\big>^{(2)}_{\rm CS}&= &
- \Big(\frac{\Gamma(d-1)}{2} \frac{\pi^{2-d}}{8}\Big) \Big(4\pi \log(2) \sum_{i=1}^n \frac{(x_{i,i+2}^2 \mu^2)^{2\epsilon}}{\epsilon} +  I_{\rm vertex}^{\rm finite} \Big) +  \Big(\frac{\Gamma\big(\frac{d}{2}\big)}{\pi^{\frac{d-2}{2}}} \Big)^2 \sum_{i>j>k>l} I^{i,j,k,l}_{\rm ladder} \nonumber \\
&=&
- \Big(\frac{\Gamma(d-1)}{2} \frac{\pi^{2-d}}{8}\Big) \Big(4\pi \log(2) \sum_{i=1}^n \frac{(x_{i,i+2}^2 \mu^2)^{2\epsilon}}{\epsilon}  \Big) + I_{\rm CS} + \mathcal{O}(\epsilon) \nonumber \\
&=&
-  {\log(2) \over 2} \sum_{i=1}^n \frac{(x_{i,i+2}^2 \pi e^{\gamma_E} \mu^2)^{2\epsilon}}{2\epsilon} +{\rm Rem}^{(2)}_{n, \rm CS}(u) + \mathcal{O}(\epsilon)  \label{CSWvev}
\eea
%
%In last expression, the numerical factor $\frac{n}{2} \mbox{Log}(2)$ term arose from rescaling the UV scale $\mu^2$ to $\tilde\mu^2 = \pi e^{\gamma_E} \mu^2$. Recall that the rescaling was necessary to combine the UV-divergent contributions into the form of the corresponding contributions in ${\cal N}=4$ SYM theory. Denote this constant as ${R}_{n, \mu}$.
In second line, we used the fact that $I_{\rm vertex}^{\rm finite}$ and $\sum_{i>j>k>l} I^{i,j,k,l}_{\rm ladder}$ are finite quantity. For convenience, we defined here
\bea
I_{\rm CS} = - \frac{1}{16\pi}  I_{\rm vertex}^{\rm finite} + \frac{1}{4} \sum_{i>j>k>l} I^{i,j,k,l}_{\rm ladder}.
\eea
Explicit expansion of the last line in (\ref{CSWvev}) yields relation between $I_{\rm CS}$ and ${\rm Rem}_{n, \rm CS}(u)$:
\bea
{\rm Rem}^{(2)}_{n, \rm CS}(u) = I_{\rm CS} + \frac{n}{2} \Log(2).
\eea
We will evaluate $I_{\rm CS}$ numerically. Before proceeding, we will need to digress to general consideration of free kinematic variables in light-like polygon, viz. the moduli space of light-like polygon. For the result of the remainder function, the reader may skip to the end of section 7.

%\color{red}
%THE ABOVE PARAGRAPH SOUNDS VERY COMPLICATED, EASILY LOST WHILE READING. WE SHOULD BETTER ORGANIZE VARIOUS TERMS IN THE FINITE PARTS.
%\color{black}
%Inserting (\ref{vertex_result_a}), we finally obtain
%\bea
%\big< W_\Box[C_n]\big>^{(2)}_{\rm CS}&= &- \Big(\frac{\Gamma(d-1)}{2} \frac{\pi^{2-d}}{8}\Big) \Big(4\pi \log(2) \sum_{i=1}^n \frac{(x_{i,i+2}^2 \mu^2)^{2\epsilon}}{\epsilon} + I_{\rm finite} \Big) +  \Big(\frac{\Gamma\big(\frac{d}{2}\big)}{\pi^{\frac{d-2}{2}}} \Big)^2 \sum_{i>j} I^{i,j}_{\rm ladder} \nonumber \\&=&   -  {\log(2) \over 2} \sum_{i=1}^n \frac{(x_{i,i+2}^2 \pi e^{\gamma_E} \mu^2)^{2\epsilon}}{2\epsilon} +{\rm Rem}_{n, \rm CS}(x) +\frac{n}{2} \log(2).  \label{CSWvev}
%\eea
%
%In last expression, the numerical factor $\frac{n}{2} \mbox{Log}(2)$ term arose from rescaling the UV scale $\mu^2$ to $\tilde\mu^2 = \pi e^{\gamma_E} \mu^2$. Recall that the rescaling was necessary to combine the UV-divergent contributions into the form of the corresponding contributions in ${\cal N}=4$ SYM theory. Denote this constant as ${R}_{n, \mu}$. We will evaluate the other contributions $I_{\rm finite}$ and $\sum_{i>j} I^{i,j}_{\rm ladder}$ numerically. Summing the two contributions, we obtain $R_{n, \rm CS}$. In the next section, we will provide numerical value of this $R_{n, \rm CS}$.
%\color{red}
%THE ABOVE PARAGRAPH SOUNDS VERY COMPLICATED, EASILY LOST WHILE READING. WE SHOULD BETTER ORGANIZE VARIOUS TERMS IN THE FINITE PARTS.
%\color{black}

%%%%%%%%%%%%%%%%%%%%%%%%%%%%%%%%%%%%
\section{Euclid, Mandelstam and Gram}
The first step in evaluating the remainder function is to specify the geometry of lightlike polygon. We shall call it the kinematics.
In this section, we present general considerations of the moduli space of a lightlike $n$-gon $C_n$.
\subsection{Moduli Space of Lightlike Polygon}
The contour $C_n$ is specified by the set of points $x_1, \cdots, x_n$. They are lightlike separated with adjacent neighbors, and can always be brought to
\bea
x_1 + \cdots + x_n = 0.
\label{translation}
\eea
by translation invariance \footnote{In other words, the center of mass of the polygon $C_n$ can always be put at the origin.}.
Equivalently, $C_n$ can be specified by the segment vectors $y_1, \cdots, y_n$. They are all light-like ($y_i^2=0$), and trivially satisfy the closedness condition
\bea
y_1 + \cdots + y_n = 0.
\label{conservation}
\eea
The two are discrete, polygon counterpart of the statement that a smooth curve can be described either by specifying position vectors of the curve  or by specifying tangent vectors of the curve. Either way, one finds that the moduli space ${\cal M}[C_n]$ of $n$-sided polygon $C_n$ in $d$-dimensional embedding space is given by
\bea
\mbox{dim}{\cal M}[C_n] = (dn - n) - d - {1 \over 2} d(d-1).
\label{modulispacedim}
\eea

The dimension of the moduli space (\ref{modulispacedim}) grows linearly with $n$, the number of $x$'s or $y$'s.
For instance, consider the $n=6$ hexagon. We can specify 6 position vectors, $x_1, \cdots, x_6$ subject to (\ref{translation}). Out of $6 \times 3 =18$ components, light-like conditions $x_{i, i+1}^2 =0$ eliminates 6, (\ref{translation}) eliminates 3 and $so(2,1)$ Lorentz transformation eliminates 3. The remaining 6 independent variables are the moduli of $C_6$.
Alternatively, we can also specify 6 segment vectors $y_1, \cdots, y_6$ subject to (\ref{conservation}). Out of $6 \times 3 = 18$ components,  light-like conditions $y_i^2 = 0$ eliminate 6,  (\ref{conservation}) eliminates 3 and $so(2,1)$ Lorentz transformation eliminates 3. The remaining 6 independent variables are the moduli of $C_6$.

On the other hand, by the Poincar\'e invariance, the lightlike Wilson loops are not functions of $x_i$'s or $y_i$'s themselves, but are functions of the Mandelstam invariants $x_{ij}^2$, $i, j = 1, \cdots, n$. They vanish for $j = i, i \pm 1$, so the net number of nontrivial invariants is given by
\bea
\mbox{dim} M(C_n) =  {1 \over 2} n (n - 3).
\label{mandelstamdimension}
\eea
Alternative choice of the Mandelstam invariants are $y_{ij}^2$. They range over $i,j = 1, \cdots, (n-1)$ because of the closedness condition (\ref{conservation}). They also vanish for $j=i$. Altogether, the net number of nontrivial invariants is given again by (\ref{mandelstamdimension}).
Their number grows quadratically with $n$, so would outgrow the dimension of $n$-gon moduli space (\ref{modulispacedim}). It must be that many of the Mandelstam invariants are redundant.

%Expression of diagrams contains inner product of momentum which equivalent to Mandelstam variables. Our input is these Mandelstam variables $x_{ij}^2$. Due to light-like condition, possible combinations are $\frac{n(n-3)}{2}$. At first sight it seems that distinguished variables are 9 for hexagon regardless of dimension.  However, they are not independent variables. Number of independent variables are given by Lorentz symmetry. In d-dimension, there are d-dimensional n vectors which gives total $d n$ degree of freedom. Lightlike condition restrict n of them and Lorentz rotation and translation restrict $d+\frac{d(d-1)}{2}$. Hence there are $(n -1)d-\frac{d(d-1)}{2}$ variables. For example, 3-dimensional hexagon calculation has independent 6 variables rather than 9 variables.

The projection of the space of Mandelstam invariants to the space of polygon moduli is achieved by the geometric condition that $n$ vectors in $d$ dimensional spacetime are necessarily linearly dependent for $n > d$. To this end, consider the Gram matrix $G$, whose $(i,j)$ entry is given by $y_i \cdot y_j$:
\bea
G \equiv M^{\rm T} \cdot M =
\begin{pmatrix}
y_1 \cdot y_1 && y_1 \cdot y_2 && y_1 \cdot y_3 && \cdots && y_1 \cdot y_n \\
y_2 \cdot y_1 && y_2 \cdot y_2 && y_2 \cdot y_3 && \cdots && y_2 \cdot y_n \\
y_3 \cdot y_1 && y_3 \cdot y_2 && y_3 \cdot y_3 && \cdots && y_3 \cdot y_n \\
 \vdots &&  \vdots && \vdots && \ddots && \vdots \\
y_n \cdot y_1 && y_n \cdot y_2 && y_n \cdot y_3 && \cdots && y_n \cdot y_n
\end{pmatrix}
\eea
Here, $M$ is $(d \times n)$ matrix whose entries are the segment vectors
$ M = (y^m_1, y^m_2, \cdots, {y}^m_n)$.
Determinant of $G$, called Gram determinant, is nothing but the square of the hypercube volume spanned by the segment vectors:
%Since we considering hexagon Wilson loop, $i,j$ runs from $1$ to $6$.
\begin{eqnarray}
 \mbox{Det} G(i, j)
= \vert\vert y_1 \wedge y_2 \wedge y_3 \wedge \cdots \wedge y_n \vert\vert^2.
\end{eqnarray}
Because of the closedness condition (\ref{conservation}), the Gram determinant vanishes identically. Moreover, $d$-dimensional spacetime accommodates at most $d$ many linearly independent vectors. Hence, in Gram matrix, determinant of any $(d+1) \times (d+1)$ sub-matrices ought to vanish identically. There are $(n-d-1)(n-d)/2$ many such choices, so these Gram sub-determinant conditions project the space of Mandelstam variables down to the space of independent scalar invariants of dimension
\bea
\mbox{dim} \Pi_G M (C_n) =
{1 \over 2} n (n - 3) - {1 \over 2} (n - d - 1) (n - d) = (d-1) n - {1 \over 2} d (d + 1).
\label{poincaredof}
\eea
This matches precisely with the dimension of the moduli space of $n$-sided lightlike polygon (\ref{modulispacedim}).

\subsection{Positivity Condition}
%Typically, the Gram sub-determinant conditions are too complex. Hence, it is hard to write down our diagram expressions only with independent variables. At the level of diagram calculation, we pretend there are $\frac{n(n-3)}{2}$ variables. Gram determinant constraint will be applied at the numerical evaluation by considering special kinematics.

In evaluating the lightlike polygon Wilson loop operator expectation value , the input data of $C_n$ are the vectors $x_i$'s or $y_i$'s of the polygon. On the other hand, the expectation value is Poincar\'e invariant, so it must depend on these vectors only through scalar products:
\bea
y_i \cdot y_j = {1 \over 2} \left[ x_{i,j+1}^2+x_{i+1,j}^2-x_{i,j}^2-x_{i+1,j+1}^2 \right].
\eea
This suggests it natural to take the Mandelstam variables as input parameters. This is what we shall do for numerical computations. On the other hand, as we saw above, the Mandelstam variables are not mutually independent and need to be further supplemented by the Gram sub-determinant conditions.
%\color{red}  It would be better insert here relation $2 y_i \cdot y_j = x_{i,j+1}^2+x_{i+1,j}^2-x_{i,j}^2-x_{i+1,j+1}^2$ to give explicit Gram-sub determinant condition represented by Mandelstam variables. \color{black}
A complication is that, typically, the Gram sub-determinant conditions are too involved to solve explicitly.
In evaluating the Feynman loop integrals, we shall employ the Mellin-Barnes transformations. During the evaluation, we shall provisionally assume that the Mandelstam variables are linearly independent until we perform the Mellin-Barnes transformations. We then evaluate the transformed expressions numerically, and at this stage we shall impose the Gram sub-determnant conditions by taking special kinematics of $C_n$ such that it becomes consistent with these conditions.
%For example, there are 9 Mandelstam variables in hexagon. We CANNOT give arbitrary 9-inputs on these Mandelstam variables. Since they are inherited from momentum, they should satisfy special condition : Gram determinant constraint.
%
%$\bullet$ Momentum side $\hspace{1.1mm}$: Suppose Poincare invariance. DOF of momentum vector in $d$ dimension is given by $n(d-1)-\frac{d(d+1)}{2}$. For $d=3$, $n=6$ case, there are 8 DOF, not 9. We expect 1 constraint condition from Gram determinant constraint.
%
%$\bullet$ Mandelstam variable side : However, Gram determinant constraint usually complicated and it is almost impossible to reflect them into integral expression. Alternatively, we considered all Mandelstam variables are free variable until arrives to Mellin-Barnes transformed integral expression.
%
%When insert inputs and obtain numerical result, we restrict ourselves on configuration space.(That means restricted area by Gram determinant condtraint.)
We found numerically that $\mbox{Rem}(u)$ yields physically meaningful values when Mandelstam variables are restricted by the Gram sub-determinant conditions and that, in solving the anomalous conformal Ward identity, the remainder function $\mbox{Rem}(u)$ is expressed in terms of cross ratios only after the Gram sub-determinant conditions are imposed to the Mandelstam variables.

Often, the Mellin-Barnes transformed integrals involve spurious poles. To avoid them, it is necessary
to impose all the Mandelstam variables to have the same sign. We shall call this condition as "positivity
condition". It turns out that, for the edge vectors $y_i$'s, the condition is satisfied by making timelike
components of adjacent edge vectors to have alternating signs. As the edge vectors are subject to the
closedness condition, this condition then implies that only even numbers of edges $n = 2\mathbb{N}$ are permissible. This purely geometric consideration imposes the polygons relevant for the lightlike ABJM Wilson
loops \footnote{The same restriction applies to all three dimensional conformal field theories.} restricted to those with even numbers of the edge. Though we do not have a fully general argument, we think that this is a general geometric condition.

To illustrate this, consider the case of hexagon. A choice of the edge vectors $y_1, \cdots, y_6$ satisfying the positivity condition and the closedness condition $y_1 + \cdots + y_6 = 0$ are
\begin{align}
\begin{array}{ccc}
y_1=(-\sqrt{a^2+b^2},a,b) & y_2=(+\sqrt{c^2+d^2},c,d) & y_3=(-\sqrt{e^2+f^2},e,f) \\
y_4=(+\sqrt{g^2+h^2},g,h) & y_5=(-\sqrt{p^2+q^2},p,q) & y_6=(+\sqrt{r^2+s^2},r,s)
\label{coordsetup}
\end{array}
\end{align}
First, we set the time-component of the edge vectors of alternating sign so that  Mandelstam variables are positive. Take for example the hexagon. Among the nine Mandelstam variables, six variables($x_{13}^2, x_{24}^2, x_{35}^2$, $x_{46}^2, x_{15}^2, x_{26}^2$) are inner product of consecutive segment vector, viz. $2 y_i \cdot y_{i+1}$. Then, for example,
\begin{equation}
x_{1,3}^2 = 2 y_1 \cdot y_2 = \ 2 \sqrt{a^2+b^2} \sqrt{c^2+d^2} + 2 a c + 2 b d.
\end{equation}
By triangle inequality, sign of this Mandelstam variable is determined by the first term, regardless of signature of each parameters. To make this Mandelstam variable positive, we see that the edge vectors must be chosen to have consecutively alternating signs of their time components.

%Moreover, the Euclidean condition requires other 3 Mandelstam variables($x_{14}^2, x_{25}^2, x_{36}^2$) should be also positive. This additional condition gives further constraints on number of independent parameters appearing in (\ref{coordsetup}). Although there are 9 possible Mandelstam variables $x_{i,j}^2$ that made from scalar product $y_i \cdot y_j$, arbitrarily choice of their value may cause inconsistency with both Euclidean and closedness condition. Therefore, that 9 variables are not suitable independent parameter that consistent with Euclidean condition and closedness. Following from counting (\ref{poincaredof}), we expect number of meaningful independent variable is 6, not 9.

%We have numerically checked that the configuration generated from Gram sub-determinant condition always satisfy both Euclidean and closedness condition.  Though lacking a general argument, we think that Euclidean + closedness configuration as equivalent to Gram sub-determinant configuration.

Such kinematical restrictions bear the following geometric implications to the `triple-collinear factorization' we will study in the next section. Recall that, by construction, a lightlike polygon is made of oriented edges which are all lightlike. When we take a polygon and let two non-adjacent vertices $x_i, x_j (j \ne i \pm 1)$ become lightlike, we see we can decompose the lightlike contour of the parent polygon as a sum of two lightlike contours of daughter polygons.
The absence of polygons with odd numbers of the edge also puts the constraint that the factorization must involve even number of consecutive vertices. This condition is also compatible with the requirement that the time component of edge vectors must be sign alternating. We see that such factorization gives rise to a nonlinear recursion relations among the lightlike Wilson loops.

%investigate triple collinear limit rather than (double) collinear limit.

%%%%%%%%%%%%%%%%%%%%%%%%%%%%%%%%%%%%%
\subsection{Moduli Space of Conformal Lightlike Polygon}
Up until now, in counting the moduli space of lightlike polygons, we only took into account the Poincar\'e symmetry of embedding spacetime. We now further endow the polygons with conformal symmetry. Replacing the Poincar\'e symmetry $so(d-1, 1)$ by the conformal symmetry $so(d, 2)$, we see that the dimension of moduli space of conformal lightlike polygons is modified to
\bea
\mbox{dim} {\cal M}_{\rm c} [C_n] = (d -1) n - {1 \over 2} (d+1) (d+2).
\label{dim-conformal-polygon}
\eea
On the other hand, we elaborated in the previous section that the geometry of lightlike polygons is more conveniently described in terms of Mandelstam variables but these variables are not mutually independent. The requisite projection of the Mandelstam variables is the Gram condition. Below, we explain how this can be achieved.

The dimension of parameter space of conformally invariant Mandelstam variables, viz. the conformal cross-ratios
\bea
u_{i,j} = \frac{x_{i,j+1}^2 x_{i+1,j}^2}{x_{i,j}^2 x_{i+1,j+1}^2},
\eea
is given by
\bea
\mbox{dim} M_c [C_n] = {1 \over 2} n (n-1) - n - n = {1 \over 2} n (n - 5),
\label{dim-crossratio}
\eea

\begin{center}
\fcolorbox{white}{white}{
  \begin{picture}(140,130) (69,-27)
    \SetWidth{2.0}
    \SetColor{Black}
    \Line(90,64)(162,76)
    \SetWidth{1.0}
    \Line[dash,dashsize=10](90,64)(90,-8)
    \Line[dash,dashsize=10](90,-8)(162,76)
    \Line[dash,dashsize=10](162,76)(162,-20)
    \Line[dash,dashsize=10](162,-20)(90,64)
    \SetWidth{2.0}
    \Line(90,-8)(162,-20)
    \Text(78,-20)[lb]{\Large{\Black{$x_j$}}}
    \Text(168,-32)[lb]{\Large{\Black{$x_{j+1}$}}}
    \Text(78,70)[lb]{\Large{\Black{$x_i$}}}
    \Text(66,28)[lb]{\Large{\Black{$x_{i,j}^2$}}}
    \Text(168,82)[lb]{\Large{\Black{$x_{i+1}$}}}
    \Text(174,28)[lb]{\Large{\Black{$x_{i+1,j+1}^2$}}}
    \Text(138,4)[lb]{\Large{\Black{$x_{i,j+1}^2$}}}
    \Text(138,46)[lb]{\Large{\Black{$x_{i+1,j}^2$}}}
  \end{picture}
}
\quad \quad
\fcolorbox{white}{white}{
  \begin{picture}(140,130) (69,-27)
    \SetWidth{2.0}
    \SetColor{Black}
    \Line(90,64)(162,76)
    \SetWidth{1.0}
    \Line[dash,dashsize=10](90,-8)(162,76)
    \Line[dash,dashsize=10](162,76)(162,-20)
    \Line[dash,dashsize=10](162,-20)(90,64)
    \SetWidth{2.0}
    \Line(90,-8)(162,-20)
    \Text(78,-20)[lb]{\Large{\Black{$x_j$}}}
    \Text(168,-32)[lb]{\Large{\Black{$x_{j+1}$}}}
    \Text(78,70)[lb]{\Large{\Black{$x_i$}}}
    \Text(66,28)[lb]{\Large{\Black{$x_{i,j}^2$}}}
    \Text(168,82)[lb]{\Large{\Black{$x_{i+1}$}}}
    \Text(174,28)[lb]{\Large{\Black{$x_{i+1,j+1}^2$}}}
    \Text(138,4)[lb]{\Large{\Black{$x_{i,j+1}^2$}}}
    \Text(138,46)[lb]{\Large{\Black{$x_{i+1,j}^2$}}}
    \Line(90,64)(90,-8)
  \end{picture}
  \label{figure_cross}
}
\\
(a) \hskip7cm (b)
\end{center}
Figure 2: {\sl Cross-ratios or anharmonic ratios are conformally invariant Mandelstam variables of lightlike polygon.}
\vskip0.5cm

The counting is simple. To construct a cross-ratio, we need two distinct edges as in Figure 2.  There are $n(n-1)/2$ possible pairs of edges. However, the resulting cross-ratio vanishes if the two edges chosen are nearest neighbors or next-nearest neighbors.

We are again in a situation that, in a given spacetime dimension, the dimension of the moduli space of conformal cross-ratios (\ref{dim-crossratio}) outgrows dimension of moduli space of conformal lightlike polygon (\ref{dim-conformal-polygon}) we want to describe. The requisite projection to the conformal cross-ratios is achieved by the Gram condition modulo conformal equivalence relations.

Let's be more explicit. A conformal covariant vector $x^m$ in $\mathbb{R}^{d-1, 1}$ can be equivalently described by projection of a vector $X^A = (X_1, X_0, X_1, \cdots, X_d)$ in embedding Minkowski space $\mathbb{R}^{d, 2}$ onto the lightlike hyperboloid:
\bea
\eta_{AB} X^A X^B  = - (X_{-1})^2 - (X_0)^2 + (X_1)^2 + \cdots + (X_d)^2 = \eta_{mn} X^m X^n - 2 X^+ X^- = 0 .
\label{hyperboloid}
\eea
Choosing the lightcone coordinates $X^\pm = (X_{-1} \pm X_d)/\sqrt{2}$ are the lightcone coordinates, the vector $x^m$ is projectively obtained by
\bea
x^m = {X^m \over X^+}.
\eea
The action of the conformal group $SO(d, 2)$ to the vector $x^m$ is equivalent to the action of linear transformations acting on $X^A$ lying on the lightlike hyperboloid (\ref{hyperboloid}). It is known that the space of $x$-vectors is $\mathbb{R}^{d-1, 1}$ provided the $SO(d,2)$ is gauge-fixed to $X^+ = 1$.  In this gauge,
\bea
X_{ij}^2 = - 2 X_i \cdot X_j = x^2_{ij}.
\eea
From this, it also follows that
\bea
Y_i := (X_{i+1} = X_i) = (y_i, 0, Y^-_i) \qquad (i=1, \cdots, n)
\eea
are lightlike in $\mathbb{R}^{d,2}$. We thus associated the conformal edge vectors $y_i$'s of a conformal lightlike polygon in the physical spacetime $\mathbb{R}^{d-1,1}$ with the edge vectors $Y_i$'s of a lightlike polygon in the embedding space $\mathbb{R}^{d,2}$. This then implies that the space of conformal cross-ratios in $\mathbb{R}^{d-1, 1}$ is the same as the space of Mandelstam variables in $\mathbb{R}^{d,2}$. Therefore, the dimension of the moduli space of conformal cross-ratios is given by
\bea
\mbox{dim} M_{\rm c} (C_n) = {1 \over 2} n (n - 1) - n - n
= {1 \over 2} n (n - 5).
\eea
We subtracted $n$ for choosing adjacent edge pairs, and $n$ for choosing next-adjacent edge pairs.

How do we match this moduli space to the moduli space of conformal lightlike polygons? The idea is that the Gram sub-determinants of the vectors project the cross-ratios down to the space of independent ones. The Gram determinant in the embedding space is now given by
\bea
G_c = M_c^T \cdot M_c =
\begin{pmatrix}
Y_1 \cdot Y_1 && Y_1 \cdot Y_2 && \cdots && Y_1 \cdot Y_n \\
Y_2 \cdot Y_1 && Y_2 \cdot Y_2 && \cdots && Y_2 \cdot Y_n \\
\vdots && \vdots&& \ddots && \vdots \\
Y_n \cdot Y_1 && Y_n \cdot Y_2 && \cdots && Y_n \cdot Y_n
\end{pmatrix}
\eea
Because of the closedness condition, the Gram determinant itself vanishes identically. Since the embedding space $\mathbb{R}^{d,2}$ accommodates at most $(d+2)$ many linearly independent vectors, Therefore, there are $(n - (d + 1)) (n - (d +1) -1)/2$ many Gram sub-determinant conditions. Therefore, the dimension of conformal cross-ratios is
\bea
\mbox{dim} \Pi_G M_c [C_n] = {1 \over 2} n (n - 5) - {1 \over 2} (n - d - 1)(n - d - 2) = (d-1) n - {1 \over 2} (d+1) (d+2).
\eea
This matches precisely the dimension of moduli space of conformal lightlike polygons (\ref{dim-conformal-polygon}).

\section{The Hexagon Remainder Function}
In this section, we compute the remainder function $\mbox{Rem}^{(2)}_{6, \rm CS}$, relevant for the hexagon ABJM Wilson loop expectation value. We explained in section 3 that this computation involves multi-dimensional scalar integrals. In this section, we compute them.

We expect from the anomalous conformal Ward identity that the remainder function $\mbox{Rem}_{n, \rm CS}$ depends only on the conformal cross-ratios. In setting up the computation, we can readily verify this property of the remainder function by varying shapes, equivalently, Mandelstam variables of the lightlike polygon. Not all the Mandelstam variables are independent and, as we explained in section 4, it is necessary to impose the Gram sub-determinant conditions. This condition turns out a stark difference from what were known for extracting the remainder function in the four-dimensional ${\cal N}=4$ SYM. In section 5.1,  we recall this situation in detail.

For the computation of multi-dimensional integrals, we utilize public packages. The scalar integrals we need to compute span up to 8-dimensional complex integrations. The traditional {\ttfamily{MB}} package \cite{Czakon:2005a} turns out not powerful enough to render the result with requisite numerical precisions. Instead, we utilize the package  {\ttfamily{FIESTA2}}, \cite{Smirnov:2009a}. In the following subsections, we present details of the computation. In section 5.2, we present numerical computations performed using the {\ttfamily{FIESTA2}} package. In section 5.3, for the special shapes of the hexagon discussed in the previous section, we reduce our multi-dimensional integrals to lower-dimensional integrals. The reduction facilitates to achieve high precision to the numerical computations. In section 5.4, we utilize the PSLQ algorithm and infer analytic expressions of the $\mbox{Rem}_{6, \rm CS}$ from the numerical results.

%%%%%%%%%%%%%%%%%%%%%%%%%%%%%%%%%%%%
\subsection{Remainder Function in ${\cal N}=4$ Super Yang-Mills Theory}
In section 4, we explained that the Mandelstam variables are the Lorentz scalars convenient for specifying the geometry of lightlike
polygon, they need to be further projected down to the space of conformal cross-ratios since they are not mutually independent. We alluded that such projection is achieved by the Gram sub-determinant conditions. Therefore, in the numerical computation in this section of the hexagon remainder function, we shall cover the moduli space of the lightlike hexagon by varying the Mandelstam invariants over the subspace that the Gram sub-determinant conditions are satisfied.

While our prescription is the most natural steps to take, this was not what was practiced when the hexagon remainder function was computed in the (3+1)-dimensional $\mathcal{N}=4$ SYM theory. There, the anomalous conformal Ward identities also put the remainder function to be a function of conformal cross-ratios. The lightlike Wilson loops were again specified by Mandelstam invariants. Remarkably, it was observed that two sets of Mandelstam invariants, one obeying the Gram sub-determinant conditions and another not, yielded an identical result for the remainder function. In so far as the cross-ratios are the same, any choice of the Mandelstam variable set is allowed regardless of solving the Gram sub-determinant conditions. Indeed, this explains why lightlike Wilson loops with odd numbers of edges are admissible configurations in the $\mathcal{N}=4$ SYM theory.
%This point is main difference between $\mathcal{N}=4$ SYM and ABJM theory.

As the Mandelstam variables can be chosen freely thus they can be taken `unphysical' values outside the moduli space of the hexagon, wide variety of kinematic limits become available in so far as evaluation of the remainder function is concerned. In the $(3+1)$-dimensional ${\cal N}=4$ SYM theory, this freedom was maximally taken into advantage. A particularly useful limit was the quasi multi-Regge kinematics (QMRK), since this kinematics enabled determination of the hexagon remainder function and understanding its analytic structure. For $(2+1)$-dimensional ABJM theory, we concluded in section 4 that such kinematic limits are not available and we should impose the Gram sub-determinant conditions throughout.

The Gram sub-determinant conditions essential for the ABJM theory bears further impact. In the the $(3+1)$-dimensional ${\cal N}=4$ SYM theory, another useful kinematic limit was to take the lightlike polygon to $(1+1)$-dimensional subspace. This limit brought in enormous simplification and facilitated computation of the remainder function analytically tractable. Unfortunately, for kinematical reasons again, this limit is also not available for $(2+1)$-dimensional ABJM theory. This is because the $(1+1)$-dimensional kinematics cruially relies on the positivity condition and the closedness of edge vectors. Take for instance the lightlike hexagon and restrict it to the $(1+1)$-dimensional lightlike basis,  $(1,-1)$ and $(1,1)$. The 6 edge vectors obeying the positivity condition are parametrized as
\bea
y_1 = (a,-a), \quad y_2 = (b,b), \quad y_3 = (c,-c), \quad y_4 = (d,d), \quad y_5 = (e,-e), \quad y_6 = (f, f),
\eea
where $a,b,c,d,e,f$ are restricted to be positive. To obey the closedness, both $a+b+c+d+e+f = 0$ and $a-b+c-d+e-f = 0$ should be satisfied. We see that these conditions cannot be met, since the positivity of $a,b,c,d,e,f$ violates first equation. Therefore, $(1+1)$-dimensional lightlike condition, positivity condition and closedness are not mutually compatible.

%%%%%%%%%%%%%%%%%%%%%%%%%%%%%%%%%%%%
\subsection{Scalar Invariants and Gram Sub-Determinant Conditions}
Here, we first study how the hexagon remainder function depends on the Mandelstam variables and the Gram sub-determinant conditions. We shall find that the dependence in the $(2+1)$-dimensional ABJM theory is very different from the dependence in the $(3+1)$-dimensional ${\cal N}=4$ SYM theory.

We computed numerically both the triple-vertex diagrams and the ladder diagrams listed in Figure 2. Adding them, we obtained the hexagon remainder function at two loops, $\mbox{Rem}^{(2)}_{6, \rm CS}$ as a function of 9 Mandelstam variables of the hexagon.
\setlength{\tabcolsep}{3pt}
\setlength{\extrarowheight}{1.5pt}
{
\begin{table}[ht]
\centering
\begin{tabular}{|c||ccccccccc|c|}
 \hline
   &  $x_{13}^2$ & $x_{24}^2$ & $x_{35}^2$ & $x_{46}^2$ & $x_{15}^2$ & $x_{26}^2$ & $x_{14}^2$ & $x_{25}^2$ & $x_{36}^2$  & Rem$^{(2)}_{6, \rm CS}$ \\ \hline \hline
 A &$ -1.0000$ & $-1.0000$ & $-1.0000$ & $-1.0000$ & $-1.0000$ & $-1.0000$ & $-1.0000$ & $-1.0000$ & $-1.0000$ & $-3.47537352$ \\ \hline
 B & $-6.8764$ & $-18.194$ & $-21.887$ & $-77.498$ & $-48.781$ & $-14.780$ & $-24.467$ & $-30.720$ & $-3.3327$  &  $-3.47342610$ \\ \hline
 C & $-4.8757$ & $-11.282$ &  $-6.1981$ & $-42.828$ & $-19.339$ & $-8.1903$ & $-15.616$ & $-10.007$ & $-2.5719$ &  $-3.47622947$  \\ \hline
 D & $-3.5979$ & $-7.3282$ & $-1.4275$ & $-24.543$ & $-7.9792$ & $-4.5361$ & $-10.424$ & $-2.6875$ & $-1.9989$ & $-3.47688979$    \\ \hline
 E & $-116.29$ & $-4.0000$ & $-116.29$ & $-2.0350$ & $-4.0000$ & $-2.0350$ & $-4.0000$ & $-4.0000$ & $-59.160$ & $-3.48197748$ \\ \hline
 F & $-4.0000$ & $-2.3528$ & $-9.0000$ & $-1.0000$ & $-1.3057$ & $-1.0000$ & $-1.0000$ & $-2.2500$ & $-3.6892$ & $-3.47579959$ \\ \hline
 G & $-4.0000$ & $-1.0000$ & $-8.8965$ & $-4.4482$ & $-1.0000$ & $-2.0000$ & $-1.0000$ & $-1.0000$ & $-5.5504$ & $-3.47576202$ \\ \hline
 H & $-1.2027$ & $-2.5332$ & $-2.0000$ & $-3.0000$ & $-6.2344$ & $-13.512$ & $-2.1782$ & $-3.6253$ & $-0.82827$ & $-3.47561202$ \\ \hline
\end{tabular}
\label{table1}
\caption{\sl Results of $R_{CS,6}$ for eight configurations of hexagon's Mandelstam variables. It suggests that R$^{(2)}_{6, \rm CS}$ takes a constant value over wide ranges of the conformal cross-ratios.}
\end{table}
}
\vskip0.5cm

In Table 1, we generated eight configurations (A)$\sim$(H) of the 9 Mandelstam variables $x_{13}^2, \cdots, x_{36}^2$,  subject to the Gram sub-determinant conditions. Equivalently, these configurations are generated by lightlike segment vectors $y_1, \cdots, y_6$ subject to the $SO(3,2)$ conformal invariance. The results indicates that the hexagon remainder function Rem$^{(2)}_{6, \rm CS}$ is a constant number, independent of the Mandelstam variables and hence the conformal cross-ratios.

To test neessity of the Gram sub-determinant conditions, we chose a configuration, say (D), and permuted subset of the nine Mandelstam variables while keeping their conformal cross-ratios fixed. Obviously, permuting the Mandelstam variables so violates the Gram sub-determinant conditions. We computed the hexagon remainder function $\mbox{Rem}^{(2)}_{6, \rm CS}$ and the results are tabulated in Table 2.
\setlength{\tabcolsep}{3pt}
\setlength{\extrarowheight}{1.5pt}
{
\begin{table}[ht]
\centering
\begin{tabular}{|c||ccccccccc|c|}
 \hline
   &  $x_{13}^2$ & $x_{24}^2 $ & $x_{35}^2$ & $x_{46}^2$ & $x_{15}^2$ & $x_{26}^2$ & $x_{14}^2$ & $x_{25}^2$ & $x_{36}^2$ & $\mbox{Rem}_{6, \rm CS}^{(2)}$ \\ \hline \hline
 D1 & $-3.5979$ & $-7.9792$ & $-1.4274$ & $-24.543$ & $-7.3282$ &
$-4.5361$ & $-10.424$ & $-2.6875$ & $-1.9989$ & $-3.70845563$ \\ \hline
 D2 & $-3.5979$ & $-7.9792$ & $-4.5361$ & $-24.543$ & $-7.3282$ & $-14.780$ & $-10.424$ & $-2.6875$ & $-1.9989$ & $-3.99210938$ \\ \hline
\end{tabular}
\caption{\sl We examined whether $\mbox{Rem}^{(2)}_{6, \rm CS}$ maintain the same values for the above Mandelstam variables. They all  have same conformal cross ratios. One remarkable observation on the remainder function in $\mathcal{N}=4$ SYM was that it has the same value for all Mandelstam variables so long as their conformal cross ratios are the same, even if the Gram sub-determinant conditions were not satisfied. This result suggests that such feature no long holds in the ABJM theory.}
\end{table}
}
\vskip0.5cm

In Table 2, we generated configurations (D1) and (D2) that have the same conformal ratios as (D) but violates the Gram sub-determinant condition \footnote{This is equivalent to saying that there is no suitable choice of $x_i$'s or $y_i$'s vectors which generate (D1) and (D2) configurations.}. We observe that the remainder function at (D),(D1),(D2) do not agree one another even though all three sets have one and the same conformal cross-ratio.  We thus conclude that, in stark contrast to the $(3+1)$-dimensional ${\cal N}=4$ SYM theory, the lightlike hexagon Wilson loop expectation value in the ABJM theory is consistent with the anomalous conformal Ward identity only if the Mandelstam variables were to satisfy the Gram sub-determinant conditions. Therefore, if two sets of the hexagon Mandelstam variables satisfy the Gram sub-determinant conditions and yield the same conformal cross-ratio, then their values of the remainder function should be the same. In our numerical computations, we have confirmed this.

%check-point, we computed the remainder function $\mbox{Rem}^{(2)}_{6, \rm CS}$ for several different sets of the Mandelstam variables that does not satisfy the Gram sub-determinant conditions.
%
\setlength{\tabcolsep}{3pt}
\setlength{\extrarowheight}{1.5pt}
{
\begin{table}[ht]
\centering
\begin{tabular}{|c||ccccccccc|c|}
 \hline
   &  $x_{13}^2$ & $x_{24}^2$ & $x_{35}^2$ & $x_{46}^2$ & $x_{15}^2$ & $ x_{26}^2$ & $ x_{14}^2$ & $ x_{25}^2 $ & $ x_{36}^2$ & $\mbox{Rem}^{(2)}_{6, \rm CS}$ \\ \hline \hline
 X & $-6.0000$ & $-2.0000$ & $-3.0000$ & $-9.0000$ & $-5.0000$ & $-7.0000$ & $-1.0000$ & $-4.0000$ & $-8.0000$ & $-3.99713002$ \\ \hline
 Y & $-1.0000$ & $-5.0000$  & $-\frac{3}{16}$ & $-27.000$ & $-1.0000$ & $-7.0000$ & $-2.0000$ & $-1.0000$ & $-2.0000$ & $-3.84236164$ \\ \hline
 Z & $-1.0000$ & $-\frac{1}{3}$ & $-1.0000$ & $-1.0000$ & $-2.0000$ & $-\frac{2}{3}$ & $-1.0000$ & $-\frac{2}{3}$ & $-1.0000$ & $-3.41789832$
\\ \hline
\end{tabular}
\caption{\sl Three random choices of the Mandelstam variables that do not obey the Gram sub-determinant conditions. The values of the remainder function $R_{CS,6}$ do not agrees with the values in Table 1 for `physical' Mandelstam variables.}
\end{table}
}
As another check, in Table 3, we considered randomly chosen configuration (X) and another configuration (Y) having the same conformal cross-ratios as (X). The two configurations yield different values for the remainder function. This affirms that configurations violating the Gram sub-determinant condition do not obey the anomalous conformal Ward identitie since these identities put the remainder function to a function only of conformal cross-ratios. As such, we call them 'unphysical' configurations.  We also considered the configuration (Z) whose cross-ratios all have value 1 and hence relevant for the $(1+1)$-dimensional configuration of the hexagon. Result (Z), however, shows that it does not yield the physical result, because the closedness, $(1+1)$-dimensional lightlikeness, and the positivity conditions cannot be satisfied simultaneously.

Summarizing, we learned that although the polygon kinematics is most conveniently described in terms of the Mandelstam invariants, they are subject to various restrictions to correspond to physical configurations. Some of these restrictions are universal, independent of spacetime dimensions,  while some other restrictions are specific to $(2+1)$-dimensional spacetime. Unfortunately, the latter restrictions were stringent enough not to allow the QMRK that played powerful role in understanding the analytic structure of the Wilson loop expectation values in the $(3+1)$-dimensional $\mathcal{N}=4$ SYM theory.

To avoid such difficulty, we identified alternative special kinematics that satisfy the Gram sub-determinant conditions and also permit continuous deformation within the moduli space of the lightlike polygon. The idea is to take the deformation parameters to asymptotic limit and reduce Mellin-Barnes integrals as simple as possible. We shall study these kinematic limits in the next sections.  For now, we present numerical result for several configurations that turn out representative of 1- and 2-parameter subspaces.
\newline
\setlength{\tabcolsep}{3pt}
\setlength{\extrarowheight}{1.5pt}
{
\begin{table}[ht]
\centering
\begin{tabular}{|c||ccccccccc|c|}
 \hline
   &  $x_{13}^2$ & $x_{24}^2$ & $x_{35}^2$ & $x_{46}^2$ & $x_{15}^2$ & $x_{26}^2$ & $x_{14}^2$ & $x_{25}^2$ & $x_{36}^2$ & $\mbox{Rem}^{(2)}_{6, \rm CS}$ \\ \hline \hline
 J1 & $-100.00$ & $-1.0000$ & $-1.0000$ & $-\frac{1}{100}$ & $-1.0000$ & $-1.0000$ & $-1.0000$ & $-1.0000$ & $-1.0000$ & $-3.4857518$ \\ \hline
 J2 & $-100.00$ & $-2000.0$ & $-100.00$ & $-100.00$ & $-5.0000$ &
$-100.00$ & $-100.00$ & $-100.00$ & $-100.00$ & $-3.4778556$ \\ \hline
\end{tabular}
\caption{\sl Mandelstam variable choices satisfying the Gram sub-determinant conditions. We checked (J1) and (J2) could be generated from suitably chosen edge vectors $y_i$'s. Since we want to take asymptotic limit while maintaining the Gram sub-determinant conditions, suitable special kinematics were considered. These results provide numerical evidence that both one-parameter family and two parameter family indeed yield satisfactory results for the remainder function. See Table 1 for comparison.}
\end{table}
}

\noindent
The configuration (J1) belongs to 1-parameter group, while (J2) belongs to 2-parameter group. We examined numerically the effect of changing these free parameters. As seen in Table 4, the remainder function $\mbox{Rem}^{(2)}_{6, \rm CS}$ takes a constant value over the ranges we changed these parameters. The result hints that we can take certain asymptotic limits of these moduli parameters and simplify the Mellin-Barnes transformation integrals.

\subsection{Special Shapes and Asymptotic Limits}
\subsubsection{The strategy}
Our goal is to compute the hexagon remainder function with high precision and infer from it analytic result. In the previous subsection, we presented the remainder function computed using the package {\ttfamily{FIESTA2}}. The numerical error is rather large, ${\cal O}(10^{-2})$. Here, we propose an alternative strategy for computing the remainder function with better numerical precision than  {\ttfamily{FIESTA2}}. We begin with the Mellin-Barns transformation to our 2-loop integrals, for which we used the Mathematica package {\ttfamily{MB}}. The problem of this transformation is that it results in multi-dimensional scalar integrals, for which numerical precision is difficult to attain. The idea is to lower the dimension of numerical integral maximally so that higher numerical precision can be achieved. The way we achieve this is as follows. Recall that the dimension of numerical integral is closely related to the number of independent terms inside the denominator $\Delta_y$ in the two-loop integral involving the gauge boson triple-vertex diagrams (See Appendix G). By choosing judiciously a set of the Mandelstam variables that satisfy the Gram sub-determinant conditions and that reduce the number of terms in $\Delta_y$, we can bring down the dimension of numerical integrals and obtain the result with high numerical precisions. Below, we explain how we performed high precision numerical computation for the gauge boson triple-vertex diagrams. The integrals are relegated in the Appendix G. Attentive readers may skip technical details and go directly to the final results (\ref{res_hex}) and (\ref{result_hex}).

\subsubsection{Computational Details}
Our strategy for the numerical computation is as follows. We apply the Mellin-Barnes transformation to every loop integrals resulting from the gauge boson triple-vertex and gauge boson ladder diagrams derived in section 3. We then take special limits of the polygon shape deformed by one- or two-moduli parameters. The integrals are defined in the complex domain. Utilizing the packages {\ttfamily{MB}} and {\ttfamily{MBresolve}}\cite{Smirnov:2009b},  we resolve singularity structure of each complex integrals. We then apply the Barnes lemma to reduce the integrals to lower-dimensional integrals. We made this procedure automatic using the package {\ttfamily{barnesroutines}}\cite{Kosower}. Next, we apply the package {\ttfamily{MBasymptotics}}\cite{Czakon} to the chosen moduli parameters and obtain simpler expressions for the integrals. We find that these expressions are reducible to at most three-dimensional complex integrals.  Finally, we evaluate them using the {\ttfamily{MB}} package and obtain numerical result with high precision.

What special limits can we choose for the Mandelstam variables of the lightlike hexagon? Subject to the Gram sub-determinant conditions, let's consider the following two special limits:  the first one has 1-moduli parameter, while second one has 2-moduli parameters.
\begin{itemize}
\item
{\bf one-parameter hexagon}
\setlength{\tabcolsep}{14.5pt}
\setlength{\extrarowheight}{1.5pt}
{
\begin{table}[ht]
\centering
\begin{tabular}{|ccccccccc|}
 \hline
$x_{13}^2$ & $x_{24}^2$ & $x_{35}^2$ & $x_{46}^2$ & $x_{15}^2$ & $x_{26}^2$ & $x_{14}^2$ & $x_{25}^2$ & $x_{36}^2$      \\ \hline \hline
$- e^a$ & $-1$ & $- 1$ & $- e^{-a}$ & $- 1$ & $- 1$ & $- 1$ & $-1$ & $- 1$  \\ \hline
\end{tabular}
\end{table}
}
%\Rightarrow &\{x_{14}^2 \rightarrow -1, \hspace{1.8mm} x_{25}^2  \rightarrow -1, \hspace{1.8mm} x_{36}^2 \rightarrow -1, \hspace{1.8mm} x_{26}^2 \rightarrow -1, \hspace{1.8mm} x_{35}^2  \rightarrow -1,\hspace{1.8mm} x_{15}^2  \rightarrow -1, \hspace{1.8mm}  \nonumber \\
%& \quad \quad \quad \quad \quad \quad \quad \quad \quad \quad \quad \quad \quad \quad \quad \quad \quad \quad \quad \quad \quad \quad \quad \quad x_{24}^2 \rightarrow -1, x_{13}^2  \rightarrow -a, \hspace{1.8mm} x_{46}^2  \rightarrow -\frac{1}{a} \} \\
\newline
The moduli parameter $a$ ranges over $- \infty < a < + \infty$. We take the configuration that obeys the positivity condition, and this puts all the Mandelstam variables to positive definite values.

\item {\bf two-parameter hexagon}
\setlength{\tabcolsep}{12pt}
\setlength{\extrarowheight}{1.5pt}
{
\begin{table}[ht]
\centering
\begin{tabular}{|ccccccccc|}
 \hline
 $x_{13}^2$ & $x_{24}^2$ & $x_{35}^2$ & $x_{46}^2$ & $x_{15}^2$ & $x_{26}^2$ & $x_{14}^2$ & $x_{25}^2$ & $x_{36}^2$ \\ \hline \hline
$-e^{\alpha}$ & $-e^{\alpha + \beta}$ & $-e^\alpha$ & $-e^\alpha$ & $-e^{\alpha - \beta}$ & $-e^\alpha$ & $-e^\alpha$ & $-e^\alpha$ & $-e^\alpha$
\\ \hline
\end{tabular}
\end{table}
}
\newline
The moduli parameters $\alpha, \beta$ range over $-\infty < \alpha, \beta < + \infty$. Again, taking the configuration obeying positivity condition, all the Mandelstam variables are positive definite.

%\mbox{Two}& \mbox{-parameter set} \nonumber \\
%\Rightarrow &\{ x_{14}^2 \rightarrow -f,\hspace{1.8mm} x_{25}^2  \rightarrow -f,\hspace{1.8mm} x_{36}^2 \rightarrow -f,\hspace{1.8mm} x_{26}^2 \rightarrow -f,\hspace{1.8mm} x_{35}^2  \rightarrow -f,\hspace{1.8mm} x_{15}^2  \rightarrow -e, \hspace{1.8mm} \nonumber \\
%& \quad \quad \quad \quad \quad \quad \quad \quad \quad \quad \quad \quad \quad \quad \quad \quad \quad \quad \quad \quad \quad \quad \quad \quad x_{24}^2 \rightarrow -\frac{f^2}{e},\hspace{1.8mm} x_{13}^2  \rightarrow -f,\hspace{1.8mm} x_{46}^2  \rightarrow -f \}

\end{itemize}

We found that the hexagon remainder function $\mbox{Rem}^{(2)}_{6, \rm CS}$ remains constant-valued for a wide range of the moduli parameters $a, \alpha, \beta$. In the previous subsection, we already presented one such choice in the result for the configurations (J1) and (J2) in the previous subsection. The result suggested that the hexagon remainder function $\mbox{Rem}^{(2)}_{6, \rm CS}$ is indeed a constant up to two loops in the ABJM theory. We performed numerical computation for both configurations and found that the two-parameter configuration yields the result with better numerical precision. Hereafter, we will exclusively discuss the two-parameter configuration results. The simplest integral is $I_{321}$ in (\ref{321_delta}). Inserting the two-moduli parameter contour to (\ref{321_delta}), we observe that the four-fold integration is reduced to three-fold integration. For instance, the denominator is reduced to
\begin{equation}
\Delta_y \Big\vert_{\mbox{2-parameter}}\quad \longrightarrow \quad e^\alpha \cdot x\bar{x}y\bar{s_1}s_2+e^{\alpha - \beta} \cdot \bar{x}\bar{y}\bar{s_2}s_3+e^{\alpha - \beta} \cdot x\bar{y}s_1s_3+ e^\alpha \cdot x\bar{y}\bar{s_1}.
\end{equation}
By itself, five terms in the denominator $\Delta_y$ are reduced to three terms, so the two-parameter configuration does not appear to simplify the multi-dimensional integrals considerably. It turned out the two-parameter configuration is more effective for other triple-vertex diagrams involving higher-dimensional integrals. The most complicated integrals resulted from the contribution $I_{531}$. The Mellin-Barns transformation of this contribution yielded 8-dimensional complex integrals. With the two-parameter special kinematics, we were able to reduce these integrals to five-dimensional integrals. We could do even better. By taking the asymptotic limits for $\alpha, \beta$ sequentially,
\bea
\alpha \rightarrow - \infty \quad \mbox{then} \quad \beta \longrightarrow +\infty.
\eea
we were able to reduce the five-dimensional integrals down to at most three-dimensional integrals.

\subsubsection{Result}
The high precision computation yielded
\begin{align}
\mbox{Rem}^{(2)}_{6, \rm CS} = - 3.470168804. \quad \Big(0.000489814\Big).
\end{align}
Utilized the PSLQ algorithm, we converted thi to an analytic expression. The result is
\begin{equation}
\mbox{Rem}^{(2)}_{6, \rm CS}= -\frac{17}{4} \zeta(2)+ 3\mbox{Log}(2)+ 3\mbox{Log}^2(2). \label{res_hex}
\end{equation}
Numerical value of the right-hand side is $-3.470169200670522$, and this agrees to our numerical result $-3.47016880435048$ within the ${\cal O}(10^{-6})$ precision.

The final result for the two-loop, lightlike hexagon ABJM Wilson loop expecation value is obtained by combining the purely abelian, matter-dependent contribution (\ref{matterWvev}) and the pure Chern-Simons contribution (\ref{CSWvev}) for $n=6$. It takes the form:
\bea
\big< W_\Box [C_6]\big>^{(2)}_{\rm ABJM}&=& -
\Big[\frac{1}{2} \mbox{Log}(2) \sum_{i=1}^6 \frac{(x_{i,i+2}^2 \pi e^{\gamma_E} \mu^2)^{2\epsilon}}{2\epsilon} +\mbox{Rem}^{(2)}_{6, \rm CS} -3 \mbox{Log}(2) \nonumber \\
&& \hskip0.1cm + {1 \over 2}\sum_{i=1}^n \frac{(x_{i,i+2}^2 4\pi e^{\gamma_E} \mu^2)^{2\epsilon}}{(2\epsilon)^2}-\mbox{BDS}^{(2)}_6 +\frac{3}{8} \pi^2 \Big] \nonumber \\
&=& - \Big[ {1 \over 2} \sum_{i=1}^6 \frac{(x_{i,i+2}^2 8 \pi e^{\gamma_E} \mu^2)^{2\epsilon}}{(2\epsilon)^2}-\mbox{BDS}^{(2)}_6 + \mbox{Rem}^{(2)}_{6, \rm CS}-3 \mbox{Log}(2)+\frac{3 \pi^2}{8} -\frac{15}{2} \mbox{Log}^2(2) \Big] \nonumber \\
&=& -  {1 \over 2} \sum_{i=1}^6 \frac{(x_{i,i+2}^2 8 \pi e^{\gamma_E} \tilde\mu^2)^{2\epsilon}}{(2\epsilon)^2} +\mbox{BDS}^{(2)}_6  + \Big(\frac{9}{2} \mbox{Log}^2(2) + \frac{ \pi^2}{3}\Big), \label{result_hex}
\eea
where the BDS contribution $\mbox{BDS}^{(2)}_6$ is already known. This is one of the main results of this paper. Like the lightlike tetragon Wilson loop expectation value, the UV finite part in (\ref{result_hex}) exhibits the  uniform transcendentality.

While we have succeeded in obtaining two-loop analytic result for the hexagon Wilson loop expectation value, we have yet no clue for the structure of the remainder function $\mbox{Rem}^{(2)}_{n, \rm CS}$ for polygons of $n\ge 8$. To crack down its structure, we will need to understand further configurational structures of the lightlike polygon Wilson loop expectation value. This is what we will undertake in the next section.

%%%%%%%%%%%%%%%%%%%%%%%%%%%%%%%%%%%%
\section{Lightlike Factorization and Antenna Function}
Conformal field theories are subject to infrared divergences due to collinear and soft bremsstrahlung partons. These divergences then allow universal factorization and scaling behavior of physical processes. A class of such processes is the parton scattering amplitudes in gauge theories. The universal factorization and scaling behavior allowed accurate prediction at fixed order perturbation theory and resummation of dominant logarithms.

Our goal in this section is to demonstrate that universal factorization and scaling behavior are also present in the lightlike polygon Wilson loops. We then introduce `universal antenna function' for a certain limit of the polygon shape, which we will utilize it in the next section to solve for the ABJM Wilson loop expectation value for arbitrary $n$.

%%%%%%%%%%%%%%%%%%%%%%%%%%%%%%%%%%%%%%%
\subsection{Infrared Factorization in Gauge Theories}
Let us recall the IR factorization in gauge theories and draw intuitions for what we may expect for the lightlike Wilson loops.
The color-ordered scattering amplitude in gauge theories has the factorization property with respect to the IR divergence. First, consider the collinear limit that lightlike momenta $k_i$ and $k_{i+1}$ of two partons $i$ and $i+1$ become parallel and coalesce to a new lightlike momentum $k_P$. Kinematically, this situation described by
\bea
k_{i} \longrightarrow h k_P \quad \mbox{and} \quad  k_{i+1} \longrightarrow (1-h) k_P, \hskip2cm ( 0 \le h \le 1)
\eea
so the two collinear partons carry the fraction $h, (1  - h)$ of the momentum $k_P$.

For the $L$-loop $n$-point scattering amplitude $A_n^{(L)} (k_1, \cdots, k_n)$, the collinear limit exhibits factorization \cite{Bern:1999a}
\begin{equation}
A_{n}^{(L)}(k_1, \cdots, k_n) \longrightarrow \sum_{\lambda=\pm} \sum_{\ell =0}^L \mbox{Split}_{-\lambda}^{(L-\ell)}(h;k_{i},\lambda_{i};k_{i+1},\lambda_{i+1}) A_{n-1}^{(\ell)}(k_P,\lambda, \cdots k_n))
\label{split}
\end{equation}
Here $\lambda$ labels the polarization state of the factorizing parton. In the summation, $L, \ell=0$ denote the tree-level amplitude.
Helicity structure is fixed by the Poincar\'e invariance, so both the scattering amplitudes and the splitting functions can be decomposed to their tree-level counterparts times scalar functions summarizing loop corrections.

We define reduced scattering amplitudes $M_{n}^{(L)}$ for the ratio of the $L$-loop scattering amplitude to the tree-level scattering amplitude:
\begin{align}
A_{n}^{(L)} (k_1, \lambda_1, \cdots, k_n, \lambda_n) = A_{n}^{(0)} (k_1, \lambda_1, \cdots, k_n \lambda_n) \cdot M_{n}^{(L)} (k_1, \cdots, k_n).
\end{align}
Similarly, we define the reduced splitting functions $R_s^{(L)}(\epsilon,z,k_P)$ for the ratio of the $L$-loop splitting function to the tree-level splitting function
\begin{align}
\mbox{Split}_{-\lambda}^{(L)}(h;k_{i},\lambda_{i};k_{i+1},\lambda_{i+1})= \mbox{Split}_{-\lambda}^{(0)}(h;k_{i},\lambda_{i};k_{i+1},\lambda_{i+1}) \cdot R_s^{(L)}(\epsilon,h;k_P),
 \label{rs}
\end{align}
where we use the dimensional regularization for the IR divergences.
In the collinear limit, the tree-level scattering amplitudes are expected to factorize as follows:
\begin{equation}
A_{n}^{(0)} \longrightarrow \sum_{\lambda=\pm} \mbox{Split}_{-\lambda}^{(0)}(h;k_{i},\lambda_{i};k_{i+1},\lambda_{i+1}) A_{n-1}^{(0)}(k_P,\lambda). \label{tree}
\end{equation}
This is illustrated in next Figure.
%------------------------------------------------------------------------------------------------------------
\begin{center}
\fcolorbox{white}{white}{
  \begin{picture}(104,80) (71,-41)
    \SetWidth{1.0}
    \SetColor{Black}
    \Arc(120,2)(26.077,148,508)
    \Line[arrow,arrowpos=0.5,arrowlength=5,arrowwidth=2,arrowinset=0.2](96,14)(72,26)
    \Line[arrow,arrowpos=0.5,arrowlength=5,arrowwidth=2,arrowinset=0.2](96,-10)(72,-22)
    \Line[arrow,arrowpos=0.5,arrowlength=5,arrowwidth=2,arrowinset=0.2](138,20)(156,38)
    \Line[arrow,arrowpos=0.5,arrowlength=5,arrowwidth=2,arrowinset=0.2](144,14)(168,26)
    \Line[arrow,arrowpos=0.5,arrowlength=5,arrowwidth=2,arrowinset=0.2](144,2)(174,2)
    \Line[arrow,arrowpos=0.5,arrowlength=5,arrowwidth=2,arrowinset=0.2](144,-10)(168,-22)
    \Line[arrow,arrowpos=0.5,arrowlength=5,arrowwidth=2,arrowinset=0.2](132,-22)(144,-40)
    \Line[arrow,arrowpos=0.5,arrowlength=5,arrowwidth=2,arrowinset=0.2](138,-16)(162,-34)
  \end{picture}
}
\\
\fcolorbox{white}{white}{
  \begin{picture}(220,88) (63,-9)
    \SetWidth{1.0}
    \SetColor{Black}
    \Arc(132,28)(18.974,162,522)
    \Arc(228,28)(18.974,162,522)
    \Line[arrow,arrowpos=0.5,arrowlength=5,arrowwidth=2,arrowinset=0.2](114,34)(78,58)
    \Line[arrow,arrowpos=0.5,arrowlength=5,arrowwidth=2,arrowinset=0.2](114,22)(78,4)
    \Line[arrow,arrowpos=0.5,arrowlength=5,arrowwidth=2,arrowinset=0.2](150,28)(174,28)
    \Line[arrow,arrowpos=0.5,arrowlength=5,arrowwidth=2,arrowinset=0.2](186,28)(210,28)
    \Line[arrow,arrowpos=0.5,arrowlength=5,arrowwidth=2,arrowinset=0.2](240,46)(264,64)
    \Line[arrow,arrowpos=0.5,arrowlength=5,arrowwidth=2,arrowinset=0.2](246,40)(276,58)
    \Line[arrow,arrowpos=0.5,arrowlength=5,arrowwidth=2,arrowinset=0.2](252,34)(282,46)
    \Line[arrow,arrowpos=0.5,arrowlength=5,arrowwidth=2,arrowinset=0.2](252,28)(282,28)
    \Line[arrow,arrowpos=0.5,arrowlength=5,arrowwidth=2,arrowinset=0.2](252,22)(282,10)
    \Line[arrow,arrowpos=0.5,arrowlength=5,arrowwidth=2,arrowinset=0.2](252,10)(276,-8)
    \Text(60,58)[lb]{\Large{\Black{$k_i$}}}
    \Text(60,4)[lb]{\Large{\Black{$k_{i+1}$}}}
    \Text(162,40)[lb]{\Large{\Black{$-$}}}
    \Text(192,40)[lb]{\Large{\Black{$-$}}}
  \end{picture}
  }
  \fcolorbox{white}{white}{
  \begin{picture}(220,88) (63,-9)
    \SetWidth{1.0}
    \SetColor{Black}
    \Arc(132,28)(18.974,162,522)
    \Arc(228,28)(18.974,162,522)
    \Line[arrow,arrowpos=0.5,arrowlength=5,arrowwidth=2,arrowinset=0.2](114,34)(78,58)
    \Line[arrow,arrowpos=0.5,arrowlength=5,arrowwidth=2,arrowinset=0.2](114,22)(78,4)
    \Line[arrow,arrowpos=0.5,arrowlength=5,arrowwidth=2,arrowinset=0.2](150,28)(174,28)
    \Line[arrow,arrowpos=0.5,arrowlength=5,arrowwidth=2,arrowinset=0.2](186,28)(210,28)
    \Line[arrow,arrowpos=0.5,arrowlength=5,arrowwidth=2,arrowinset=0.2](240,46)(264,64)
    \Line[arrow,arrowpos=0.5,arrowlength=5,arrowwidth=2,arrowinset=0.2](246,40)(276,58)
    \Line[arrow,arrowpos=0.5,arrowlength=5,arrowwidth=2,arrowinset=0.2](252,34)(282,46)
    \Line[arrow,arrowpos=0.5,arrowlength=5,arrowwidth=2,arrowinset=0.2](252,28)(282,28)
    \Line[arrow,arrowpos=0.5,arrowlength=5,arrowwidth=2,arrowinset=0.2](252,22)(282,10)
    \Line[arrow,arrowpos=0.5,arrowlength=5,arrowwidth=2,arrowinset=0.2](252,10)(276,-8)
    \Text(60,58)[lb]{\Large{\Black{$k_i$}}}
    \Text(60,4)[lb]{\Large{\Black{$k_{i+1}$}}}
    \Text(162,40)[lb]{\Large{\Black{$+$}}}
    \Text(192,40)[lb]{\Large{\Black{$+$}}}
  \end{picture}
  }
\end{center}
Figure 3. {\sl Factorization of scattering amplitudes in gauge theory. The $\lambda = \pm$ refers to the polarization of the intermediate, factorized particle state.}
\vskip0.5cm
%---------------------------------------------------------------------------------------
Inserting the relation (\ref{split}) to (\ref{tree}), we get
\begin{equation}
M_n^{(L)}\longrightarrow \sum_{\ell =0}^L R_s^{(\ell)} M_{n-1}^{(L-\ell)} \label{split_M}
\end{equation}
By definition, $R_s^{(0)}=1$ and $M_{n}^{(0)}=1$. The reduced amplitudes $M_n^{(L)}$ at one- and two-loops factorize to
\begin{align}
M_n^{(1)} &\longrightarrow M_{n-1}^{(1)}+R_s^{(1)} \label{ratio1} \\
M_n^{(2)} &\longrightarrow M_{n-1}^{(2)}+R_s^{(1)}M_{n-1}^{(1)}+R_s^{(2)}. \label{ratio2}
\end{align}
In the $(3+1)$-dimensional SYM theory, it is known that (\ref{ratio1}) and (\ref{ratio2}) are related each other by the collinear relation \cite{Anastasiou:2003kj}:
%it is easy to verify the $R_s^{(1)}$ and $R_s^{(2)}$ also satisfies the same recursive relations:
\bea
M_n^{(2)}(\epsilon)=\frac{1}{2}\big( M_n^{(1)}(\epsilon)\big)^2+f^{(2)}(\epsilon) M_n^{(1)}(2\epsilon) + C^{(2)}.
\label{m}
\eea
Here, $C^{(2)}$ is a finite constant, equal to $-{1 \over 2} \zeta_2^2$. Also,
$f^{(2)}(\epsilon) = - (\zeta_2 + \zeta_3 \epsilon + \zeta_4 \epsilon^2 + \cdots )$.
Substituting (\ref{ratio1}) to (\ref{m}),
\bea
M_n^{(2)}(\epsilon) &=& {1 \over 2} (M^{(1)}_{n-1}(\epsilon) + R^{(1)}_s(\epsilon))^2 + f^{(2)}(\epsilon) (M^{(1)}_{n-1}(2 \epsilon) + R_s^{(1)}(2 \epsilon)) + C^{(2)} \nonumber \\
&=& M^{(2)}_{n-1} (\epsilon) + R^{(1)}_s (\epsilon) M^{(1)}_{n-1} + {1 \over 2} (R_s^{(1)}(\epsilon))^2 + f^{(2)}(\epsilon) R^{(1)}_s (2 \epsilon).
\eea
In the second line, we utlized the above collinear relation for $M^{(2)}_{n-1}$. Comparing this with (\ref{ratio2}), we obtain recursive relation for the splitting function:
\bea
R^{(2)}_s(\epsilon) = {1 \over 2} (R_s^{(1)})^2 + f^{(2)}(\epsilon) R^{(1)}_s(2 \epsilon) + {\cal O}(\epsilon).
\eea
More generally, the scalar splitting function $R_s^{(\ell)}$ also follows from the BDS-like relation for all higher $\ell > 1$ loops. Indeed, for QCD and $(3+1)$-dimensional ${\cal N}=4$ SYM theory, the scalar splitting function $R_s^{(1)}$ was calculated explicitly and its universality was established \cite{Kosower:1999rx}, \cite{Anastasiou:2003kj}.

Another source of the IR divergences in gauge theories is emission of the soft partons. These divergences also provide another kind of factorization. More explicitly, in the limit of one parton becomes soft, the scattering amplitudes exhibit an abelian factorization that it becomes a product of an eikonal factor with a lower-point scattering amplitude. At tree-level, when $b$-parton becomes soft, $k_b \simeq 0$, the soft factorization is given by
\begin{equation}
A_{n}^{(0)}(k_1, \cdots, k_a, k_b, k_c, \cdots k_n) \quad \longrightarrow \quad S^{(0)} (k_a, k_b, k_c) A_{n-1}^{(0)}(k_1, \cdots, k_a, k_c, \cdots, k_n) \qquad \mbox{for} \qquad k_b \rightarrow 0,
\end{equation}
where $S^{(0)}(k_a, k_b, k_c)$ denotes the tree-level eikonal factor,
\bea
S^{(0)}(k_a, k_b, k_c) = −{ 1 \over \sqrt{2}} \left[ {\epsilon_b^\pm \cdot k_a \over k_a \cdot k_b} - {\epsilon_b^\pm \cdot k_c \over k_b \cdot k_c} \right].
\eea
%

%which is given in helicity-spinor notation by
%\begin{equation}
%S_{a,b,c}(s_{ab},s_{bc},s_{ac})=\frac{\big<a c\big>}{\big<a b\big>\big<b c\big>}.
%\end{equation}
%\color{blue} It was derived from tree-level amplitude in $\mathcal{N}=4$ SYM,
%
%\bea
%A^{\rm{tree}} = \frac{\big< 1 2\big>^4}{\big< 1 2\big> \big< 2 3\big> \cdots \big< n 1\big>}
%\eea
%
The soft bremsstrahlung factorization has the feature that this eikonal factor does not depend on the helicity of external particles.
The soft factorization also holds at higher loops. For example, at one loop, the scattering amplitude factorizes in the soft limit as
\bea
&& A^{(1)}_n (k_1, \cdots, k_a, k_b, k_c, \cdots, k_n)
\nonumber \\
&& \hskip0.5cm \longrightarrow S^{(0)}(k_a, k_b, k_c) A_{n-1}^{(1)}(k_1, \cdots, k_a, k_c, \cdots, k_n) + S^{(1)}(k_a,k_b, k_c) A^{(0)}_{n-1} (k_1, \cdots, k_a, k_c, \cdots, k_n). \qquad \qquad
\eea
Here, $S^{(1)}$ is the one-loop eikonal function. In dimensional regularization, it reads \cite{Bern:1998sc}:
\bea
S^{(1)}(k_a, k_b, k_c) = - S^{(0)}(k_a, k_b, k_c) {1 \over (4 \pi)^{2 - \epsilon}} {\Gamma(1 + \epsilon) \Gamma^2 (1 - \epsilon) \over \Gamma (1 - 2 \epsilon)} {1 \over \epsilon^2} \left( {(-s_{ac}) \mu^2 \over (- s_{ab}) (-s_{bc})} \right)^{\epsilon} {\pi \epsilon \over \sin (\pi \epsilon)},
\eea
where $s_{ab} = 2 k_a \cdot k_b$, etc. So, the soft factorization behavior is analogous to that of the collinear limit, just replacing the splitting function of the latter to the eikonal function.

%This eikonal factor do not depends on the helicity of external particles. This is the distinguished property of the soft bremsstrahlung factorization.
%\color{red} IS THE SOFT FACTORIZATION ONLY AT TREE-LEVEL? IF IT HOLDS ALSO AT HIGHER LOOPS, LIKE THE SPLITTING FUNCTION, WE SHOULD CONSIDER THE MOST GENERAL SITUATIONS IN THE ABOVE EQUATIONS. \color{black}
%The soft factorization also holds at higher loops. For example, at one loop, the scattering amplitude factorizes in the soft limit as
%
%\bea
%&& A^{(1)}_n (k_1, \cdots, k_a, k_b, k_c, \cdots, k_n)  \nonumber \\&& \hskip1cm \longrightarrow S^{(0)}(k_a, k_b, k_c) A_{n-1}^{(1)}(k_1, \cdots, k_a, k_c, \cdots, k_n) + S^{(1)}(k_a,k_b, k_c) A^{(0)}_{n-1} (k_1, \cdots, k_a, k_c, \cdots, k_n).\eea
%
%The factorization behavior is quite similar to that of the collinear limit, just replacing the splitting function of the latter to the eikonal function.

%Here, $S^{(1)}$ is the one-loop eikonal function (hep-ph/9810409)
%
%\bea
%S^{(1)}(k_a, k_b, k_c) = - S^{(0)}(k_a, k_b, k_c) {1 \over (4 \pi)^{2 - \epsilon}} {\Gamma(1 + \epsilon) \Gamma^2 (1 - \epsilon) \over \Gamma (1 - 2 \epsilon)} {1 \over \epsilon^2} \left( {(-s_{ac}) \mu^2 \over (- s_{ab}) (-s_{bc})} \right)^{\epsilon} {\pi \epsilon \over \sin (\pi \epsilon)}.
%\eea
%

The antenna function is a universal function introduced to describe in a unified manner all leading infrared singularities of tree-level scattering amplitudes as the color-connected set of momenta becomes collinear or soft. Consider, in color-order scattering amplitude, two hard momenta $k_a, k_b$ and one momentum $k_c$ in between. The unified factorization then takes the form
\begin{equation}
A_{n}^{(0)}(k_1, \cdots, k_a, k_c, k_b, \cdots, k_n) \quad \rightarrow \quad \sum_{\lambda} \mbox{Ant}(\hat{a},\hat{b} \leftarrow a,c,b)A_{n-1}(k_1, \cdots, -k_{\hat{a}},-k_{\hat{b}}, \cdots, k_n),
\end{equation}
where the antenna function Ant contains information of the parton $c$:
\begin{itemize}
\item collinear splitting function for $k_c \cdot k_a \rightarrow 0$ and $k_c \cdot k_b = $ finite  ($k_{\hat{a}}=-(k_a+k_c), k_{\hat{b}}=-k_b$)
\item collinear splitting function for $k_c \cdot k_b \rightarrow 0$ and $k_c \cdot k_a = $ finite ($k_{\hat{a}} = - k_a, k_{\hat{b}} = - (k_c + k_b)$)
\item soft eikonal function for both $k_c \cdot k_a \rightarrow 0$ and $k_c \cdot k_b \rightarrow 0$
($k_{\hat{a}}=-k_a, k_{\hat{b}}=-k_b$).
\end{itemize}
The momentua $k_{\hat{a}}, k_{\hat{b}}$ are reconstructed from the original momenta via the reconstruction function \cite{Kosower:2002su}. The antenna function can also be extended to higher loops in terms of parton currents $J$ that was used in the Berends-Giele recursion relations \cite{Berends:1987me}. At $L$-loops,
\bea
\mbox{Ant}^{(L)}(\hat{a}, \hat{b} \leftarrow a, c, \cdots, m, b)
= \sum_{\ell = 0}^L \sum_{i=1}^m J^{(\ell)}(a, c, \cdots, i; \hat{a}) J^{(L-\ell)}(i+1, \cdots, m, b; \hat{b})
\eea
Then, the factorization of the leading-color contribution to higher-loop scattering amplitudes can be derived by matching to known purely collinear limit or purely soft bremsstrahlung limit. This leads to
\begin{equation}
A_{n}^{(L)}(k_1, \cdots, k_n) \longrightarrow \sum_{\ell=0}^L \sum_{\lambda} \mbox{Ant}^{(\ell)} (\hat{a},\hat{b} \leftarrow a,1,b) \cdot A_{n-1}^{(L-\ell)} (k_1, \cdots, -k_{\hat{a}},-k_{\hat{b}}, \cdots, k_n).
\label{antenna-loop}
\end{equation}
This can be generalized to multiple collinear singularities that involve simultaneous vanishing of Mandelstam invariants in these collinear momenta and one of the two hard momenta $a$ or $b$. We can also generalize this to multiple collinear-soft or purely multiple soft singularities that arise from vanishing of additional Mandelstam invariants involving other hard momenta as well.

Note that the leading singularities in the additional Mandelstam invariants are already incorporated to the antenna function. Therefore, these singularities also capture the leading behavior in the multiple collinear-soft or multiple soft singularities. Indeed, the $h \rightarrow 0$ limit of the collinear splitting function must also describe the soft bremsstrahlung eikonal. As such, (\ref{antenna-loop}) describes the leading singularity behavior of $L$-loop leading-color scattering amplitudes in all singular limits involving the color-connected singular set of momenta $k_1, \cdots, k_m$.

One can generalize the factorization to multi-parton kinematics. The next level of factorization involves two unresolved parton kinematics. The factorization in doubly unresolved limit is given at $L$ loops by
\begin{equation}
A_n^{(L)}(k_1, \cdots, k_n)  \longrightarrow  \sum_{\ell=0}^L \sum_{\lambda} \mbox{\rm Ant}^{(\ell)}(\hat{a},\hat{b} \leftarrow a,1,2,b) \cdot A_{n-2}^{(\ell)}(k_1, \cdots,-k_{\hat{a}},-k_{\hat{b}},\cdots, k_n). \label{antftn}
\end{equation}
This antenna function have various channels, for instance, triple collinear, double collinear, collinear soft and double soft. Among them, we will focus on the first case that $s_{a1},s_{12},t_{a12}$ goes to 0.

Much the way the splitting function or the eikonal function are universal, we expect the antenna function also have universal structures.

%%%%%%%%%%%%%%%%%%%%%%%%%%%%%%%%%%%%%
\subsection{Lightlike Factorization of Wilson Loop}
One expects that the lightlike polygon Wilson loops provides another class of processes that exhibit IR divergences and factorizations thereof. Indeed shape or geometry of the lightlike contour $C_n$ exhibits two types of move that can be viewed as the soft and the collinear limits. The soft bremsstrahlung limit takes place when two adjacent vertex points coalesce. The collinear limit takes place when two adjacent edges coalesce, equivalently, when three consecutive vertices become lightlike arrayed.

\vskip0.5cm
\begin{center}
\bigskip
\fcolorbox{white}{white}{
  \begin{picture}(148,157) (142,-53)
    \SetWidth{2.0}
    \SetColor{Black}
    \Line(208,93)(160,69)
    \Line(160,69)(144,29)
    \Line(144,29)(152,-19)
    \Line(152,-19)(200,-51)
    \Line(200,-51)(280,-19)
    \Line(288,29)(280,-19)
    \SetWidth{0.3}
    \Line(208,93)(288,29)
    \SetWidth{1.0}
    \Line[dash,dashsize=10](208,93)(280,85)
    \Line[dash,dashsize=10](280,85)(288,29)
    \Vertex(224,93){2.828}
    \Vertex(280,85){2.828}
    \SetWidth{2.0}
    \Line(224,93)(288,29)
    \Line(224,93)(208,93)
    \SetWidth{1.0}
    \SetColor{Red}
    \Line[arrow,arrowpos=1,arrowlength=5,arrowwidth=2,arrowinset=0.2](280,101)(224,101)
  \end{picture}
}
\qquad
\qquad
\qquad
\fcolorbox{white}{white}{
  \begin{picture}(148,148) (142,-62)
    \SetWidth{2.0}
    \SetColor{Black}
    \Line(208,84)(160,60)
    \Line(160,60)(144,20)
    \Line(144,20)(152,-28)
    \Line(152,-28)(200,-60)
    \Line(200,-60)(280,-28)
    \Line(288,20)(280,-28)
    \SetWidth{0.3}
    \Line(208,84)(288,20)
    \SetWidth{1.0}
    \Line[dash,dashsize=10](208,84)(280,76)
    \Line[dash,dashsize=10](280,76)(288,20)
    \Vertex(256,52){2.828}
    \Vertex(280,76){2.828}
    \SetWidth{2.0}
    \Line(208,84)(256,52)
    \Line(256,52)(288,20)
    \SetWidth{1.0}
    \SetColor{Red}
    \Line[arrow,arrowpos=1,arrowlength=5,arrowwidth=2,arrowinset=0.2](288,68)(272,52)
  \end{picture}
}
\\
\medskip
\end{center}
Figure 4. {\sl Infrared singularities of lightlike polygon. There are two limits a contour $C_{n}$ can be reduced to $C_{n-1}$. The left figure describes the soft bremsstrahlung limit. The vertex $x_{i+1}$ coalesces to the adjacent vertex $x_i$, equivalently, the edge vector $y_i$ approaches to $0$. The right figure describes the collinear limit. The vertex $x_{i}$ approaches the lightlike edge connecting the vertices, $x_{i-1}$ and $x_{i+1}$, equivalently, two adjacent edge vectors $y_{i-1}$ and $y_i$ coalesce to a new lightlike vector. Although resulting topologies of are the same, the two limits should be distinguished. By analyticity, this is possible only if the limits are singular.}
\vskip0.5cm
The significance of these two processes is evident from geometric considerations among the vertex points $x_1, \cdots, x_n$. Generically, two non-adjacent vertex points are not lightlike separated. From either configurations, if we take succession of the above two processes for either vertex vectors or edge vectors within a lightlike polygon, we see that two non-adjacent cusp points of the polygon can be made lightlike separated. The limiting configuration is a lightlike polygon split to two lightlike polygons. Hereafter, this kinematic limit will be referred as {\sl lightlike factorization}. The classification is purely geomeric, so it must hold for observables defined for general quantum field theories of arbitrary spacetime dimensions.

In applying the above infrared factorizations of lightlike contour to ABJM Wilson loops, there is one further issue to be considered. We have shown in the last section that the lightlike ABJM Wilson loop cannot be defined on a polygon of odd numbers of edges since it does not permit configuration obeying the positivity condition \footnote{Recall also that this parallels to the fact that the ABJM scattering amplitudes involve even number of partons, though reasons are entirely different.}.
Whereas infrared factorization of the $(3+1)$-dimensional ${\cal N}=4$ SYM theory requires a single parton to fuse to other hard partons, infrared factorization of the $(2+1)$-dimensional ABJM theory requires two partons to fuse to other hard partons.
%This kinematics is referred to as doubly unresolved limit.
Thus, in the ABJM theory, we need to define an antenna function for double parton emissions.  \footnote{Antenna function was studied for scattering amplitudes in QCD and other gauge theories. We are adopting the same terminology to lightlike Wilson loop expectation values. In $(3+1)$-dimensional ${\cal N}=4$ SYM theory, the scattering amplitude - Wilson loop duality relates the universal splitting function for collinear limit in scattering amplitudes to the universal factorization for lightlike limit of Wilson loop expectation values.}.

Intuitively, the above discussion makes it clear that the lightlike factorization is {\sl universal} --- the factorization should be independent of geometric details of spectator vertices or edges in the rest of the polygon. In the ABJM theory, we explained in the previous section that the positivity condition of the Mandelstam variables and the closedness condition of the edge vectors restricted the contour to even number of vertices, equivalently, even number of edges. Consistency with these conditions require that the infrared singularity must involve odd numbers of consecutive edges fusing to a single edge and consecutive vertices pairing up to dimerized configuration. Therefore, the basic building block of the lightlike factorization of a polygon Wilson loop is the collinear-soft-collinear limit among 4 consecutive vertices, equivalently, 3 consecutive edges. We shall introduce the ABJM {\sl antenna function} that describes in a unified way all leading singularities of such processes.

Incidentally, we do not consider the limit where three consecutive edges are purely collinear. This is because the corresponding edge vectors in general violate the positivity condition. We will further discuss this restriction below. We also do not consider the limit where two consecutive edges are purely soft. Although kinematically permitted, this limit requires to take several Mandelstam invariants to zero simultaneously. Numerically, such a limit is technically involved and difficult to handle. In this paper, we will not study this corner of the moduli space and simply contend that the universal antenna function we derive below be reduced to the correct double eikonal function of such processes once relevant factorization is taken judicially.

Our next goal is to explicitly check the universality of the antenna function for the lightlike Wilson loop. By definition, the Wilson loop operator is color-ordered. Therefore, we can describe its collinear factorization in a manner similar to the color-ordered scattering amplitudes in QCD exhibits the factorization with respect to the collinear divergences \cite{Bern:1999a}.  So, take the collinear limit that three adjacent edge vectors $y_{i}, y_{i+1}$ and $y_{i+2}$ become lightlike parallel and coalesce to a new lightlike edge vector $y_P$. This situation is described by what we call `doubly unresolved limit' of $C_n \rightarrow C_{n-2}$:
\bea
y_{i}  \rightarrow (1 - h_1 - h_2) y_P,  \qquad y_{i+1} \rightarrow h_1 y_P, \quad y_{i+2} \rightarrow h_2 y_P \qquad
\mbox{where} \qquad y_P^2 = 0.
\label{kinematics}
\eea
In the doubly unresolved limit, we expect the lightlike polygon Wilson loop expectation value at $L$-loop factorizes universally as
\bea
\big< W_\Box [C_n] \big>^{(L)} \longrightarrow
\sum_{\ell=0}^L \mbox{Ant}^{(\ell)}(h_1, h_2; y_i, y_{i+1}, y_{i+2}) \big< W_\Box [C_{n-2}]
\big>^{(L-\ell)}.
\eea
We will abbreviate the antenna function that arises from factorization of the polygon $C_n$ Wilson loop as $\mbox{Ant}[C_n]$.
In the kinematics Eq.(\ref{kinematics}), the antenna function is closely related to the Wilson loop expectation value for the collapsing tetragon made of the edges $y_i, y_{i+1}, y_{i+2}, - y_P$. Our goal is to show that this antenna function is actually independent of the number of edges $n$ of the contour $C_n$ and hence universal. Note that the tree-level factorization for $L, \ell =0$:
\bea
\big< W_\Box[C_n] \big>^{(0)} \quad \longrightarrow \quad
\mbox{Ant}^{(0)}[C_n] \cdot \big< W_\Box [C_{n-2}] \big>^{(0)}
\eea
is actually a trivial statement since, in our normalization, all the quantities involved are 1.

At two-loop order, the doubly-unresolved configuration leads to the factorization:
\bea
\big< W_\Box [C_n] \big>^{(2)} \quad \longrightarrow \quad && \mbox{Ant}^{(2)}[C_4] \cdot \big< W_\Box [C_{n-2}] \rangle^{(0)} + \mbox{Ant}^{(0)} [C_4]  \cdot \big< W_\Box [C_n] \big>^{(2)}
\nonumber \\
=&& \big< W_\Box [C_{n-2}]\big>^{(2)} + \mbox{Ant}^{(2)}[C_4].
\eea
The antenna function $\mbox{Ant}[C_4]$ is local in color-ordered contour geometry, so it is independent of $n$ and universal:
\bea
\boxed{
\mbox{Ant}^{(2)}[C_4] = \big< W_\Box[C_n] \big>^{(2)} - \big< W_\Box[C_{n-2}] \big>^{(2)} \quad \mbox{for \ all} \ \  n.
}
\label{antftn}
\eea
%

%%%%%%%%%%%%%%%%%%%%%%%%%%%%%%%%%%%%%
\section{Antenna Function for the ABJM  Wilson Loops}
%%%%%%%%%%%%%%%%%%%%%%%%%%%%%%%%%%%%%

%To find Antenna function, we can approach by two channel. One way is using known analytic result of n-point scattering amplitude/n-gon Wilson loop. From (\ref{antftn}), Antenna function expected to be appeared by subtracting n-2 point amplitude from n point amplitude(with special kinematics). Since matter diagrams are equivalent to 1-loop $\mathcal{N}$=4 SYM, we have analytic n-point result inherited from BDS ansatz.

%Another way is calculating partial diagrams explicitly. Some diagrams are vanishing under special kinematics, someone cancelled by lower point amplitude and remaining diagrams will generates Antenna function. Regardless number of external legs, number of these partial diagrams which contributing Antenna function remains constant. This is another signal of universality.

%As indicated in last section, we will consider a situation that $s_{23},s_{34},s_{234} \rightarrow 0$. This special kinematics could be considered triple collinear limit($p_2 // p_3 // p_4$) or soft + collinear limit($p_3=0, p_2//p_4$).\footnote{Indeed, this soft+collinear limit is distinguished from collinear soft limit which appears in several literature. In ordinary case, collinear soft indicates $p_2=0, p_3//p_4$, in other words  $s_{23},s_{34},s_{234},s_{12} \rightarrow 0$.}

Built upon the idea of the previous section, we now construct the antenna function for the lightlike polygon Wilson loops in the ABJM theory. From (\ref{antftn}), the antenna function is obtained by subtracting $\big< W_\Box [C_{n-2}] \big>^{(2)}$ from $\big< W_\Box[C_n] \big>^{(2)}$. At first sight, it appears imperative to calculate $\big< W_\Box [C_{n-2}] \big>^{(2)}$ and $\big< W_\Box[C_n] \big>^{(2)}$. This turns out not the case, as most of the Feynman diagrams cancel each other. Eventually, only a small subset of Feynman diagrams contributes to the antenna function. In fact, the number of these diagrams are fixed regardless of $n$, which again is an indicative of the universality of the antenna function.

%%%%%%%%%%%%%%%%%%%%%%%%%%%%%%%%%%%
\subsection{Moduli Space of Lightlike Polygon Factorization}
In section 2, we already learned that anomalous conformal Ward identity offers hints on the analytic structure of the Wilson loop expectation value, separately for matter contribution and gauge boson contribution. As such, we shall consider the factorization limit for each contribution. In this subsection, we focus on lightlike factorization of matter contribution.

%%%%%%%%%%%%%%%%%%%%%%%%%%%%%%%%%%
\subsubsection{triple collinear and soft-collinear kinematics}
As alluded above, we will need to deal with doubly-unresolved configuration involving four consecutive vertices, say, $x_2, x_3, x_4, x_5$ on a polygon $C_n$. The lightlike factorization takes place when $x_2$ and $x_5$ are lightlike separated. To reach this configuration, take two step. First, take $x_2$ and $x_5$ lightlike seperated. This does not yet put the edge vectors $y_2,y_3,y_4$ %or $y_5,y_6,y_1$
parallel, nor the contour $C_n$ factorized into two parts. Next, take $x_{24}^2$ and $x_{35}^2$ to 0. This gives the triple collinear / soft collinear limit of the edge vectors $y_2, y_3, y_4$. Upon taking these limits, the upper tetragon flattened, reducing $C_n$ to $C_{n-2}$. Note that this limit still leaves the vertices $x_3$ and $x_4$ as unrestricted moduli parameters. A corner of this moduli space where the 3 Mandelstam variables $x_{24}^2,x_{35}^2,x_{25}^2$ go to zero. There are two ways to approach this corner:
\begin{itemize}
\item {\bf triple-collinear limit}
\bea
y_2 \parallel y_3 \parallel y_4
\eea
\item {\bf  soft-collinear limit}
\bea
y_3=0, \quad  y_2 \parallel y_4
\eea
\end{itemize}
%Let first observe about triple collinear limit. %\footnote{Indeed, this soft+collinear limit is distinguished from collinear soft limit which appears in several literature. In ordinary case, collinear soft indicates $p_2=0, p_3//p_4$, in other words  $s_{23},s_{34},s_{234},s_{12} \rightarrow 0$.}
%
\vskip0.5cm
\begin{center}
\bigskip
\fcolorbox{white}{white}{
  \begin{picture}(124,84) (64,-38)
    \SetWidth{2.0}
    \SetColor{Black}
    \Line(66,0)(72,-36)
    \SetWidth{0.4}
    \Line(66,0)(186,-6)
    \SetWidth{2.0}
    \Line(186,-6)(180,-36)
    \SetWidth{1.0}
    \Line[dash,dashsize=10](66,0)(90,42)
    \Line[dash,dashsize=10](90,42)(162,42)
    \Line[dash,dashsize=10](162,42)(186,-6)
    \SetWidth{2.0}
    \Line(66,0)(84,6)
    \Line(84,6)(168,6)
    \Line(168,6)(186,-6)
    \SetWidth{1.0}
    \SetColor{Red}
    \Line[arrow,arrowpos=1,arrowlength=5,arrowwidth=2,arrowinset=0.2](96,36)(90,12)
    \Line[arrow,arrowpos=1,arrowlength=5,arrowwidth=2,arrowinset=0.2](156,36)(162,12)
    \SetColor{Black}
    \Vertex(168,6){2.828}
    \Vertex(90,42){2.828}
    \Vertex(84,6){2.828}
    \Vertex(162,42){2.828}
  \end{picture}
}
\quad \quad \qquad \qquad
\fcolorbox{white}{white}{
  \begin{picture}(124,84) (64,-38)
    \SetWidth{2.0}
    \SetColor{Black}
    \Line(66,0)(72,-36)
    \SetWidth{0.4}
    \Line(66,0)(186,-6)
    \SetWidth{2.0}
    \Line(186,-6)(180,-36)
    \SetWidth{1.0}
    \Line[dash,dashsize=10](66,0)(90,42)
    \Line[dash,dashsize=10](90,42)(162,42)
    \Line[dash,dashsize=10](162,42)(186,-6)
    \Vertex(162,42){2.828}
    \Vertex(90,42){2.828}
    \Vertex(96,6){2.828}
    \Vertex(84,6){2.828}
    \SetWidth{2.0}
    \Line(66,0)(84,6)
    \Line(84,6)(96,6)
    \Line(96,6)(186,-6)
    \SetWidth{1.0}
    \SetColor{Red}
    \Line[arrow,arrowpos=1,arrowlength=5,arrowwidth=2,arrowinset=0.2](96,36)(90,12)
    \Line[arrow,arrowpos=1,arrowlength=5,arrowwidth=2,arrowinset=0.2](156,36)(108,12)
  \end{picture}
}
\end{center}
 Figure 5. {\sl There are two different special limits in doubly-unresolved geometries. Left figure describes the triple-collinear limit. Two vertices $x_{i+1}$ and $x_{i+2}$ approach a point on the edge connecting $x_i$ and $x_{i+3}$. Right figure describes the soft-collinear limit. The vertices $x_{i+1}$, $x_{i+2}$ come close each other and the edge vectors $y_{i-1}$ and $y_{i+1}$ become parallel.}
\vskip0.5cm
The triple-collinear limit is described by the contour geometry
%\footnote{ $s_{ab}\equiv (y_a+y_b)^2$, $y_{abc} \equiv (y_a+y_b+y_c)^2$ etc. Hence, $s_{23}$ equivalent to $x_{24}^2$ and $s_{234}$ equivalent to $x_{25}^2$.}
%
\bea
y_2 \equiv h_1 y_C, \quad y_3 \equiv h_2 y_C, \quad y_4=h_3 y_C \quad \mbox{where} \quad h_1, h_2, h_3 \ge 0, \quad h_1+h_2+h_3=1 \quad \mbox{and} \quad y_C^2=0,
\eea
where the 'parton fraction' $h_1, h_2, h_2$ spans the local chart of the moduli space. Naively, the dimension of this moduli space is $[0, 1] \times [0, 1]$. The actual moduli space turns out $[0, 1]$, as we now explain. The kinematics describes the limit that three consecutive segment vectors are parallel one another. The corresponding Mandelstam invariants are
\bea
x_{13}^2= x_{14}^2=h_1 x_{15}^2, \quad x_{46}^2= x_{36}^2=h_3 x_{26}^2 \quad
x_{24}^2=(1-h_3)^2 y_C^2, \quad x_{35}^2=(1-h_1)^2 y_C^2, \quad x_{25}^2=y_C^2 \rightarrow 0. \label{hextc_profile}
\eea
So we see that, as the three edge vectors $y_2, y_3, y_4$ become parallel one another, the three Mandelstam invariants $x_{25}^2, x_{24}^2, x_{35}^2$ goes to 0. Their ratios are fixed with respect to the parton fractions $h_1, h_2, h_3$. We then recall that the Mandelstam invariants of physical configuration must satisfy the Gram sub-determinant conditions. For the above triple-collinear configuration, the Gram sub-determinant condition requires $h_2^2 x_{15}^2 x_{26}^2 =0 $ and is solved by $h_2 \rightarrow 0$. Therefore, for all $n$, we must set $h_2 = 0$. This then leads to the moduli space of the triple-collinear limit to be the domain of $h_1 = - h_3$, viz. $[0, 1]$.
% Eventually, its topology approaches the soft-collinear limit. However, this cannot satisfy the Euclidean condition. Also, $x^2_{25}, x^2_{24}, x^2_{35}$ go to zero in both cases, but their ratios are different.

The soft-collinear limit is described by the contour geometry
\bea
y_2 \equiv h_1 y_C, \quad y_3 \equiv y_{S}, \quad y_4=h_3 y_C \quad \mbox{where} \quad h_1, h_3 \ge 0 , \quad h_1+h_3=1, \quad y_C^2=0, \quad y_{S} \simeq 0.
\eea
The moduli space of this configuration is given by the domain of $h_1 = - h_3$, viz. $[0, 1]$. This can be checked straightforwardly.
The contour geometry describes the limit that a diminishing edge vector is squeezed between two collinear edge vectors. The corresponding Mandelstam invariants are
\bea
x_{13}^2 = %z_1 x_{15}^2, \quad
x_{14}^2=h_1x_{15}^2, \quad
x_{46}^2= %z_3 x_{26}^2, \quad
x_{36}^2=h_3 x_{26}^2, \quad
x_{24}^2=h_1 x_{25}^2, \quad x_{35}^2=h_3 x_{25}^2, \quad x_{25}^2=2y_C \cdot y_{S} \label{hexsc_profile_a}
\eea
It is straightforward to check that this kinematics automatically satisfy the Gram sub-determinant conditions provided $y_C^2=0$ and $y_{S} \rightarrow 0$. The four Mandelstam invariants $x_{13}^2$, $x_{14}^2$, $x_{46}^2$ and $x_{36}^2$ coincides with the triple-collinear limit invariants if $h_2$ is taken to 0. However, the ratios among $x_{25}^2, x_{24}^2, x_{35}^2$ are different from the triple-collinear limit, so should be considered separately.

\subsubsection{Factorization and Positivity Condition}
When computing the antenna function in perturbation theory, we need to impose two conditions to the moduli parameters of the polygon $C_n$: the Gram sub-determinant condition and the positivity condition. We identified that the Gram sub-determinant conditions is satisfied by both the triple collinear geometry on the subspace $h_2=0$ and the soft-collinear geometry. What about the positivity condition? Here, we show that the soft-collinear geometry of the polygon is uniquely singled out as the configuration that satisfy the positivity condition.

The triple collinear geometry is inconsistent with the positivity condition. To see this, start from
\bea
y_2 = h_1 y_C, \quad y_3 = h_2 y_C, \quad y_4 = h_3 y_C \quad \mbox{with} \quad h_1, h_2, h_3 \ge 0, \quad h_1+h_2+h_3=1, \quad y_C^2=0.
\nonumber
\eea
For a given lightlike vector $y_C$, the time-component of the $y_2,y_3,y_4$ vectors have the same sign since $h_1, h_2, h_3$ are all positive. On the other hand, the positivity condition requires alternating sign flip of the time-component. As such, the triple-collinear geometry contradicts this condition. We discard the triple collinear limit hereafter.

On the other hand, the soft-collinear geometry turns out to satisfy the positivity condition. To illustrate this, take the hexagon and consider the following parametrization of the edge vectors
\begin{align}
y_2&=h_1(\sqrt{2},1,1), \quad \quad y_3=a(-\sqrt{2},1,1), \quad \quad y_4=h_3(\sqrt{2},1,1) \label{p2p3p4}
\end{align}
We introduced a small parameter $a \simeq 0$ to render the vector $y_3$ soft. Fusion of these three edge vectors result in a new lightlike edge vector $y_C = (\sqrt{2}, 1, 1)$. The other edge vectors $y_5, y_6$ and $y_1$ are set to
\begin{align}
y_5&=(-\sqrt{f^2+g^2},f,g), \quad \quad y_6=(\sqrt{r^2+s^2},r,s), \quad \quad y_1=(-\sqrt{b^2+d^2},b,d),
\end{align}
where $g,r,s$ are free parameters while other three parameters $b,d,f$ will be fixed by the closedness condition. The Mandelstam variables $x^2_{ij}$ can then be read from these edge vectors from the identities $- 2 y_i \cdot y_j = x^2_{i+1, j} + x^2_{i, j+1} - x^2_{i, j} - x^2_{i+1, j+1}$. This configuration satisfies the positivity condition.

\subsection{Matter Contribution to Antenna Function}
Let's begin with the ABJM matter contribution to the antenna function. By (\ref{antftn}), this contribution to the antenna function is extracted from
\bea
\mbox{Ant}^{(2)}_{\rm matter} [C_4] = \big< \widetilde{W}_\Box[C_n] \big>^{(2)}_{\rm matter} - \big< \widetilde{W}_\Box [C_{n-2}] \big>^{(2)}_{\rm matter}.
\eea
Here, $\big< \widetilde{W}_\Box[C_n] \big>^{(2)}_{\rm matter}$ refers to Wilson loop expectation value for the polygon $C_n$ of soft-collinear geometry.
%THE NOTATION IS NOT CONSISTENT. THE N-2 POLYGON DOES NOT INVOLVE ANY SOFT-COLLINEAR VECTORS. IT IS ONLY THE FIRST N POLYGON. SO, WE NEED TO PUT THE TILDE ONLY FOR THE FIRST ONE?
%\color{blue}
%The notation emphasizes that the reduced $(n-2)$-gon is not part of the original $n$-gon. For example, consider the tetragon $C_4$ with segment vectors $y_1, \cdots, y_4$. Mandelstam invariants are $x_{13}^2, x_{24}^2$. But here, $C_6 \rightarrow C_4$ after taking soft-collinear limit, segment vectors of reduced $C_4$ is $y_1, y_P, y_5, y_6$, not $y_1, y_2, y_3, y_4$. Related Mandelstam variables are $x^2_{15}, x^2_{26}$. Therefore, what I want to express here by tilde notation is tetragon result with such replacement.
%\color{black}
After the soft-collinear limit is taken, the reduced polygon $C_{n-2}$ consists of $(n-2)$ edge vectors $y_1,y_C,y_4, \cdots y_{n}$. Again, the new set of Mandelstam invariants $x_{ij}^2$ are obtained by inner product of these edge vectors.
The Wilson loop expectation value $\big< \widetilde{W}_\Box [C_{n-2}] \big>^{(2)}_{\rm matter}$ is obtained by inserting this new Mandelstam invariants to $\big< W_\Box [C_{n-2}] \big>^{(2)}_{\rm matter}$.

In case the three edge vectors $y_2, \ y_3, \ y_4$ coalesce in the soft-collinear limit, we shall call the vectors $y_1,y_2,y_3,y_4,y_5$ as `relevant edges'. We can then classify the matter-dependent 2-loop diagrams according to the locations the one-loop gauge propagator is attached:
\begin{align}
\mbox{Group A} \quad &: \quad \mbox{Neither end is attached to the relevant edges} \nonumber \\
\mbox{Group B} \quad &: \quad \mbox{One end is attached to the relevant edges while the other end is attached elsewhere} \nonumber \\
\mbox{Group C} \quad &: \quad \mbox{Both ends are attached to the relevant edges} \nonumber
\end{align}
We claim that Feynman diagrams belonging to Group A and Group B do not contribute to the antenna function. In other words,
\bea
&& \big< \widetilde{W}_\Box[C_n] \big>^{(2)}_{\rm matter}\Big|_{\mbox{\tiny{Group A}}} = \big< \widetilde{W}_\Box [C_{n-2}] \big>^{(2)}_{\rm matter}\Big|_{\mbox{\tiny{Group A}}}
\nonumber \\
&& \big< \widetilde{W}_\Box[C_n] \big>^{(2)}_{\rm matter} \Big|_{\mbox{\tiny{Group B}}} = \big< \widetilde{W}_\Box [C_{n-2}] \big>^{(2)}_{\rm matter} \Big|_{\mbox{\tiny{Group B}}}
\eea
Nontrivial contributions to the antenna function stem entirely from 9 Feynman diagrams belonging to the Group C.
\vskip0.5cm
\begin{center}
\bigskip
\fcolorbox{white}{white}{
  \begin{picture}(94,94) (76,-34)
    \SetWidth{2.0}
    \SetColor{Black}
    \Line(78,28)(96,52)
    \Line(96,52)(138,58)
    \Line(138,58)(168,40)
    \Line(168,40)(162,-8)
    \Line(162,-8)(138,-32)
    \Line(138,-32)(108,-32)
    \Line(108,-32)(78,-8)
    \Line(78,-8)(78,28)
    \SetWidth{1.0}
    \PhotonArc(133.637,60.726)(26.776,-163.23,-33.365){4}{5.5}
    \Vertex(132,34){6}
    \Line[dash,dashsize=10](78,-8)(162,-8)
  \end{picture}
}
\quad \quad
\fcolorbox{white}{white}{
  \begin{picture}(94,93) (76,-34)
    \SetWidth{2.0}
    \SetColor{Black}
    \Line(78,27)(96,51)
    \Line(96,51)(138,57)
    \Line(138,57)(168,39)
    \Line(168,39)(162,-9)
    \Line(78,-9)(78,27)
    \SetWidth{1.0}
    \PhotonArc(133.637,59.726)(26.776,-163.23,-33.365){4}{5.5}
    \Vertex(132,33){6}
    \SetWidth{2.0}
    \Line(78,-9)(162,-9)
    \SetWidth{0.1}
    \Line[dash,dashsize=10](78,-9)(108,-33)
    \Line[dash,dashsize=10](108,-33)(144,-33)
    \Line[dash,dashsize=10](144,-33)(162,-9)
  \end{picture}
}
\\
\bigskip
(a) \qquad \qquad \qquad \qquad \qquad \qquad (b)
\\
\bigskip
\fcolorbox{white}{white}{
  \begin{picture}(94,94) (76,-34)
    \SetWidth{2.0}
    \SetColor{Black}
    \Line(78,28)(96,52)
    \Line(96,52)(138,58)
    \Line(138,58)(168,40)
    \Line(168,40)(162,-8)
    \Line(78,-8)(78,28)
    \SetWidth{1.0}
    \PhotonArc[clock](177,-26)(75,-180,-253.74){4}{8.5}
    \Vertex(120,22){6}
    \SetWidth{2.0}
    \Line(78,-8)(108,-32)
    \Line(108,-32)(144,-32)
    \Line(144,-32)(162,-8)
    \SetWidth{1.0}
    \Line[dash,dashsize=10](78,-8)(162,-8)
  \end{picture}
}
\fcolorbox{white}{white}{
  \begin{picture}(94,94) (76,-34)
    \SetWidth{2.0}
    \SetColor{Black}
    \Line(78,28)(96,52)
    \Line(96,52)(138,58)
    \Line(138,58)(168,40)
    \Line(168,40)(162,-8)
    \Line(78,-8)(78,28)
    \SetWidth{1.0}
    \PhotonArc[clock](165.143,-2.286)(49.144,-142.797,-259.278){4}{8.5}
    \Vertex(120,16){6}
    \SetWidth{2.0}
    \Line(78,-8)(108,-32)
    \Line(108,-32)(144,-32)
    \Line(144,-32)(162,-8)
    \SetWidth{1.0}
    \Line[dash,dashsize=10](78,-8)(162,-8)
  \end{picture}
}
\fcolorbox{white}{white}{
  \begin{picture}(94,94) (76,-34)
    \SetWidth{2.0}
    \SetColor{Black}
    \Line(78,28)(96,52)
    \Line(96,52)(138,58)
    \Line(138,58)(168,40)
    \Line(168,40)(162,-8)
    \Line(78,-8)(78,28)
    \SetWidth{1.0}
    \PhotonArc[clock](156,16)(30,-90,-270){4}{8.5}
    \Vertex(126,16){6}
    \SetWidth{2.0}
    \Line(78,-8)(108,-32)
    \Line(108,-32)(144,-32)
    \Line(144,-32)(162,-8)
    \SetWidth{1.0}
    \Line[dash,dashsize=10](78,-8)(162,-8)
  \end{picture}
}
\fcolorbox{white}{white}{
  \begin{picture}(94,93) (76,-34)
    \SetWidth{2.0}
    \SetColor{Black}
    \Line(78,27)(96,51)
    \Line(96,51)(138,57)
    \Line(138,57)(168,39)
    \Line(168,39)(162,-9)
    \Line(78,-9)(78,27)
    \SetWidth{1.0}
    \PhotonArc[clock](162.75,1.5)(44.021,-166.201,-261.18){4}{6.5}
    \Vertex(126,27){6}
    \SetWidth{2.0}
    \Line(78,-9)(162,-9)
    \SetWidth{0.1}
    \Line[dash,dashsize=10](78,-9)(108,-33)
    \Line[dash,dashsize=10](108,-33)(144,-33)
    \Line[dash,dashsize=10](144,-33)(162,-9)
  \end{picture}
}
\\
\bigskip
(c) \qquad \qquad \qquad \qquad \qquad (d) \qquad \qquad \qquad \qquad \qquad (e) \qquad \qquad \qquad \qquad \qquad (f)
\\
\end{center}
Figure 6. {\sl The Feynman diagrams belonging to Group A and Group B. Upper diagrams belong to the Group A. (a) is equivalent to (b) when collinear limit taken. The diagrams in the second line belong to the Group B. (c)+(d)+(e) is equivalent to (f) after collinear limit. They are cancelled, therefore do not contribute to Antenna function.}
\vskip0.5cm
%---------------------------------------------------------------------------------------------------------------------------------------
We found that, for any $n$, there are always 9 types of diagram that contribute to the antenna function:
\begin{align}
\mbox{Group C} = \Big\{  I_{21}, \ \ I_{32}, \ \ I_{43}, \ \ I_{54}, \ \ I_{31}, \ \ I_{41}, \ \ I_{52}, \ \ I_{53}, \ \ I_{42} \Big\}
\end{align}
In other words, all nontrivial contributions to the antenna function are from `local moves' around the three edge vectors fusing one another. This features a heuristic and intuitive explanation for the universality.

We calculated these 9 diagrams and computed the antenna function. From the known analytic results of these diagrams, we took the soft-collinear limit and subtracted the relevant $C_{n-2}$ diagrams. Up to ${\cal O}(\epsilon^0)$, we found the result as
\bea
\mbox{Ant}^{(2)}_{\rm matter} &=& \frac{1}{4\epsilon^2} + \frac{1}{4\epsilon}\big(\mbox{Log} h_1 + \mbox{Log} h_3 + \mbox{Log}(x_{24}^2 + \mbox{Log}(x_{35}^2) \big) \nonumber \\
&+& \frac{1}{2} \mbox{Log} h_1 \mbox{Log}(x_{24}^2) +\frac{1}{2} \mbox{Log} h_3 \mbox{Log}(x_{35}^2) +\frac{1}{2} \mbox{Log} (x_{35}^2) \mbox{Log}(x_{24}^2) -\frac{1}{2} \mbox{Log}h_1 \mbox{Log}h_3 \nonumber \\
&-& \frac{\pi^2}{6}.
\eea
In obtaining this result, we used the Abel's identity for the dilogarithms :
\begin{equation}
\mbox{Li}_2(u)+\mbox{Li}_2(v)-\mbox{Li}_2(u v)=\mbox{Li}_2\Big(\frac{u-uv}{1-uv}\Big)+\mbox{Li}_2\Big(\frac{v-uv}{1-uv}\Big)-\mbox{log}\Big(\frac{1-u}{1-u v}\Big) \mbox{log}\Big(\frac{1-v}{1-u v}\Big)
\end{equation}
and the Landen's identity:
\begin{equation}
\mbox{Li}_2(x)+\mbox{Li}_2(\frac{1}{x})=\frac{\pi^2}{3}-\frac{1}{2}\mbox{Log}^2(x)-i\pi \mbox{Log}(x).
\end{equation}

%\begin{figure}
%\centering
%\includegraphics[scale=0.5]{vertex612613614basic}
%\caption{Diagrams $I_{614},I_{613}$ and $I_{612}$ are considered. After taking triple collinear limit, we numerically checked this agrees to vertex diagram in tetragon. Hence, this type diagrams are not contributed to splitting function.}
%\label{figure2}
%\end{figure}

%%%%%%%%%%%%%%%%%%%%%%%%%%%%%%%%%%%%%%%%%%%%%%%%%%%%%%%%%%%%%
\subsection{Chern-Simons Contribution to Antenna function}
We now turn to the contribution of the pure Chern-Simons sector to the antenna function.
%Our strategy for finding the antenna function for pure Chern-Simons sector is very similar to that of matter-dependent sector.
Again, we expect that
\bea
\mbox{Ant}^{(2)}_{\rm CS}[C_4] = \big< \widetilde{W}_\Box[C_n] \big>^{(2)}_{\rm CS} - \big< \widetilde{W}_\Box [C_{n-2}] \big>^{(2)}_{\rm CS}
\eea
Here, the Wilson loops $ \big< \widetilde{W}_\Box[C_n] \big>^{(2)}_{CS}$ and $\big< \widetilde{W}_\Box [C_{n-2}] \big>^{(2)}_{CS}$ are defined the same way as we defined for the matter contributions.

As explained in the previous section, the pure Chern-Simons contribution consists of the ladder diagrams and the triple-vertex diagrams. We found that the ladder diagrams does not give rise to infrared divergences, so they do not contribute to the antenna function. %Based on observation of matter-dependent Antenna function, we extend their argument to here.
We thus focus on the triple-vertex diagrams. We can again classify the relevant Feynman diagrams according to the combinatorics the triple gauge bosons are attached to the polygon $C_n$. As for the matter contributions, we showed in the last subsection that only `local moves' to the relevant edges contribute to the leading IR singularities. This turns out also the case for the Chern-Simons part: the contribution is completely determined by the triple-vertex diagrams whose gauge bosons are all attached to the relevant edges, $y_1, y_2, y_3, y_4, y_5$.

There are also IR divergences arising from `semi-local moves'. For instance, $I_{654}$ in $C_6$ is divergent. However, this divergence is cancelled by the diagram $I_{65P}$ in $C_{4}$ and $y_2 \parallel y_3 \parallel y_4$ when we compute the antenna function as the difference between the Wilson loop of $C_n$ and the Wilson loop of $C_{n-2}$. One readily notes that nontrivial contributions to the antenna function come from (1) the process that is divergent in $C_n$ but finite in $C_{n-2}$ and (2) the process that is finite in $C_n$ but divergent in $C_{n-2}$. These two processes are completely captured by the local moves to the relevant edges.

Recall the pure Chern-Simons result of the lightlike Wilson loop expectation value for hexagon and tetragon:
\begin{align}
\big< W_{\Box}[C_6]\big>^{(2)}_{\rm CS}&=-  \Big[ \frac{\Log(2)}{2} \frac{\sum_{i=1}^6(x_{i,i+2}^2  \pi e^{\gamma_E} \mu^2)^{2\epsilon}}{2\epsilon} - \frac{17}{16}\zeta_2 + {3 \over 4} \mbox{Log}^2(2) \Big] \nonumber \\
\big< W_{\Box}[C_4]\big>^{(2)}_{\rm CS}&=- \Big[\frac{\Log(2)}{2} \frac{\sum_{i=1}^4(x_{i,i+2}^2  \pi e^{\gamma_E} \mu^2)^{2\epsilon}}{2\epsilon} - \ \frac{5}{8}\ \zeta_2  + {1 \over 2}\mbox{Log}^2(2) \Big]. \label{4and6}
\end{align}
Then, $\big< \widetilde{W}_\Box[C_6] \big>^{(2)}_{\rm CS}$ is obtained by taking the soft-collinear geometry (\ref{hexsc_profile_a}) to $\big< W_\Box [C_6]\big>^{(2)}_{\rm CS}$. For $\big< \widetilde{W}_\Box [C_{4}] \big>^{(2)}_{\rm CS}$, we replace $x_{13}^2$ and $x_{24}^2$ in (\ref{4and6}) by $x_{15}^2$ and $x_{26}^2$. The contribution to the antenna function is then obtained from the difference
\begin{align}
\mbox{Ant}^{(2)}_{\rm CS}[C_6]&= \big< \widetilde{W}_\Box[C_6] \big>^{(2)}_{\rm CS} - \big< \widetilde{W}_\Box [C_{4}] \big>^{(2)}_{\rm CS} \nonumber \\
&=\Big[\frac{\mbox{Log}(2)}{2\epsilon} +\frac{1}{2}\mbox{Log}(2) \  \mbox{Log}( h_1) +\frac{1}{2}\mbox{Log}(2) \ \mbox{Log}( h_3) +\frac{1}{2}\mbox{Log}(2) \ \mbox{Log} (x_{24}^2) +\frac{1}{2}\mbox{Log}(2) \ \mbox{Log} (x_{35}^2) \nonumber \\
&+ \mbox{Ant}^{(2)}[C_6]\Big|_{\rm finite} \Big]. \label{ant64exp}
\end{align}
Here,
\bea
\mbox{Ant}^{(2)}[C_6] \Big|_{\rm finite} = -\frac{7}{4}\zeta_2 + \mbox{Log}^2(2).
\eea
%
%This is what we obtained from explicit calculation of the ten diagrams involving the relevant moves.

It turned out we need to numerically evaluate the Chern-Simons contribution to the antenna function. Hereafter,
we shall explicitly compute the contribution from the soft-collinear factorization of hexagon $C_6$ and octagon $C_8$ contours.
%%%%%%%%%%%%%%%%%%%%%%%%%%%%%%%%%
\subsubsection{hexagon $\longrightarrow$ tetragon}
For computational simplicity, let's first consider the soft-collinear factorization of hexagon $C_6$ to tetragon $C_4$. We computed contribution of the Chern-Simions contribution to the antenna function $\mbox{Ant}^{(2)}_{\rm CS}[C_6]$. Earlier, we alluded that
the gauge boson ladder diagrams do not contribute to the antenna function, though they do exhibit leading IR
singularities. We can classify the ladder diagrams into three groups:
\begin{align}
\mbox{Group A} &: \quad \{I_{3366},\ \ I_{6634},\ \ I_{6623}, \ \ I_{6624},\ \ I_{4466},\ \ I_{6622} \} \nonumber \\
\mbox{Group B} &: \quad \{ I_{5511} \}\nonumber \\
\mbox{Group C} &: \quad \mbox{all \ other \ diagrams} \nonumber
\end{align}
We found numerically that diagrams belonging to Group C vanishes in the soft-collinear limit.
We also checked numerically that the following identity holds:
\bea
\big< \widetilde{W}_\Box[C_n] \big>^{(2)}_{\rm ladder}\Big\vert_{A} - \big< \widetilde{W}_\Box [C_{n-2}] \big>^{(2)}_{\rm ladder} \Big\vert_{A} &=0 \nonumber \\
\big< \widetilde{W}_\Box[C_n] \big>^{(2)}_{\rm ladder} \Big\vert_{B} - \big< \widetilde{W}_\Box [C_{n-2}] \big>^{(2)}_{\rm ladder} \Big\vert_{B} &=0. \nonumber
\eea
We conclude that the ladder diagrams do not contribute to the antenna function.
\vskip0.5cm
\begin{center}
\bigskip
\fcolorbox{white}{white}{
  \begin{picture}(292,100) (76,-28)
    \SetWidth{2.0}
    \SetColor{Black}
    \Line(84,46)(114,70)
    \Line(114,70)(150,70)
    \Line(150,70)(174,40)
    \Line(174,16)(174,40)
    \Line(78,4)(84,46)
    \Line(78,4)(108,-26)
    \Line(108,-26)(144,-26)
    \Line(144,-26)(174,16)
    \SetWidth{1.0}
    \SetColor{Red}
    \Line[dash,dashsize=10](84,4)(174,16)
    \SetColor{Black}
    \Photon(84,34)(126,40){4}{4}
    \Photon(126,40)(90,-8){4}{6}
    \Photon(126,40)(162,-2){4}{5}
    \SetWidth{2.0}
    \Line(276,46)(306,70)
    \Line(306,70)(342,70)
    \Line(342,70)(366,40)
    \Line(366,16)(366,40)
    \SetWidth{1.0}
    \Photon(318,40)(318,10){4}{3}
    \SetWidth{2.0}
    \Line(270,4)(276,46)
    \SetWidth{1.0}
    \Line[arrow,arrowpos=1,arrowlength=5,arrowwidth=2,arrowinset=0.2](204,22)(246,22)
    \SetWidth{2.0}
    \SetColor{Red}
    \Line(270,4)(366,16)
    \SetWidth{1.0}
    \SetColor{Black}
    \Photon(276,34)(318,40){4}{4}
    \Photon(318,40)(354,16){4}{4}
  \end{picture}
}
\\
\bigskip
\bigskip
\fcolorbox{white}{white}{
  \begin{picture}(292,100) (76,-28)
    \SetWidth{2.0}
    \SetColor{Black}
    \Line(84,46)(114,70)
    \Line(114,70)(150,70)
    \Line(150,70)(174,40)
    \Line(174,16)(174,40)
    \Line(78,4)(84,46)
    \Line(78,4)(108,-26)
    \Line(108,-26)(144,-26)
    \Line(144,-26)(174,16)
    \SetWidth{1.0}
    \SetColor{Red}
    \Line[dash,dashsize=10](84,4)(174,16)
    \SetWidth{2.0}
    \SetColor{Black}
    \Line(276,46)(306,70)
    \Line(306,70)(342,70)
    \Line(342,70)(366,40)
    \Line(366,16)(366,40)
    \Line(270,4)(276,46)
    \SetWidth{1.0}
    \Line[arrow,arrowpos=1,arrowlength=5,arrowwidth=2,arrowinset=0.2](204,22)(246,22)
    \SetWidth{2.0}
    \SetColor{Red}
    \Line(270,4)(366,16)
    \SetWidth{1.0}
    \SetColor{Black}
    \PhotonArc[clock](133.764,-65.182)(69.204,127.12,65.92){4}{5.5}
    \Photon(126,-2)(132,-26){4}{3}
    \PhotonArc[clock](318.62,10.145)(35.861,-173.363,-350.603){4}{8.5}
    \Photon(324,46)(318,10){4}{3}
  \end{picture}
}
\\
\end{center}
Figure 7. {\sl Examples of two-loop Feynman diagrams that belong to the set $TV$. These diagrams are `local' and hence yields nontrivial contribution to the antenna function. The red edge denotes the  segment vector $y_C$ resulting from taking the soft-collinear limit.}
\vskip0.5cm
This brings us to the contribution of the triple-vertex diagrams.
We found that only the following 10 diagrams give rise to leading IR singularities
and hence can contribute to the antenna function:
\begin{equation}
TV =\{I_{321}, \ \ I_{432}, \ \ I_{543}, \ \ I_{421}, \ \ I_{532},  \ \ I_{431}, \ \ I_{542}, \ \ I_{521}, \ \ I_{531}, \ \ I_{541} \}
\end{equation}
We computed these diagrams numerically using the Mathematica package {\tt FIESTA}.
Table 6 summarizes inputs and numerical results of the IR finite part of the antenna function
derived from the hexagon Wilson loop.
\setlength{\tabcolsep}{1.8pt}
\setlength{\extrarowheight}{1.5pt}
{
\begin{table}[ht]
\centering
\begin{tabular}{|c||ccccccccc|c|c|}
 \hline
    &$x_{13}^2 $ &  $x_{24}^2$ & $x_{35}^2$ & $x_{46}^2$ & $x_{15}^2$ & $x_{26}^2$ & $x_{14}^2$ & $x_{25}^2$ & $x_{36}^2$ & $\big< W \big>_{\rm finite}$ & $\mbox{Ant}_{\rm finite}$ \\ \hline \hline
 (1) & -1.95797 & -0.0005 & -0.0005 & -2.13973 & -3.91357 & -4.28024 &
-1.95511 & -0.00100 & -2.14001 &-3.47673 & -2.38762   \\ \hline
(2) &-5.02424 & -0.00100 & -0.00100 & -5.09307 & -10.048 & -10.188
       &  -5.02275 & -0.00200 & -5.09389 & -3.47701 & -2.38886 \\ \hline
(3) &-8.83791 &-0.00100 & -0.00100 & -13.6207 & -17.6764 & -27.2428
       & -8.83749 & -0.00200 & -13.6211 &  -3.47126 & -2.39603 \\ \hline
(4) &-11.4515 & -0.00199 &-0.00100 &  -5.23415 & -22.9042 & -10.4691
       & -11.4517 & -0.00200 & -5.23392 & -3.4768 & -2.39128 \\ \hline
\end{tabular}
\caption{Table 6.  \sl Numerical result for the IR finite part of the lightlike hexagon Wilson loop expectation value and the antenna function. Notice that, for different configurations of the Mandelstam invariants, the results suggest that the IR finite part of the antenna function maintains a constant value.}
\end{table}
}
\vskip0.5cm

We also have an alternative method for calculating the antenna function. As in the two-parameter configuration of the hexagon, we can reduce the number of terms in the denominator of the Mellin-Barnes integrals by taking the soft-collinear configuration.
Moreover, we take a hint from the previous numerical results that the IR finite part of
the antenna function is independent of the polygon geometry.
This allows us to take the asymptotic limits for 3 of the Mandelstam invariants
$x_{25}^2$, $x_{15}^2$, $x_{26}^2$ and also for the the parton fractions
$h_1 \rightarrow 0, h_3 \rightarrow 1$. Taking these limits,
we succeeded in reducing to maximally 2-dimensional complex integrals.
Evaluating these integrals numerically,  we find that
\bea
\mbox{Ant}^{(2)}_{\rm CS} &=& \frac{0.346574}{\epsilon} \label{ant64} \\
&+&\frac{1}{2}\Log(2) \  \Log (z_1) +\frac{1}{2} \Log (2) \ \Log (z_3) +\frac{1}{2} \Log(2) \ \Log(x_{24}^2) +\frac{1}{2}\Log (2) \
\Log (x_{35}^2) \nonumber \\
&-& 2.398181603. \nonumber
\eea
This result fits to what we expect from (\ref{ant64exp}). The numerical constant in (\ref{ant64}) can be identified with
\bea
-2.398181603 := -\frac{7}{4}\zeta_2 + \log^2(2) = -2.398181603066195
\nonumber
\eea
within the precision of $\mathcal{O}(10^{-7})$. This result reassures our intuitive picture that only those Feynman diagrams that are local move to the soft-collinear fusion contribute to the antenna function.

\subsubsection{octagon $\longrightarrow$ hexagon}
To convince that the antenna function we derived is universal for all $n$, we also computed the Chern-Simons part of the antenna function $\mbox{Ant}^{(2)}_{\rm CS}[C_8]$ for the factorization of octagon $C_8$ to hexagon $C_6$. Again, the set of Feynman diagrams that contribute to the antenna function comes only from the triple-vertex diagrams and consists of the 10 diagrams $TV$. This is because there are 5 relevant edge vectors for the soft-collinear kinematics. In fact, upon careful diagrammatic considerations, we confirmed that this argument holds for arbitrary $n$.
%\begin{equation}
%X=\{I_{321}, \ I_{432}, \ I_{543}, \ I_{421}, \ I_{532},  \ I_{431}, \ I_{542}, \ I_{521}, \ I_{531}, \ I_{541}\}
%\end{equation}
%
\vskip0.5cm
\begin{center}
\bigskip
\fcolorbox{white}{white}{
  \begin{picture}(106,58) (64,-52)
    \SetWidth{2.0}
    \SetColor{Black}
    \Line(66,4)(72,-26)
    \Line(72,-26)(96,-50)
    \Line(96,-50)(138,-50)
    \Line(138,-50)(162,-26)
    \Line(162,-26)(168,4)
    \SetWidth{1.0}
    \Line[dash,dashsize=10](72,-26)(162,-26)
    \PhotonArc[clock](83.871,-43.198)(42.675,113.287,-9.171){4}{7.5}
    \Photon(108,-14)(84,-38){4}{3}
  \end{picture}
}
\quad
\fcolorbox{white}{white}{
  \begin{picture}(106,58) (64,-52)
    \SetWidth{2.0}
    \SetColor{Black}
    \Line(66,4)(72,-26)
    \Line(72,-26)(96,-50)
    \Line(96,-50)(138,-50)
    \Line(138,-50)(162,-26)
    \Line(162,-26)(168,4)
    \SetWidth{1.0}
    \Line[dash,dashsize=10](72,-26)(162,-26)
    \PhotonArc[clock](117,-39.8)(33.049,176.878,3.122){4}{8.5}
    \Photon(114,-8)(114,-50){4}{4}
  \end{picture}
}
\quad
\fcolorbox{white}{white}{
  \begin{picture}(106,58) (64,-52)
    \SetWidth{2.0}
    \SetColor{Black}
    \Line(66,4)(72,-26)
    \Line(72,-26)(96,-50)
    \Line(96,-50)(138,-50)
    \Line(138,-50)(162,-26)
    \Line(162,-26)(168,4)
    \SetWidth{1.0}
    \Line[dash,dashsize=10](72,-26)(162,-26)
    \Photon(126,-20)(150,-38){4}{3}
    \PhotonArc(145.364,-47.545)(37.444,63.622,183.759){4}{7.5}
  \end{picture}
}
\\
\bigskip
\bigskip
\fcolorbox{white}{white}{
  \begin{picture}(106,58) (64,-52)
    \SetWidth{2.0}
    \SetColor{Black}
    \Line(66,4)(72,-26)
    \Line(72,-26)(96,-50)
    \Line(96,-50)(138,-50)
    \Line(138,-50)(162,-26)
    \Line(162,-26)(168,4)
    \SetWidth{1.0}
    \Line[dash,dashsize=10](72,-26)(162,-26)
    \Photon(114,-8)(114,-50){4}{4}
    \PhotonArc(104.872,-55.553)(48.421,21.254,132.756){4}{8.5}
  \end{picture}
}
\quad
\fcolorbox{white}{white}{
  \begin{picture}(106,58) (64,-52)
    \SetWidth{2.0}
    \SetColor{Black}
    \Line(66,4)(72,-26)
    \Line(72,-26)(96,-50)
    \Line(96,-50)(138,-50)
    \Line(138,-50)(162,-26)
    \Line(162,-26)(168,4)
    \SetWidth{1.0}
    \Line[dash,dashsize=10](72,-26)(162,-26)
    \PhotonArc[clock](176.4,-137.6)(135.86,132.852,93.545){4}{7.5}
    \Photon(126,-14)(126,-50){4}{4}
  \end{picture}
}
\quad
\fcolorbox{white}{white}{
  \begin{picture}(106,58) (64,-52)
    \SetWidth{2.0}
    \SetColor{Black}
    \Line(66,4)(72,-26)
    \Line(72,-26)(96,-50)
    \Line(96,-50)(138,-50)
    \Line(138,-50)(162,-26)
    \Line(162,-26)(168,4)
    \SetWidth{1.0}
    \Line[dash,dashsize=10](72,-26)(162,-26)
    \PhotonArc[clock](79.543,-84.457)(84.394,98.548,33.399){4}{6.5}
    \Photon(120,-14)(114,-50){4}{3}
  \end{picture}
}
\\
\bigskip
\bigskip
\fcolorbox{white}{white}{
  \begin{picture}(106,58) (64,-52)
    \SetWidth{2.0}
    \SetColor{Black}
    \Line(66,4)(72,-26)
    \Line(72,-26)(96,-50)
    \Line(96,-50)(138,-50)
    \Line(138,-50)(162,-26)
    \Line(162,-26)(168,4)
    \SetWidth{1.0}
    \Line[dash,dashsize=10](72,-26)(162,-26)
    \PhotonArc[clock](196.616,-181.657)(181.922,127.846,99.05){4.5}{6.5}
    \Photon(120,-14)(150,-38){5}{3}
  \end{picture}
}
\quad
\fcolorbox{white}{white}{
  \begin{picture}(106,58) (64,-52)
    \SetWidth{2.0}
    \SetColor{Black}
    \Line(66,4)(72,-26)
    \Line(72,-26)(96,-50)
    \Line(96,-50)(138,-50)
    \Line(138,-50)(162,-26)
    \Line(162,-26)(168,4)
    \SetWidth{1.0}
    \Line[dash,dashsize=10](72,-26)(162,-26)
    \PhotonArc[clock](117.87,-88.957)(86.983,124.183,59.513){4.5}{6.5}
    \Photon(108,-8)(84,-38){4.5}{3}
  \end{picture}
}
\quad
\fcolorbox{white}{white}{
  \begin{picture}(106,58) (64,-52)
    \SetWidth{2.0}
    \SetColor{Black}
    \Line(66,4)(72,-26)
    \Line(72,-26)(96,-50)
    \Line(96,-50)(138,-50)
    \Line(138,-50)(162,-26)
    \Line(162,-26)(168,4)
    \SetWidth{1.0}
    \Line[dash,dashsize=10](72,-26)(162,-26)
    \PhotonArc[clock](117.87,-88.957)(86.983,124.183,59.513){4.5}{6.5}
    \Photon(108,-8)(120,-50){4.5}{4}
  \end{picture}
}
\\
\bigskip
\bigskip
\fcolorbox{white}{white}{
  \begin{picture}(106,58) (64,-52)
    \SetWidth{2.0}
    \SetColor{Black}
    \Line(66,4)(72,-26)
    \Line(72,-26)(96,-50)
    \Line(96,-50)(138,-50)
    \Line(138,-50)(162,-26)
    \Line(162,-26)(168,4)
    \SetWidth{1.0}
    \Line[dash,dashsize=10](72,-26)(162,-26)
    \PhotonArc[clock](117.87,-88.957)(86.983,124.183,59.513){4.5}{6.5}
    \Photon(108,-8)(150,-38){4.5}{3}
  \end{picture}
}
\\
\end{center}
Figure 8. {\sl Elements of $TV$. Regardless of $n$, there are always 5 relevant segment vectors for the soft-collinear kinematics. These 10 diagrams are expected to contribute to the leading IR singularities and hence to the antenna function.}
\vskip0.5cm
Fortuitously, the seven Feynman diagrams $\{I_{321}, \ I_{432}, \ I_{543}, \ I_{421}, \ I_{532},  \ I_{431}, \ I_{542}\}$ for the octagon $C_8$ are exactly the same as those for the hexagon $C_6$. The remaining three diagrams $\{ I_{521}, \ I_{541}, \ I_{531} \}$ depend on the invariant $y_1 \cdot y_5$. We also need to modify this invariant according to the substitution
\bea
- 2y_1 \cdot y_5 = x_{25}^2 - x_{15}^2 - x_{26}^2  \quad \mbox{(Hexagon)} \quad \longrightarrow \quad - 2y_1 \cdot y_5 = x_{16}^2 + x_{25}^2 - x_{15}^2 - x_{26}^2 \quad \mbox{(Octagon)},
\eea
%
%Because $x_{16}^2$ is no more 0.
from which we see that it generates an additional Mandelstam invariant $x_{16}^2$.
For details, see appendix H.

Such change of the Mandelstam invariant is an exception for the hexagon to octagon and are not needed for the polygon with $n \ge 8$. That is, the sum of diagrams in the set $TV$ yields the same result for all $n \ge 8$.
Therefore, we expect the Chern-Simons contribution to the antenna function is the same for all $n \ge 8$. Of course, this is the feature we expect from the universality of the antenna function.

To evaluate the ten Feynman diagrams belonging to the set $TV$, we start with the soft-collinear geometry of the octagon $C_8$:
\bea
y_2 \equiv h_1 y_C, \quad y_3 \equiv y_{S}, \quad y_4=h_3 y_C&, \quad h_1, h_3 \ge 0, \quad h_1+h_3=1, \quad y_C^2=0 \quad y_{S} \simeq 0 .
\nonumber
\eea
In this limit, the Mandelstam invariants scale as
\bea
x_{13}^2= %z_1 x_{15}^2, \quad
x_{14}^2=h_1 x_{15}^2, \quad x_{46}^2= %z_3 x_{26}^2, \quad
x_{36}^2=h_3 x_{26}^2, \quad
x_{24}^2=h_1 x_{25}^2, \quad x_{35}^2=h_3 x_{25}^2, \quad x_{25}^2=2 y_C\cdot y_{S} \nonumber
\eea
\bea
x_{47}^2=h_3 x_{27}^2+ h_1 x_{57}^2, \quad x_{38}^2 = h_1 x_{58}^2 + h_3 x_{28}^2, \quad
x_{37}^2=h_1 x_{57}^2+ h_3 x_{27}^2, \quad x_{48}^2= h_3 x_{28}^2+ h_1 x_{58}^2.
\label{octtc_profile}
\eea
Algorithmically, we can generate this configuration starting from the hexagon by adding two edge vectors $y_7$ and $y_8$ and then imposing the positivity condition. The results of our numerical computation are summarized in Table 6.  The results suggest that the finite part of the antenna function is independent of the choice of input Mandelstam invariants and that its numerical value is consistent with the numerical value extracted from the hexagon counterpart $\mbox{Ant}^{(2)}_{\rm CS, \ finite}$.

\setlength{\tabcolsep}{10pt}
\setlength{\extrarowheight}{1.5pt}
{
\begin{table}[!h]
\centering
\begin{tabular}{|c||c|c|}
 \hline
       & \{$x_{13}^2, x_{24}^2, x_{35}^2, x_{46}^2, x_{57}^2, x_{68}^2, x_{17}^2, x_{28}^2, x_{14}^2, x_{25}^2, x_{36}^2, x_{47}^2, x_{58}^2,$ &  $\mbox{Ant}_{\rm CS}[C_8]_{\rm finite}$ \\
       &  $x_{16}^2, x_{27}^2, x_{38}^2, x_{15}^2, x_{26}^2, x_{37}^2, x_{48}^2$\} & \\ \hline \hline
       & \{-0.53645, -0.00010, -0.00010, -0.47705, -0.25192, -11.3322 & \\
(1) &   -2.27493, -7.89600, -0.53633, -0.00020, -0.47688, -0.53960, -0.87527, &  -2.39734  \\
       & -2.00140, -0.82752, -4.38564, -1.07288, -0.95403, -0.53974, -4.38553  \} & \\ \hline
       %%%%%%%%%%%%%%%%%%%%%%%%%%%%%%%%%%%%%%%%%%%%%%%%%
       & \{-2.35947, -0.00020, -0.00020, -3.04945, -12.433, -4.83899 & \\
 (2) &  -15.7136, -14.7954, -2.35936, -0.00040, -3.04955, -15.0670, -8.89897, &  -2.39495  \\
       & -0.450812, -17.7019, -11.8471, -4.71903, -6.09920, -15.0677, -11.8471  \} & \\ \hline
       %%%%%%%%%%%%%%%%%%%%%%%%%%%%%%%%%%%%%%%%%%%%%%%%%
       & \{-1.25316, -0.00020, -0.00020, -2.83305, -25.2163, -5.38963 & \\
 (3) &   -40.8014, -13.3704, -1.25260, -0.00040, -2.83315, -27.9871, -5.72270, & -2.39414  \\
       & -1.94647, -30.7599, -9.54667, -2.50596, -5.66640, -27.9889, -9.5462  \} & \\ \hline
       %%%%%%%%%%%%%%%%%%%%%%%%%%%%%%%%%%%%%%%%%%%%%%%%%
       & \{-2.04042, -0.00020, -0.00020, -2.85248, -25.2976, -0.636597, & \\
 (4) &  -47.2655, -8.46627, -2.04047, -0.00040, -2.85243, -28.1002, -2.39373, & -2.39518  \\
       & -0.349137, -30.9049, -5.42984, -4.08109,-5.70511, -28.1021, -5.42996 \} & \\ \hline
       %%%%%%%%%%%%%%%%%%%%%%%%%%%%%%%%%%%%%%%%%%%%%%%%%
       & \{-2.28437, -0.00010, -0.00010, -6.81235, -63.2938, -9.61232, & \\
 (5) &  -91.7177, -14.8614, -2.28411, -0.00020, -6.81221, -69.9258, -0.960108,  & -2.39859  \\
       & -11.3101, -76.5589, -7.91075, -4.56858, -13.6247, -69.9268, -7.91066 \} & \\ \hline
       %%%%%%%%%%%%%%%%%%%%%%%%%%%%%%%%%%%%%%%%%%%%%%%%%
       & \{-0.654001, -0.00006, -0.00034, -5.93175, -30.7878, -3.90942, & \\
 (6) &  -60.0102, -8.71534, -0.653408, -0.00040, -5.93188, -36.6137, -0.131069, &  -2.38945 \\
       & -7.86413, -37.6436, -7.42770, -4.35975, -6.97872, -36.6154, -7.42736  \} & \\ \hline
\end{tabular}
\caption{\sl Numerical result for the Chern-Simons contribution to the two-loop antenna function $\mbox{Ant}^{(2)}_{\rm CS}[C_8]_{\rm finite}$.
%As we indicated, $\mbox{Ant}^{(2)}_{CS}[C_n]|_{Fin}$ also gives same result.
In numerical computation, we took $x_{25}^2 \simeq 0$ as a small quantity but not exactly zero. As we decrease $x_{25}^2$, we observed that $\mbox{Ant}^{(2)}_{\rm CS}[C_8]_{\rm finite}$ approaches to -2.39818.}
\end{table}
}
\vskip0.5cm
%In soft+collinear limit, we setted three Mandelstam variables $x_{24}^2,x_{35}^2,x_{25}^2$ approach to 0. But explicit replacement it to 0 lead to divergence. To avoid that, we treat them as small value, not 0. Starting from $-\frac{1}{5}$ we gradually decrease its value up to $-\frac{1}{2500}$. As expected it stably approach constant value as we decreasing $x_{25}^2$. This result presented in table 8.
%\setlength{\tabcolsep}{10pt}
%\setlength{\extrarowheight}{1.5pt}
%{
%\begin{table}[!h]
%\centering
%\begin{tabular}{|c||c|}
% \hline
%   & $\mbox{Ant}_{\small{\mbox{IR finite}}}$ \\ \hline \hline
% $x_{25}^2=-\frac{1}{5}$ & -2.25935 \\ \hline
% $x_{25}^2=-\frac{1}{10}$ & -2.28817\\ \hline
% $x_{25}^2=-\frac{1}{25}$ & -2.35553 \\ \hline
% $x_{25}^2=-\frac{1}{50}$ & -2.36843 \\ \hline
% $x_{25}^2=-\frac{1}{100}$ & -2.38061 \\ \hline
% $x_{25}^2=-\frac{1}{250}$ & -2.38801 \\ \hline
% $x_{25}^2=-\frac{1}{500}$ & -2.39183 \\ \hline
% $x_{25}^2=-\frac{1}{1000}$ & -2.394 \\ \hline
% $x_{25}^2=-\frac{1}{2500}$ & -2.39414 \\ \hline
%\end{tabular}
%\caption{Value of $\mbox{Ant}_{\small{\mbox{IR finite}}}$. As sending $x_{25}^2$ to 0, IR finite part of Antenna function goes to specific value -1.01189 stably.}
%\end{table}
%}
%\bigskip

From the numerical results based on the hexagon and octagon Wilson loops, we find that the Chern-Simons contribution to the two-loop antenna function is given by
\bea
\mbox{Ant}^{(2)}_{\rm CS}[C_n] &=& \frac{\Log(2)}{2\epsilon} \nonumber \\
&+& \frac{1}{2}\Log(2) \  \Log (z_1) +\frac{1}{2}\Log (2) \ \Log (h_3) +\frac{1}{2}\Log(2) \ \Log(x_{24}^2) +\frac{1}{2}\Log(2) \ \Log (x_{35}^2) %-2.39818
\nonumber \\
%&= \frac{\mbox{Log}2}{2\epsilon}+\frac{1}{2}\mbox{Log}2 \  \mbox{Log} h_1 +\frac{1}{2}\mbox{Log}2 \ \mbox{Log} h_3 +\frac{1}{2}\mbox{Log}2 \ \mbox{Log} (x_{24}^2) +\frac{1}{2}\mbox{Log}2 \ \mbox{Log} (x_{35}^2)
&-&\frac{7}{4}\zeta(2) + \Log^2(2).
\eea

\subsection{ABJM Antenna Function}
The two-loop antenna function of the ABJM theory is then obtained by adding the matter-dependent contribution $\mbox{Ant}^{(2)}_{\rm matter}[C_n]$ and the pure Chern-Simons contribution $\mbox{Ant}^{(2)}_{\rm CS}[C_n]$ and suitably rescaling the regulator energy scale. The result reads
\begin{align}
& \mbox{Ant}^{(2)}_{\rm ABJM}\nonumber \\
&=\frac{1}{4\epsilon^2} + \frac{1}{4\epsilon}\big(\Log (z_1) + \Log (z_3) + \Log(x_{24}^2 \hat{\mu}^2) + \Log(x_{35}^2 \hat{\mu}^2) \big) \nonumber \\
&+\frac{1}{2} \Log(z_1) \Log(x_{24}^2 \hat{\mu}^2) +\frac{1}{2} \Log( h_3) \Log(x_{35}^2 \hat{\mu}^2) +\frac{1}{2} \Log (x_{35}^2 \hat{\mu}^2) \Log(x_{24}^2 \hat{\mu}^2) -\frac{1}{2} \Log (z_1) \Log (z_3) \nonumber \\
&+ \frac{1}{2}\Log^2(2) -\frac{11}{4}\zeta(2) .
\end{align}
Here, $\hat{\mu}^2 = 2 \mu^2$ is the rescaled regularization scale. The result is independent of $n$, confirming our intuition that the IR factorization is a local move and hence the antenna function should be a universal quantity.

%%%%%%%%%%%%%%%%%%%%%%%%%%%%%%%%%%%%%%%%%%%%%%%%%%%%%%%%%%%%%
\section{Recursion Relations and ABJM Wilson Loop Expecation Value}
Having obtained the universal antenna function, in this section, we shall obtain the Wilson loop expectation value for arbitrary polygon with $n \ge 8$. The strategy is to utilize the lightlike factorization and derive recursion relations between Wilson loops for polygon contours $C_n$ of different $n$.

Let's start with the Chern-Simons contribution. The two-loop antenna function takes the form:
\begin{align}
\mbox{Ant}^{(2)}_{\rm CS}[C_n] &= \mbox{Ant}^{(2)}_{CS}[C_n]\Big|_{\rm div} + \mbox{Ant}^{(2)}_{\rm CS}[C_n] \Big|_{\rm finite}
\end{align}
Here, $\mbox{Ant}^{(2)}_{\rm CS}[C_n] \Big|_{\rm div}$ and $\mbox{Ant}^{(2)}_{\rm CS}[C_n] \Big|_{\rm finite}$ are IR divergent, respectively, IR finite parts:
\begin{align}
&\mbox{Ant}^{(2)}_{\rm CS}[C_n]\Big|_{\rm div} = \frac{\Log(2)}{2\epsilon} + \frac{1}{2}\Log(2) \left[  \Log (h_1) +  \Log (h_3) +  \Log (x_{24}^2) +  \Log (x_{35}^2) \right] \nonumber \\
&\mbox{Ant}^{(2)}_{\rm CS}[C_n] \Big|_{\rm finite}=-\frac{7}{4}\zeta(2) + \Log^2(2).
\end{align}
%
%Recall that $\mbox{Ant}^{(2)}_{\rm CS}[C_n] \Big|_{\rm div}$ arose  from the difference of IR divergences between Wilson loops of $C_n$ and $C_{n-2}$.
In deriving the Wilson loop expectation value, we are primarily interested in the analytic structure of the remainder function Rem$_{n,\rm CS}^{(2)}$. Therefore, it suffices to concentrate on the finite part, $\mbox{Ant}^{(2)}_{\rm CS}[C_n] \Big|_{\rm finite}$.

Intuitively, we can guess for the IR finite part of the remainder function,  Rem$_{n,\rm CS}^{(2)}-\frac{n}{2} \Log(2)$. As a first step, consider $n=8$ octagon. Before imposing the Gram sub-determinant conditions, there are twelve conformal cross-ratios for the octagon. The remainder function for the octagon is a function of these cross-ratios:
\bea
\mbox{Rem}_{8,\rm CS}^{(2)} = \mbox{Rem}_{8, \rm CS}^{(2)}\big(u_{14},u_{25},u_{36},u_{47},u_{58},u_{16},u_{27},u_{38},u_{15},u_{26},u_{37},u_{48}
\big)
\eea
In the soft-collinear limit,  $y_2 \parallel y_4$ and $y_3 \sim 0$, these cross-ratios are restricted accordingly:
\bea
u_{14} = u_{25} = u_{36} = u_{38} = u_{37} = 1, \quad u_{15} = 0 \nonumber
\eea
\bea
u_{47} = \frac{1-u_{26}}{u_{48}}, \quad u_{27} = \frac{1-u_{48}}{u_{26}}.
\eea
On the other hands, in the soft-collinear limit, the octagon $C_8$ is reduced to the hexagon $C_6$, for which the following nine Mandelstam invariants are relevant:
\bea
x_{16}^2, \ \ x_{15}^2, \ \ x_{26}^2, \ \ x_{17}^2, \ \ x_{27}^2, \ \  x_{28}^2,  \ \ x_{57}^2, \ \ x_{58}^2, \ \ x_{68}^2.
\eea
From these invariants, we can form the following three conformal cross-ratios:
\bea
u_{58} = \frac{x_{15}^2 x_{68}^2}{x_{58}^2 x_{16}^2}, \qquad u_{16} = \frac{x_{17}^2 x_{26}^2}{x_{16}^2 x_{27}^2}, \qquad u_{27} u_{37} u_{47} = \frac{x_{28}^2 x_{57}^2}{x_{27}^2 x_{58}^2}
\eea
Therefore, by the lightlike factorization, the finite part of the antenna function $\mbox{Ant}^{(2)}_{\rm CS}[C_8] \Big|_{\rm finite}$ must be reduced to
\bea
&&\mbox{Rem}_{8, \rm CS}^{(2)} \big(1,1,1, (1-u_{26})/u_{48},u_{58},u_{16}, (1-u_{48})/u_{26},1,0,u_{26},1,u_{48}
\big) - 4 \Log(2) \nonumber \\
&=& \mbox{Rem}_{6,\rm CS}^{(2)} \big(u_{58},u_{16},u_{27} u_{37} u_{47} \big) -3 \Log(2) + \mbox{Ant}^{(2)}_{\rm CS}[C_8] \Big|_{\rm finite} \nonumber \\
&=& \mbox{Rem}_{6,\rm CS}^{(2)} \big(u_{58},u_{16},u_{27} u_{37} u_{47} \big) -3 \Log(2) -\frac{7}{4}\zeta(2) + \Log^2(2) \nonumber \\
&=& -6\zeta(2) + 4\Log^2(2).
\eea
In the last expression, we used the numerical result that $\mbox{Rem}_{6, \rm CS}^{(2)} \big(u_{58},u_{16},u_{27} u_{37} u_{47} \big)$ is a  constant, independent of the input values of the conformal cross-ratios.

With such restricted information, it is impossible to determine general structure of the remainder function $\mbox{Rem}_{8, \rm CS}^{(2)}(u_{14},u_{25},u_{36},\cdots,u_{48})$ . However, for a given analytic structure of the remainder function, its soft-collinear limit should be controlled by the universal antenna function. This enables us to draw a conjecture that is consistent with the soft-collinear geometry to be
\bea
\mbox{Rem}_{8, \rm CS}^{(2)} = -6\zeta(2) + 4\mbox{Log}^2(2) + 4 \Log(2).
\eea
In other words, our conjecture is that the remainder function is independent of the twelve conformal cross-ratios. Moreover, utilizing numerical evidence that $\mbox{Rem}_{n, \rm CS}^{(2)}$ is constant-valued for arbitrary $n$, we find that the remainder function obeys the recursion relation
\bea
\boxed{
\mbox{Rem}_{n,\rm CS}^{(2)} - \frac{n}{2} \Log(2) = \mbox{Rem}_{n-2, \rm CS}^{(2)} -\frac{n-2}{2} \Log(2) + \mbox{Ant}^{(2)}_{\rm CS}[C_n]\Big|_{\rm finite}.
}
\eea
We can now iteratively solve this recursion relation along with our conjecture as the input:
\bea
\mbox{Rem}_{n,\rm CS}^{(2)} - \frac{n}{2} \Log(2) &=& \mbox{Rem}_{n-2, \rm CS}^{(2)} -\frac{n-2}{2} \Log(2) + \mbox{Ant}^{(2)}_{\rm CS}[C_n]\Big|_{\rm finite} \nonumber \\
&=& \mbox{Rem}_{n-2,\rm CS}^{(2)} -\frac{n-2}{2} \Log(2) -\frac{7}{4}\zeta(2) + \Log^2(2) \nonumber \\
&=& \cdots \nonumber \\
&=& \mbox{Rem}_{6, \rm CS}^{(2)} -3\Log(2) + \frac{n-6}{2} \big(-\frac{7}{4}\zeta(2) + \Log^2(2) \big) \nonumber \\
&=& \big[ \frac{1}{2} \Log^2(2) - \frac{7 \pi^2}{48}\big] n + \frac{\pi^2}{6}.
\eea
Our conjecture is further supported by numerical estimation of $\mbox{Rem}_{n, \rm CS}^{(2)}$ for $n=8,10, 12, \cdots, 20$ \cite{Wiegandt:2011a}.

Putting together, we now have the analytic result for the Chern-Simons contribution to the lightlike polygon Wilson loop expectation value as
\begin{equation}
\big< W_\Box [C_n]\big>^{(2)}_{\rm CS}= - \frac{\mbox{Log}(2)}{2} \sum_{i=1}^n \frac{(x_{i,i+2}^2 \pi e^{\gamma_E} \mu^2)^{2\epsilon}}{2\epsilon}+n \left( \frac{7 \pi^2}{48} - \frac{1}{2} \mbox{Log}^2(2)\right) - \frac{\pi^2}{6}. \label{w_result}
\end{equation}
Combining with the matter contribution (\ref{remainder-matter}), we finally arrives at the central result of this paper:
\begin{align}
\boxed{
\big< W_\Box [C_n]\big>^{(2)}_{\rm ABJM}= - {1 \over 2} \sum_{i=1}^n \frac{(x_{i,i+2}^2 8 \pi e^{\gamma_E} \mu^2)^{2\epsilon}}{(2\epsilon)^2} + \mbox{BDS}_n^{(2)}  + n \left( \frac{\pi^2}{12} +\frac{3}{4} \Log^2(2) \right) - \frac{\pi^2}{6}.
}
\end{align}

%%%%%%%%%%%%%%%%%%%%%%%%%%%%%%%%%%%%
%%%%%%%%%%%%%%%%%%%%%%%%%%%%%%%%%%%%
\section{Test: Spacelike Circular Wilson Loop}
Having obtained the ABJM Wilson loop expectation value for lightlike polygon contour, we now would like to put the result to a test.
In this section, we shall consider a specific thermodynamic limit of $n\rightarrow \infty$ so that $C_n$ approaches a spacelike circle or ellipse.
Exact result of the Wilson loop expectation value for the circular contour is already known for pure Chern-Simons theory and for the ABJM theory by localization methods. We compare our result in this specific $n \rightarrow \infty$ limit with these exact results.

In Euclidean space, a circle (more generally an ellipse) is obtainable from a lightlike polygon $C_n$ by inscribing its vertices to touch the circle and taking the continuum limit $n \rightarrow \infty$. In Lorentzian spacetime, a spacelike circle (more generally an ellipse) is obtainable from a lightlike polygon $C_n$ by inscribing its edges to cross the circle and taking the continuum limit $n \rightarrow \infty$. It should therefore be possible to obtain a spacelike circular Wilson loop expectation value from the specified continuum limit of the lightlike polygon Wilson loop expectation value. Such test was first studied in \cite{Maldacena:2009a}.  %Therefore, one can naively expect that expectation value of regular polygon Wilson loop implies information of circular Wilson loop expectation value at $n\rightarrow \infty$ limit.

Geometrically, the biggest difference of the lightlike polygon from the spacelike circular loop is the existence of vertices, where the polygon contour forms sharp cusps. Associated with these cusps are the UV divergences of the Wilson loop. However, these divergent parts are essentially abelian and they can be exponentiated and factored out. The remaining finite part should then be relatable to the spacelike circular Wilson loop expectation value in the thermodynamic limit. Roughly speaking, each vertices can be viewed as an elementary excitation along the loop and the $n \rightarrow \infty$ limit will populate the excitations such that the Wilson loop can be treated as a statistical system.

To obtain the circular Wilson loop, we shall arrange the geometry of the $n$ vertices as \cite{Heslop:2010a}
\bea
x_{2k} = \left(2 \ \mbox{sin} \frac{\pi}{2n}, \quad \mbox{cos} \frac{(2k+1)\pi}{n}, \quad \mbox{sin} \frac{(2k+1)\pi}{n} \right), \qquad x_{2k+1} = \left(0, \quad \mbox{cos} \frac{2k\pi}{n}, \quad \mbox{sin} \frac{2k\pi}{n} \right).
\label{heslop}
\eea
This configuration yields a polygon whose contour is sandwiched between a stack of two spacelike circles, all lying within $(1+1)$-dimensional subspace. Mandelstam invariants of the polygon are given by
\begin{align}
x_{2k,2j}^2&=-4\mbox{sin}^2\frac{(k-j)\pi}{n} \label{man_a} \\
x_{2k,2j+1}^2&= - 4\Big(\mbox{sin}^2\frac{(k-j-\frac{1}{2})\pi}{n} - \mbox{sin}^2\frac{\pi}{2n} \Big) \label{man_c} \\
x_{2k+1,2j+1}^2&=x_{2k,2j}^2 = -4\mbox{sin}^2\frac{(k-j)\pi}{n}. \label{man_b}
\end{align}
Evidently, the polygon satisfies the positivity condition and the closedness condition. Also, $(x_{2k+1}-x_{2k})^2 = 0$ holds.\\
\begin{center}
\fcolorbox{white}{white}{
  \begin{picture}(440,156) (55,-15)
    \SetWidth{1.0}
    \SetColor{Black}
    \Line[arrow,arrowpos=1,arrowlength=5,arrowwidth=2,arrowinset=0.2](176,36)(176,124)
    \Line[arrow,arrowpos=1,arrowlength=5,arrowwidth=2,arrowinset=0.2](176,36)(120,-12)
    \Line[arrow,arrowpos=1,arrowlength=5,arrowwidth=2,arrowinset=0.2](176,36)(264,36)
    \Line[arrow,arrowpos=1,arrowlength=5,arrowwidth=2,arrowinset=0.2](276,76)(312,76)
    \Line[arrow,arrowpos=1,arrowlength=5,arrowwidth=2,arrowinset=0.2](372,36)(372,124)
    \Line[arrow,arrowpos=1,arrowlength=5,arrowwidth=2,arrowinset=0.2](372,36)(460,36)
    \Line[arrow,arrowpos=1,arrowlength=5,arrowwidth=2,arrowinset=0.2](372,36)(316,-12)
    \SetWidth{2.0}
    \Oval(176,36)(16,56)(0)
    \Oval(176,76)(16,56)(0)
    \Oval(372,36)(16,56)(0)
    \SetWidth{1.0}
    \SetColor{Red}
    \Line(120,36)(136,68)
    \Line(136,68)(152,20)
    \Line(152,20)(176,60)
    \Line(176,60)(192,20)
    \Line(192,20)(216,68)
    \Line(216,68)(232,36)
    \Line(232,36)(216,84)
    \Line(120,36)(136,84)
    \Line[dash,dashsize=10](136,84)(152,52)
    \Line[dash,dashsize=10](152,52)(168,92)
    \Line[dash,dashsize=10](168,92)(192,52)
    \Line[dash,dashsize=10](192,52)(216,84)
    \Line(324,44)(316,36)
    \Line(316,36)(328,28)
    \Line(328,28)(352,20)
    \Line(352,20)(368,20)
    \Line(368,20)(380,20)
    \Line(380,20)(396,24)
    \Line(396,24)(416,28)
    \Line(416,28)(428,36)
    \Line(428,36)(424,44)
    \Line(424,44)(412,48)
    \Line(408,48)(396,52)
    \Line(396,52)(384,52)
    \Line(384,52)(368,52)
    \Line(368,52)(356,52)
    \Line(360,52)(344,48)
    \Line(344,48)(340,48)
    \Line(340,48)(324,44)
    \SetColor{Black}
    \Line[arrow,arrowpos=1,arrowlength=5,arrowwidth=2,arrowinset=0.2](100,56)(100,36)
    \Line[arrow,arrowpos=1,arrowlength=5,arrowwidth=2,arrowinset=0.2](100,56)(100,80)
    \Text(52,56)[lb]{\Large{\Black{$2 \mbox{sin} \frac{\pi}{2n}$}}}
    \Text(276,88)[lb]{\Large{\Black{$n \rightarrow \infty$}}}
    \Text(184,120)[lb]{\Large{\Black{$t$}}}
    \Text(108,-4)[lb]{\Large{\Black{$x$}}}
    \Text(460,48)[lb]{\Large{\Black{$y$}}}
    \Text(380,120)[lb]{\Large{\Black{$t$}}}
    \Text(304,-4)[lb]{\Large{\Black{$x$}}}
    \Text(264,48)[lb]{\Large{\Black{$y$}}}
  \end{picture}
}
\label{largen}
\end{center}
Figure 9. {\sl The geometry of thermodynamic limit of lightlike polygon inscribed between two $(1+1)$-dimensional spacelike circular loops. }

%----------------------------------------------------------------------------------------------------------------------------------------
\vskip0.5cm
There is a slight complication. As mentioned in Section 3, this $(1+1)$-dimensional kinematics would not satisfy the Gram sub-determinant conditions. While this is true for finite $n$, we now argue that the conditions are fully satisfied in the $n \rightarrow \infty$ thermodynamic limit.

First, focus on the pure Chern-Simons contribution to the Wilson loop expectation value. From the results in the section 8, we expect the result to be
\begin{equation}
\big< W_\Box [C_n]\big>^{(2)}_{\rm CS}= - \frac{\mbox{Log}(2)}{2} \sum_{i=1}^n \frac{(x_{i,i+2}^2 \pi e^{\gamma_E} \mu^2)^{2\epsilon}}{2\epsilon}+n \left( \frac{7 \pi^2}{48} - \frac{1}{2} \mbox{Log}^2(2)\right) - \frac{\pi^2}{6}. \label{CSstructure}
\end{equation}
were if the configuration to satisfy the Gram sub-determinant conditions. As said, the configuration (\ref{heslop}) does not satisfy them. However, we note that ($k \times k$) Gram sub-determinant conditions consist of $k$-th power of $\mbox{sin}^2 \frac{\pi}{n}$, because of all the Mandelstam invariants (\ref{man_c}) are accompanied by the factor of $\mbox{sin}^2 \frac{\pi}{n}$. Therefore, all the Gram sub-determinant conditions would vanish in the thermodynamic limit  $n \rightarrow \infty$ and (\ref{CSstructure}) would approach the correct result. Geometrically, distance between the two enveloping circles $2 \mbox{sin} \frac{\pi}{2n}$ goes to zero, and the polygon collapses to a spacelike circle. This is illustrated in Figure \ref{largen}.
%Again, for the pure Chern-Simons part, the lightlike polygon Wilson loop expectation value was
%
%\begin{equation}
%\big< W_\Box [C_n]\big>^{(2)}_{\rm CS}= - \frac{\mbox{log}(2)}{2} \sum_{i=1}^n \frac{(x_{i,i+2}^2 \pi e^{\gamma_E} \mu^2)^{2\epsilon}}{2\epsilon}+n \left( \frac{7 \pi^2}{48} - \frac{1}{2} \mbox{Log}^2(2)\right) - \frac{\pi^2}{6}.
%\end{equation}
%
Dropping off the UV divergent parts, the Chern-Simons contribution reads
\begin{equation}
\big< W_\Box [C_n]\big>^{(2)}_{\rm CS}\Big|_{\rm{finite}}= n \left( \frac{7 \pi^2}{48} - \frac{1}{2} \mbox{Log}^2(2)\right) - \frac{\pi^2}{6}.
\end{equation}
The result is independent of the shape or the geometry of $C_n$. This fits well with the expectation that the pure Chern-Simons theory is topological. The thermodynamic limit $n \rightarrow \infty$ gives rise to a linear divergence, proportional to the perimeter $n := 2 \pi R/a$ for a circle of radius $R$ and short-distance defining scale $a$. It would be very interesting to understand this from the viewpoint of polygon regularization of the topological link invariants.

For the matter contribution part, the lightlike polygon Wilson loop expectation value was
\bea
\big< W_\Box[C_n]\big>^{(2)}_{\rm matter}
&=& - {1 \over 2}\sum_{i=1}^n \frac{1}{(2\epsilon)^2}(x_{i,i+2}^2 4\pi e^{\gamma_E} \mu^2)^{2\epsilon}+\mbox{BDS}^{(2)}_n-\frac{1}{16}n \pi^2 +{\cal O}(\epsilon).
\eea
Again, dropping off the UV divergent part of the result of Section 3.1, we have
\bea
\big< W_\Box[C_n]\big>^{(2)}_{\rm matter}\Big|_{\rm finite} &=& \sum_{i>j+1}^n I_{i,j}(x) ,
\eea
where
\bea
I_{i,j}(x) \equiv -\mbox{Li}_2(1-as)-\mbox{Li}_2(1-at)+\mbox{Li}_2(1-aP^2)+\mbox{Li}_2(1-aQ^2)-\frac{1}{16}n \pi^2.
\eea
\label{reg_pola}
%
%
%\bea
%\big< W_\Box[C_n]\big>^{(2)}_{\rm matter}\Big|_{\rm finite} = -\mbox{Li}_2(1-as)-\mbox{Li}_2(1-at)+\mbox{Li}_2(1-aP^2)+\mbox{Li}_2(1-aQ^2)-\frac{1}{16}n \pi^2.
%\eea
%\label{reg_pola}
%
We now recall from Section 3.1 that the parameters were defined by
\bea
a=\frac{s+t-P^2-Q^2}{st-P^2 Q^2}, \quad P^2=x_{i,j+2}^2, \quad Q^2=x_{i+1,j}^2, \quad s=x_{i,j}^2, \quad t=x_{i+1,j+1}^2.
\eea
Unlike the pure Chern-Simons part, the matter contribution part is sensitive to the shape or the geometry of the polygon $C_n$. Again,
this fits to our intuition that, once the matter is coupled, the Chern-Simons theory is no longer topological.

The next step is to consider a suitable geometrical limit so that the lightlike polygon $C_n$ asymptotes to the spacelike circle $O$. This regular polygon geometry is specified by (\ref{man_a}), (\ref{man_b}) and (\ref{man_c}) and the thermodynamic limit $n \rightarrow \infty$. We took this configuration to $\langle W_\Box[C_n] \rangle^{(2)}_{\rm matter}\Big\vert_{\rm finite}$ and evaluated its value numerically with respect to $n$.

\begin{figure}[!h]
\centering
\includegraphics[scale=0.6]{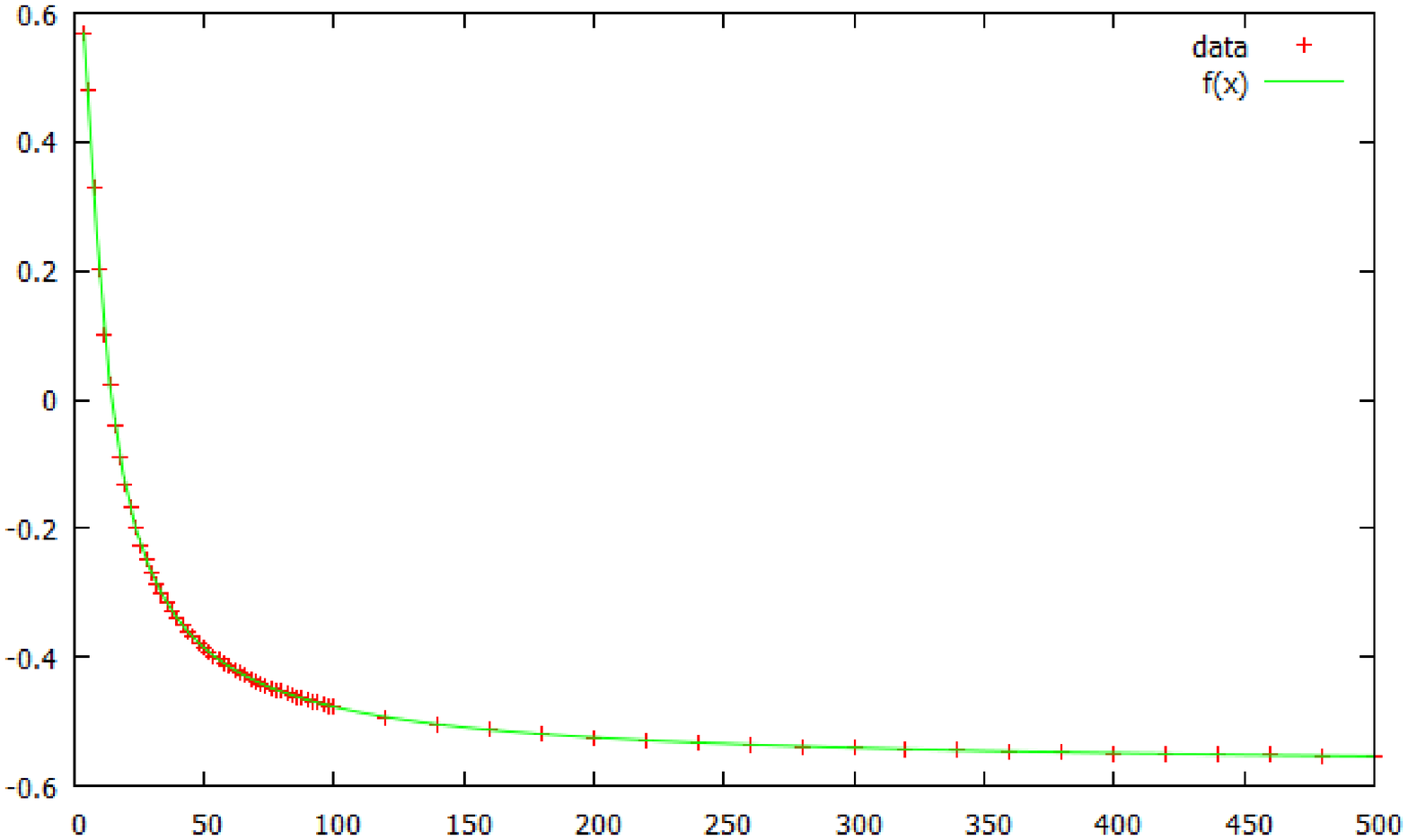}
\label{figure_regular}
\end{figure}

Geometrically, the regular polygon has the property $x_{i,j}^2 = x_{i+1, j+1}^2 = \cdots = x_{i-1,j-1}^2$. This leads to the relation
\bea
I_{i,j}(x) = I_{i+1,j+1}(x) = \cdots = I_{i-1,j-1}(x) . \label{perm}
\eea
Define $F(n)$ and $f(n)$ by
\bea
F(n) &\equiv& \sum_{\substack{i>j \\ i \neq j+1}} I_{i,j} = \sum_{\substack{\rm{permutation} \\ \rm{of} \ j_*}} \sum_{\substack{i=j_*+2}}^{n+j_*-2} I_{i,j=j_*} \nonumber \\
f(n) &\equiv& \sum_{\substack{i=j_*+2}}^{n+j_*-2} I_{i,j=j_*},
\eea
where $j_*$ is some arbitrary reference number. By definition, $F(n)$ is equal to $\big< W_\Box[C_n]\big>^{(2)}_{\rm matter}|_{\rm finite}$. Also, (\ref{perm}) ensures that $f(n)$ is equal to ${1 \over n} F(n)$. We numerically computed the function $f(n)$ with high precision and fitted with 3-parameter $f(n)=\frac{a}{n^2}+\frac{b}{n}+c$. As we seen in Figure \ref{figure_regular}, this fitting works extremely well over two-orders of magnitudes, $6 \le n \le 500$. We found that the least chi-square fit of the coefficients $a,b,c$ to be
\bea
a = -21.1513 (\pm 0.005399), \qquad b = 9.86953(\pm 0.001135), \qquad c = -0.573233 (\pm 0.0002458)
\eea
Note that the value of $b$ very close to $\pi^2 := 9.869604401$.

With this result, we have
\bea
\big< W_\Box[C_n]\big>^{(2)}_{\rm matter}|_{\rm finite} = F(n) = \frac{a}{n} + \pi^2 + c n.
\eea
The first term dies out in the thermodynamic limit $n \rightarrow \infty$.
Figure 10. {\sl The least chi-square fit of the function $f(n)$ with respect to $n$. Red crosses are numerical data obtained from (\ref{reg_pola}). The green line is fitting function $f(n)=\frac{a}{n^2}+\frac{b}{n}+c$.}
\vskip0.5cm

Ultimately, we expect the UV finite part of the ABJM Wilson loop for the regular polygon asymptotes at large $n$ limit to
\begin{align}
\big< W_{\Box}[C_n^{\rm{regular}}]\big>^{(2)}_{\rm{ABJM}}|_{\rm{finite}} &= \big< W_{\Box}C_n^{\rm{regular}}]\big>^{(2)}_{\rm{matter}}|_{\rm{finite}} + \big< W_{\Box}[C_n^{\rm{regular}}]\big>^{(2)}_{\rm{CS}}|_{\rm{finite}}\nonumber \\
&=\rho n + \big(\pi^2 -\frac{\pi^2}{6} \big)  \label{result}
\end{align}
We replaced the $n$-independent constant by $\pi^2$, as evidenced by the above numerical fitting, and included $-\frac{\pi^2}{6}$ inherited from the $\big< W_\Box [C_n]\big>^{(2)}_{\rm CS}|_{\rm{finite}}$.

Let us compare this result with the previous results known for a spacelike circular ABJM Wilson loop.
In \cite{Rey:2008a}, the circular Wilson loop expectation value in ABJM theory was first computed and the result can be expanded in perturbative series of $\lambda$ was
\begin{equation}
\big< W_{\Box}[\bigcirc]\big>_{\rm{ABJM}}=1+\lambda^2 \big(\pi^2 -\frac{\pi^2}{6} \big)+ {\cal O}(\lambda^4).
\end{equation}
Also, using exact supersymmetry localization technique, the circular Wilson loop expectation value in ABJM theory was computed \cite{Marino:2009a}. The result, expanded in perturbative series of $\lambda$, is given by
\begin{equation}
\big< W_\Box[\bigcirc] \big>_{\rm ABJM} = 1+\frac{5\pi^2}{6} \lambda^2 + \mathcal{O}(\lambda^3)
\end{equation}
Our results in the $n \rightarrow \infty$ thermodynamic limit (\ref{result}) reproduces these previous results, thus passing a nontrivial consistency check.

\section{Comments on ABJ Theory}

In this paper, we studied quantum properties of the lightlike polygon Wilson loop expectation value in the ABJM theory. By invoking lightlike factorization and recursion relations thereof, we obtained the result for arbitrary polygon $C_n$ up to two loops. As a checkpoint, we took the continuum limit $n \rightarrow \infty$ and extracted the expectation value of spacelike circular Wilson loop. The result matched perfectly with previously known analytic results.

There are also variant conformal field theories related to the ABJM theory. The so-called ABJ theory has gauge group of different rank. As most of our considerations were based on the conformal invariance, it is of interest to explore the expectation value of the lightlike Wilson loop in this theory.

We already explained that the only difference of the ABJ theory from the ABJM theory is that the rank of gauge group is unequal: $G = U(N_1) \times U(N_2)$, but with the same Chern-Simons level. In the limit of infinite number of colors, $N_1, N_2 \rightarrow \infty$ , there are now two `t Hooft coupling constants, $\lambda_1 \equiv \frac{N_1}{k}$ and $\lambda_2 \equiv \frac{N_2}{k}$. The contribution of pure Chern-Simons system come from loops of gauge fields in the adjoint representations. The corresponding diagrams have weights $N_1^2$ or $N_2^2$. On the other hands, the contribution of matter system come from loops of bosons and fermions in the bi-fundamental representations. The corresponding diagrams have weight $N_1 N_2$. The one-loop corrections vanish for the same reasons as the ABJM theory. The two-loop corrections are nonzero and have weights  $\frac{1}{2}(\lambda_1^2+\lambda_2^2)$ for the pure Chern-Simons contribution and $\lambda_1 \lambda_2$ for the matter contribution. Reflecting this fact, we obtain Wilson loop expectation value in ABJ theory.
\begin{align}
\big< W_{\Box}[C_n]\big>^{(2)}_{\mbox{\small{ABJ}}}&=\lambda_1 \lambda_2  \Big[ - {1 \over 2} \sum_{i=1}^n \frac{(x_{i,i+2}^2 4\pi e^{\gamma_E} \mu^2)^{2\epsilon}}{(2 \epsilon)^2}
+ \mbox{BDS}^{(2)}_n - \frac{1}{16}n \pi^2 \Big]
\nonumber \\
& \frac{\lambda_1^2+\lambda_2^2}{2}\Big[  - \frac{\log(2)}{2} \sum_{i=1}^n \frac{(x_{i,i+2}^2 \pi e^{\gamma_E} \mu^2)^{2\epsilon}}{2\epsilon} +\frac{7n\pi^2}{48} - \frac{n}{2}\log^2(2) - \frac{\pi^2}{6}  \Big]
\end{align}
As for the ABJM theory, we might combine the pure Chern-Simons contribution and the matter contribution to a single expression. Doing so, one would get
\bea
\big< W_{\Box}[C_n]\big>^{(2)}_{\mbox{\small{ABJ}}}=\Delta \Big[ &-& \sum_{i=1}^n \frac{\big(x^2_{i, i+2} \mu'^{2}\big)^{2\epsilon}}{8\epsilon^2} + \mbox{BDS}_n^{(2)}
\nonumber \\
&+&\frac{1}{96} \Big( \pi^2 \big( 8(n-2)+(7n-8) \delta^2 \big) - 6n(2+\delta^2)(6+\delta^2) \log^2(2) \Big) \Big] \label{ABJ_res}
\eea
Here, $\mu'^2$ is
\begin{equation}
\mu'^2= \mu^2 \pi e^{\gamma_E} 2^{3+\frac{1}{2} \delta^2}
\end{equation}
and
\begin{equation}
\Delta^2= \lambda_1 \lambda_2, \quad \quad \quad \delta \equiv \frac{\lambda_1 - \lambda_2}{\sqrt{\lambda_1 \lambda_2}}
\end{equation}
Hence, ABJM result obtained at $\Delta \rightarrow \lambda$ and $\delta \rightarrow 0$ limit. Also, the result exhibits uniform transcendentality as for the ABJM theory.

The result (\ref{ABJ_res}), however, somewhat weird. Suppose there is exact result on Wilson loop expectation value of ABJ theory. Then, perturbative expansion with respect to $\lambda_1$ or $\lambda_2$ should reproduce (\ref{ABJ_res}). But this is impossible, shifted energy scale $\mu'^2$ contains $2^{\frac{(\lambda_1-\lambda_2)^2}{2\lambda_1 \lambda_2}}$ term. Therefore, combining matter dependent part and pure Chern-Simons part looks not consistent with perturbative expansion.

\section*{Acknowledgement}
We thank Benjamin Basso, James Drummond, Claude Duhr, Johannes Henn, Paul Heslop, Yu-Tin Huang, Hyunsoo Min and, especially, Jaesung Park for numerous helpful discussions. SJR acknowledges organizers and participants of IAS Focused Program "Scattering Amplitudes in Hong Kong" (17-21 November, 2014) for excellent scientific program, constructive inputs and stimulating discussions. This work was supported in part by the National Research Foundation of Korea grants 2005-0093843, 2010-220-C00003 and 2012K2A1A9055280.

\section*{Appendix}
\appendix
%%%%%%%%%%%%%%%%%%%%%%%%%%%%%%%%%%%%%%%%%%%%%%%%%%%%%%%%%%%%%%%%%%%%%%%%%%%%%%%%%%%%%%%%
\section{Notation and Convention}
Here, we collect notations and conventions we adopted throughout the paper. \newline
%%%%
$\bullet$ spacetime conventions:
\bea
{\cal M}_3 = \mathbb{R}^{1,2}: \qquad
(\rmd s)^2 &=& \eta_{mn} \rmd x^m \rmd x^n \hskip1.3cm (m, n = 0, 1, 2) \nonumber \\
\eta_{mn} &=& \mbox{diag}(-,+,+), \qquad \delta_m^n = \mbox{diag}(+,+,+) \nonumber \\
\epsilon^{mnp} \epsilon_{qrs}&=& -\mbox{det}
\begin{pmatrix}
\eta^m_q && \eta^m_r && \eta^m_s\\
\eta^n_q && \eta^n_r && \eta^n_s\\
\eta^p_q && \eta^p_r && \eta^p_s
\end{pmatrix}
\eea
\newline
%%%%%
$\bullet$ Symmetries of ABJM theory
\begin{align}
\mbox{gauge symmetry} \ \ {\cal G}\otimes\overline{\cal G} &: \qquad \mbox{U(N)} \times \overline{\mbox{U(N)}} \nonumber \\
\mbox{global symmetry} \ \ \ G \ \ &: \qquad \mbox{SU(4)} \nonumber \\
\mbox{parity symmetry} \ \ \ P \ \ &: \qquad k \leftrightarrow - k, \quad \mbox{U(N)} \leftrightarrow \overline{\mbox{U(N)}} \nonumber
\end{align}
We denote trace over fundamental representations of $\mbox{U(N)}$ and $\overline{\mbox{U(N)}}$ as $\mbox{Tr}$ and $\overline{\mbox{Tr}}$, respectively. The gauge algebras $\mathfrak{g} = $u$(N)$ and $\overline{\mathfrak{g}} = \overline{\mbox{u}(N)}$ are isomorphic. We denote their generators by the same notation $T^a$,$(a=0,1,\cdots,N^2-1)$. They are Hermitian and normalized to
\begin{equation}
\mbox{Tr}(T^a T^b) = \overline{\mbox{Tr}}(T^a T^b) =\frac{1}{2}\delta^{ab}.
\end{equation}
%%%%%
$\bullet$
BPS Wilson loop operators \cite{Rey:1998a, Maldacena:1998a} in ABJM theory come in two classes. The ${1 \over 6}$-BPS Wilson loop operators are
\begin{equation}
W_\Box[C]=\frac{1}{N} \mbox{Tr}\mathcal{P} \mbox{exp} \left[i \oint_C d\tau \Big(A_m \dot{x}^m(\tau) + |\dot{x}|M^J_I Y^I Y_J^\dag \Big)\right] \label{def:Wilson_a1}
\end{equation}
%This operator is invariant under transformation (\ref{gauge_tr}).\footnote{If we choose Chern-Simons action by $S_{CS}= \frac{k}{4\pi} \int d^dx \epsilon^{\mu \nu \rho} \mbox{Tr} \Big( A_\mu \partial_{\nu} A_\rho+\frac{2i}{3} A_{\mu} A_{\nu} A_{\rho} \Big)$, then we should flip sign of Wilson loop operator (\ref{def:Wilson_a}). In other words, Wilson loop operator should be defined by $W_N[C]=\frac{1}{N} \mbox{Tr}\mathcal{P} \mbox{exp} \big[-i \oint_C d\tau \Big(A_m \dot{x}^m(\tau) \Big)\big]$ in this case.}
In previous paper \cite{Rey:2008a,Plefka:2008a}, it was proved that (\ref{def:Wilson_a1}) preserves $\frac{1}{6}$ supersymmetry if we choose the SU(4) matrix $M_I^J$ acting on a path ${\cal C}$ in SU(4) internal space so that the combined path $C\otimes{\cal C}$[SU(4)] is light-like.
The light-like $C$ is the simplest since $\vert \dot{x} \vert = 0$ and ${M_I}^J = 0$ in SU(4) space. In this case, the Wilson loop operators in the fundamental representations are
\begin{align}
W_\Box[C]=\frac{1}{N} \mbox{Tr}\mathcal{P} \mbox{exp} \left[i \oint_C \rmd\tau \Big(A_m \dot{x}^m(\tau)\Big)\right] \label{def:Wilson_b1}
\end{align}
and
\begin{align}
\overline{W}_\Box[C]=\frac{1}{N} \overline{\mbox{Tr}} \mathcal{P} \mbox{exp} \left[i \oint_C \rmd\tau \Big(\overline{A}_m \dot{x}^m(\tau)\Big)\right]. \label{def:Wilson_b1}
\end{align}
For the ABJM theory, we define the parity-even and parity-odd Wilson loop operators as
\begin{equation}
\mathcal{W}_\Box [C]:=\frac{1}{2}\Big(W_\Box [C]+\overline{W}_\Box [C] \Big) \label{def:Wilson_appendix}
\end{equation}
and
\bea
\ \ T\mathcal{W}_\Box [C] := \frac{1}{2} \Big( W_\Box [C] - \overline{W}_\Box [C] \Big).
\eea
From the viewpoint of M2-branes on orbifold, they are the Wilson loop operators in untwisted and twisted sectors, respectively. In the limit of infinitely many colors, expectation value of the parity-odd Wilson loop vanishes.

There also exist $\frac{1}{2}$-BPS Wilson loop operators. They are \cite{Drukker:2009b}
\bea
{\cal W}_\Box[C] = {1 \over 2 N} \mbox{Tr}\mathcal{P} \mbox{exp} \Big[i \oint_C \rmd\tau \ \mathbb{H}\Big]
\eea
where $\mathbb{H}$ is the super-connection matrix in the $\mathbb{Z}_2$ graded space of $\mathfrak{g} \oplus \overline{\mathfrak{g}}$ given by
\bea
\mathbb{H} &\equiv
\begin{pmatrix}
A_m \dot{x}^m(\tau) + |\dot{x}|M^J_I Y^I Y_J^\dag && |\dot{x}| \eta_I^m \overline{\psi}_m^I \\
|\dot{x}| \psi^m_I \overline{\eta}_m && \overline{A}_m \dot{x}^m(\tau) + |\dot{x}|\overline{M}^J_I Y_J^\dag Y^I \\
\end{pmatrix}.
\eea
For light-like contour $C$, $\vert \dot{x} \vert =0$ and $M_I^J = \overline{M}^I_J = 0$. So, both ${1 \over 2}$-BPS and ${1 \over 6}$-BPS Wilson loop operators are reduced to the Wilson loop operators in pure Chern-Simons theory. Stated differently, when the contour $C$ is light-like, the supersymmetry preserved by the Wilson loop operator is enhanced from ${1 \over 6}$ to ${1 \over 2}$.
\hfill\break
%%%%%
$\bullet$ Lightlike Polygons $C_n$: \hfill\break
\vskip0.5cm
\begin{center}
\fcolorbox{white}{white}{
  \begin{picture}(204,190) (95,-15)
    \SetWidth{2.0}
    \SetColor{Black}
    \Line(108,76)(120,124)
    \Line(120,124)(186,148)
    \Line(186,148)(258,112)
    \Line(258,112)(258,76)
    \SetWidth{1.0}
    \Photon(96,76)(120,76){5}{1}
    \Photon(96,70)(120,70){5}{1}
    \Photon(246,76)(270,76){5}{1}
    \Photon(246,70)(270,70){5}{1}
    \SetWidth{2.0}
    \Line(108,70)(114,22)
    \Line(114,22)(186,-8)
    \Line(186,-8)(258,34)
    \Line(258,34)(258,70)
    \Text(108,130)[lb]{\Large{\Black{$x_n$}}}
    \Text(186,154)[lb]{\Large{\Black{$x_1$}}}
    \Text(264,118)[lb]{\Large{\Black{$x_2$}}}
    \Text(264,28)[lb]{\Large{\Black{$x_{k-1}$}}}
    \Text(186,-20)[lb]{\Large{\Black{$x_k$}}}
    \Text(96,4)[lb]{\Large{\Black{$x_{k+1}$}}}
    \Text(222,112)[lb]{\Large{\Black{$y_1$}}}
    \Text(150,118)[lb]{\Large{\Black{$y_n$}}}
    \Text(210,22)[lb]{\Large{\Black{$y_{k-1}$}}}
    \Text(156,16)[lb]{\Large{\Black{$y_k$}}}
  \end{picture}
}
\end{center}
Figure A1. {\sl Vertices and edges of lightlike polygon}
\vskip0.5cm
A lightlike polygon is defined by a polygon in $\mathbb{R}^{d-1,1}$ whose $n$ edges are formed by lightlike segment vectors $y_1, y_2, \cdots, y_n$. Edges meet at $n$ cusp points $x_1, x_2, \cdots, x_n$, related to segment vectors by
\bea
y_i^m \equiv x^m_{i+1} - x^m_i.
\eea
Relative vector $x_{i,j}$ is defined by
\begin{equation}
x_{i,j}^m \equiv x_i^m -x_j^m.
\end{equation}
A point $z_i$ on the $i$-th edge can be parametrized by
\begin{equation}
z^m_i(\tau) = x^m_i + \tau \hspace{0.3mm} y^m_i \quad \mbox{where} \quad \tau \in [0,1]
\end{equation}
The inner product of segment vector is related to relative coordinates.
\begin{equation}
(y_{ij})^2 = - 2 y_i \cdot y_j = x_{i,j+1}^2+x_{i+1,j}^2-x_{i,j}^2-x_{i+1,j+1}^2.
\end{equation}
%%%%%%%%%%%%%%%%%%%%%%%%%%%%%%%%%%%%%%%%%%%%%%%%%%%%%%%%%%%%%%%%%%%%%%%%%%%%%%%%%%%%%%%%%
\section{Chern-Simons sector in ABJM theory}
The ABJM theory consists of two sectors: the pure Chern-Simons sector and the matter sector interacting with the Chern-Simons sector. Hence, it is always useful to isolate the pure Chern-Simons sector first and then couple the matter sector to it.
\\
%%%%%
$\bullet$ Chern-Simons part of ABJM action is given by:
\bea
S_{CS}&=& \frac{k}{4\pi} \int \rmd^dx \epsilon^{mnp} \mbox{Tr} \Big( A_m \partial_{n} A_p+\frac{2i}{3} A_{m} A_{n} A_{p} \Big) \label{CS_action_a}\\
\overline{S}_{CS}&=& -\frac{k}{4\pi} \int \rmd^dx \epsilon^{mnp} \overline{\mbox{Tr}} \Big( \overline{A}_m \partial_{n} \overline{A}_p + \frac{2i}{3} \overline{A}_{m} \overline{A}_{n} \overline{A}_{p} \Big) \label{CS_action_b} \\
S_{\mbox{gauge-fix}}&=&\frac{k}{4\pi} \int \rmd^dx \Big[ \frac{1}{\xi} \mbox{Tr}(\partial_m A^m)^2 + \mbox{Tr}(\partial_m c^* D_m c) \Big] \label{gf_action_a}\\
\overline{S}_{\mbox{gauge-fix}}&=&
-\frac{k}{4\pi} \int \rmd^dx \Big[ \frac{1}{\xi} \overline{\mbox{Tr}}(\partial_m \overline{A}^m)^2 + \overline{\mbox{Tr}}(\partial_m \overline{c}^* D_m \overline{c}) \Big] \label{gf_action_b}
\eea
Gauge field $A_{m}$ is in adjoint representation of $\mbox{U(N)}$, while $\overline{A_{m}}$ is in adjoint representation of $\overline{\mbox{U(N)}}$. $S_{\mbox{\tiny{gf}}}$ and $\overline{S}_{\mbox{\tiny{gf}}}$ are gauge fixing term, obtained by usual Fadeev-Popov method. $c,\overline{c}$ are pair of Fadeev-Popov ghosts and star notation means their conjugate. Covariant derivative $D_m c$ defined by $\partial_m c+i[A_m,c]$.

Manifestly, (\ref{CS_action_a}) is invariant under gauge transformation. To see this, let consider following gauge transformation:
\bea
A \rightarrow g A g^{-1}-i(\partial_m g)g^{-1} \label{gauge_tr}
\eea
where $g(x)$ is element of gauge group. Under this transformation, action is changed by
\bea
S_{\rm CS} \rightarrow S_{\rm CS}-\frac{k}{12\pi}g^{-1}(\partial_m g)g^{-1}(\partial_n g)g^{-1}(\partial_p g)+i \frac{k}{4\pi}\partial_m (g^{-1}(\partial_n g)A_p). \label{var_gauge}
\eea
Second term of right-hand side is equivalent to winding number density $w(g)$. This term can be expressed by $-2 \pi k w(g)$. Definition of winding number density $w(g)$ is:
\bea
w(g)=\frac{1}{24\pi^2} \epsilon^{mnp} g^{-1}(\partial_m g)g^{-1}(\partial_n g)g^{-1}(\partial_p g).
\eea
This quantity always have integer value. We focus on $e^{iS_{CS}}$, hence there is no effect of winding number density. Third term in (\ref{var_gauge}) is just total derivative term,  it will be vanished with suitable boundary condition. \\
\newline
%%%%%
$\bullet$ Feynman rules of gauge field propagator

We can read Feynman rules of gauge field propagator from ABJM action (\ref{CS_action_a})-(\ref{gf_action_b}):
\bea
\mbox{ U(N) gauge propagator} &:& \quad \Delta_{mn}=\frac{2\pi}{k} \Big( \frac{\epsilon_{mnp} l^p}{l^2}+\xi \frac{l_m l_n}{l^4} \Big) \\
\overline{\mbox{U(N)}} \ \mbox{gauge propagator} &:& \quad \overline{\Delta}_{m n}=-\frac{2\pi}{k} \Big( \frac{\epsilon_{m n p} l^p}{l^2}+\xi \frac{l_m l_n}{l^4} \Big)
\eea
In this paper, we choose Landau gauge, that is $\xi=0$. With this gauge, Feynman rule reduced to simple monomial form.

Feynman rules in position space could be obtained by Fourier transformation. We considered gauge field propagating from $x$ to $y$.
\begin{equation}
\Delta_{mn}(x,y)=\frac{2\pi}{k} \int \frac{\rmd^d l}{(2\pi)^d} \frac{\epsilon_{mnp}  l^p}{l^2} e^{-i l \cdot (x-y)} \label{bessel}
\end{equation}
Carrying out $d-1$ angle direction integration first, we arrives to
\begin{equation}
\Delta_{mn}(x,y)=\frac{2\pi}{k} \frac{1}{(2\pi)^d} \epsilon_{mnp} \partial^p|x-y|^{2-d} \int_0^\infty \rmd t \ t^{\frac{d}{2}-2} \ J_{\frac{d}{2}-1}(t)
\end{equation}
where $J_\nu(t)$ is Bessel function and $|x-y|=((x-y)^2)^{\frac{1}{2}}$. Integration over Bessel function can be done, finally propagator in position space is given by
\begin{equation}
\Delta_{mn}(x,y)=\frac{\Gamma(\frac{d}{2})}{k\hspace{0.5mm}\pi^{(\frac{d}{2}-1)}} \frac{\epsilon_{mnp} (x-y)^p}{((x-y)^2)^{\frac{d}{2}}}
\end{equation}
\newline
%%%%%
$\bullet$ Path ordering of Wilson loop operator

Previous Feynman rules for gauge field reveals that $\big< A_m A_n \big>$ and $\big< A_n A_m \big>$ give a different sign due to presence of epsilon tensor. Hence, we need to define suitable definition of path ordering. In this paper, we choose the path ordering consistent with the charge conjugation by
\bea
\mathcal{P}(A_m(z(\tau))A_n( z(\tau^\prime))) :=
\big< A_m(z(\tau)) A_n(z(\tau^\prime)) \big> \theta(\tau - \tau\rq{}) + \big< A_n(z(\tau^\prime)) A_m(z(\tau)) \big> \theta(\tau\rq{} - \tau),
\eea
where $\tau, \tau^\prime$ are parameters running from 0 to 1.

With definition of this path ordering, expansion of Wilson loop operator along given contour $C$ is:
\bea
W_\Box [C]=\frac{1}{N}\mbox{Tr} \ \mathcal{P} %&&
\lim_{n \rightarrow \infty} \prod_{k=1}^n [1 + a^m A_m (x + a) ]
%\Big[\mathbb{I}+i \oint_C dx^\mu A_\mu(x)-\oint_C dx^\mu \int^x dy^\nu A_\mu(x)A_\nu(y)-i\oint_C dx^\mu \int^x dy^\nu \int^y dz^\rho A_\mu(x) A_\nu(y) A_\rho(z) \nonumber \\
%&&+\oint_C dx^\mu \int^x dy^\nu \int^y dz^\rho \int^z dw^\sigma A_\mu(x) A_\nu(y) A_\rho(z) A_\sigma(w)+\cdots \Big]
\eea
%%%%%%%%%%%%%%%%%%%%%%%%%%%%%%%%%%%%%%%%%%%%%%%%%%%%%%%%%%%%%%%%%%%%%%%%%%%%%%%%%%%%%
\section{Self-Energy Correction to the Gauge Fields}
Matter-dependent self-energy correction to the gauge field in ABJM theory behaves very similar to the tree-level gauge field in $\mathcal{N}$= 4 SYM. In this section, we will provide explicit calculation of the self-energy correction.
\hfill\break
\vskip0.5cm
\begin{center}
\fcolorbox{white}{white}{
  \begin{picture}(122,50) (65,-41)
    \SetWidth{1.0}
    \SetColor{Red}
    \Photon(66,-10)(102,-10){5}{3}
    \Photon(150,-10)(186,-10){5}{3}
    \Arc[clock](126,-17)(25,163.74,16.26)
    \Photon(66,-22)(102,-22){5}{3}
    \Photon(150,-22)(186,-22){5}{3}
    \Arc(126.169,-15.226)(24.775,-159.258,-15.869)
    \SetColor{Blue}
    \Arc(126,-16)(16.971,135,495)
  \end{picture}
}
\quad \quad
\fcolorbox{white}{white}{
  \begin{picture}(122,50) (65,-41)
    \SetWidth{1.0}
    \SetColor{Red}
    \Photon(66,-10)(102,-10){5}{3}
    \Photon(150,-10)(186,-10){5}{3}
    \Photon(66,-22)(102,-22){5}{3}
    \Photon(150,-22)(186,-22){5}{3}
    \Arc[dash,dashsize=6,clock](126,-17)(25,163.74,16.26)
    \Arc[dash,dashsize=6,clock](126,-15)(25,-16.26,-163.74)
    \SetColor{Blue}
    \Arc[dash,dashsize=6](126,-16)(16.971,135,495)
  \end{picture}
} \\
\bigskip
\fcolorbox{white}{white}{
  \begin{picture}(122,56) (65,-38)
    \SetWidth{1.0}
    \SetColor{Red}
    \Photon(66,-4)(102,-4){5}{3}
    \Photon(150,-4)(186,-4){5}{3}
    \Photon(66,-16)(102,-16){5}{3}
    \Photon(150,-16)(186,-16){5}{3}
    \PhotonArc[clock](126,-11)(25,163.74,16.26){4}{5.5}
    \PhotonArc(126,-9)(25,-163.74,-16.26){4}{5.5}
    \PhotonArc(126,-10)(13.416,-63,297){2}{9}
  \end{picture}
}
\quad \quad
\fcolorbox{white}{white}{
  \begin{picture}(122,50) (65,-41)
    \SetWidth{1.0}
    \SetColor{Red}
    \Photon(66,-10)(102,-10){5}{3}
    \Photon(150,-10)(186,-10){5}{3}
    \Photon(66,-22)(102,-22){5}{3}
    \Photon(150,-22)(186,-22){5}{3}
    \Arc[dash,dashsize=1.4,clock](126,-17)(25,163.74,16.26)
    \Arc[dash,dashsize=1.6](126,-15)(25,-163.74,-16.26)
    \Arc[dash,dashsize=1.6](126,-16)(16.971,135,495)
  \end{picture}
}
\end{center}
Figure A2. {\sl Feynman diagrams contributing to the one-loop self-energy of the gauge boson. The solid lines are matter fields, the wave lines are gauge fields, and the dotted lines are Faddeev-Popov ghosts.}
\vskip0.5cm
$\bullet$ Matter-dependent sector of the ABJM theory action contains
\bea
S_{ABJM} &:=&  \int \rmd^d x
\Big[{1 \over 2} \overline{\mbox{Tr}} \left( -(D_m Y)^\dagger_I D^m Y^I  + i \Psi^{\dagger I} D \hskip-0.22cm / \Psi_I  \right) + {1 \over 2} \mbox{Tr} \left(- D_m Y^I (D^m Y)^\dagger_I  +
 i \Psi_I D \hskip-0.22cm / \Psi^{\dagger I}  \right)
\nonumber \\
&&
\hskip1cm - V_{\rm F} - V_{\rm B} \, \Big]
\eea
Complexified Hermitian scalars and Majorana spinors in this action are($I=1,2,3,4$):
\bea
&& Y^I = (X^1 + i X^5, X^2 + i X^6, X^3 - i X^7, X^4 - i X^8):  \qquad ({\bf N}, \overline{\bf N}; {\bf 4}) \nonumber \\
&& Y^\dagger_I = (X^1 - i X^5, X^2 - i X^6, X^3 + i X^7, X^4 + i X^8): \hskip0.8cm ( \overline{\bf N}, {\bf N}; \overline{\bf 4})
\nonumber \\
&& \Psi_I = (\psi^2 + i \chi^2, - \psi^1 - i \chi^1, \psi_4  - i \chi_4 , - \psi_3 + i \chi_3 ) : \hskip0.4cm
({\bf N}, \overline{\bf N}; \overline{\bf 4}) \nonumber \\
&& \Psi^{\dagger I} = (\psi_2 - i \chi_2, - \psi_1 + i \chi^1, \psi^4 + i \chi^4, - \psi^3 - i \chi^3): \hskip0.3cm
(\overline{\bf N}, {\bf N}; {\bf 4})
\eea
Here, covariant derivatives are defined as
\bea
D_m Y^I = \partial_m Y^I + i A_m Y^I - i Y^I \overline{A}_m \, , \quad D_m Y^\dagger_I = \partial_m Y^\dagger_I + i \overline{A}_m Y^\dagger_I - i Y^\dagger_I A_m
\eea
and similarly for fermions $\Psi_I, \Psi^{\dagger I}$. $ V_{\rm F}$ and $V_{\rm B}$ are interaction terms, it contains sextet bosonic interaction and Yukawa interaction.
\newline
%%%%%
$\bullet$ $\mathbb{R}^{1,2}$ Majorana spinor and Dirac matrices:
\bea
&& \Psi \equiv \mbox{two-component} \,\,\, \mbox{Majorana} \,\,\, \mbox{spinor} \nonumber \\
&& \Psi^\alpha = \epsilon^{\alpha \beta}\Psi_\beta ,
\quad \Psi_\alpha = \epsilon_{\alpha \beta} \Psi^\beta \quad
\mbox{where} \quad \epsilon^{\alpha \beta} = - \epsilon_{\alpha \beta} = i \sigma^2 \nonumber \\
&& {\gamma^m_\alpha}^\beta = ( i \sigma^2, \sigma^3, \sigma^1),
\quad (\gamma^m)_{\alpha \beta} = (-\mathbb{I}, \sigma^1, -
\sigma^3) \quad \mbox{obeying} \quad \gamma^m \gamma^n = \eta^{mn} \hspace{0.7mm} \mathbb{I}_{2\times 2}-
\epsilon^{mnp} \gamma_p.\nonumber \\
&& \mbox{Hence,}\quad \mbox{Tr}(\gamma^m \gamma^n) = 2\eta^{mn} \label{gamma_in3d} , \quad \mbox{Tr}[\gamma_\mu \gamma_\rho \gamma_\nu \gamma_\sigma]=2g_{\mu \rho} g_{\nu \sigma}+2g_{\rho \nu} g_{\mu \sigma}-2g_{\mu \nu} g_{\rho \sigma}
\eea
\newline
%%%%%
$\bullet$ Feynman rules for bosons, fermions and ghosts are explicitly readable from ABJM action:
\bea
\mbox{boson propagator}: \quad && {D_I}^J (p) = \delta_I^J \,  {-i \over p^2 - i \epsilon}  \nonumber \\
\mbox{fermion propagator}: \quad && {S^I}_J (p) =\delta^I_J \, { i p \hskip-0.22cm / \over p^2 - i \epsilon}   \nonumber \\
\mbox{ghost propagator}: \quad && K(p) \, = \, {-i \over p^2 - i \epsilon}
\eea
\newline
%%%%%
The gauge fields receive self-energy corrections from gauge field self-interactions, Faddeev-Popov ghosts and matter fields. The corrections are three-dimensional counterpart of color (anti)screening by each respective spin fields: the matter fields -- both bosons and its superpartner fermions -- and the ghost fields screen the color, while the gauge fields anti-screen the color. The physics of this self-energy corrections appear less noted, so we repeat details of this computation below.

$\bullet$ self-energy corrections from matter fields: \hfill\break
As we described, there are four distinguished diagrams in matter-dependent part. Let us first consider the boson and fermion matter-dependent diagrams. Expression of these two diagrams is :
\bea
i\Delta_{mn}= - \frac{\epsilon_{mpq} k^q}{k^2} i\Pi_{pr} \frac{\epsilon_{nrs} k^s}{k^2} \label{gauge_self}
\eea
Here, $i\Pi_{mn}=i\Pi_{mn}^b+i\Pi_{mn}^f$ is self one-loop correction of gauge field due to boson and fermion matter fields, respectively.

The boson contribution is:
\bea
i \ \Pi_{mn}^{\rm boson} (k)
%4\int \frac{\rmd^dl}{(2\pi)^d} \frac{i}{(k+l)^2} i(2l+k)_p \frac{i}{l^2} i(2l+k)_r \nonumber \\
= (-)^0 \cdot 4 \cdot (2s + 1) \int \frac{\rmd^dl}{(2\pi)^d} \frac{(2l+k)_m (2l+k)_n}{l^2(k+l)^2} \Big\vert_{s=0}
\eea
The fermion contribution is
\bea
i\ \Pi_{pr}^{\rm fermion} (k)=
(-)^1 \cdot 4 \cdot (2s+1) \int \frac{\rmd^dl}{(2\pi)^d}\frac{l_m(l+k)_n +(l+k)_m l_n -\eta_{mn} l\cdot(l+k)}{l^2(k+l)^2}
%(-1)^{\tiny{\mbox{FD}}}4 \mbox{Tr}\Big[\int \frac{\rmd^d l}{(2\pi)^d} \gamma_r \frac{\not{l}}{l^2} \gamma_p \frac{(\not k+\not l)}{(k+l)^2}\Big]
\eea
In both cases, we made it explicit that $(-)^F$ counts the statistics, $(2s+1)$ counts the spin degrees of freedom, and 4 counts the  $SU(4)$ degrees of freedom.
%Numerator can be simplified by applying (\ref{gamma_in3d}).
The boson and fermion contributions add together
%
%\bea
%i\Pi_{pr}^f&=&  -8\int \frac{\rmd^dl}{(2\pi)^d}\frac{l_p(l+k)_r +(l+k)_r l_p -\eta_{p r} l\cdot(l+k)}{l^2(k+l)^2}
%\eea
%
%and,
%
\bea
i\Pi_{mn}^{\rm boson}+i\Pi_{mn}^{\rm fermion} =\int \frac{\rmd^dl}{(2\pi)^d}\frac{4k_m k_n + 8\eta_{mn} l\cdot(l+k)}{l^2(k+l)^2}.
\eea
The $k_m k_n$ term will be vanish upon contraction with the Levi-Civita tensors in (\ref{gauge_self}). We evaluate the remaining integral after Wick rotation to the $d$-dimensional Euclidean space.
%Using the master formula
%%
%\bea
%\int \frac{\rmd^d\overline{l}}{(2\pi)^d}\frac{(\overline{l}^2)^a}{(\overline{l}^2+D)^b}=\frac{\Gamma(b-a-\frac{1}{2}d)\Gamma(a+\frac{d}{2})}{(4\pi)^\frac{d}{2} \Gamma(b) \Gamma(\frac{d}{2})} D^{-(b-a-\frac{d}{2})}
%\eea
%
%Finally, we arrives at
%
The result is
\bea
i\Pi_{mn}^{\rm boson}+i\Pi_{mn}^{\rm fermion} = \left[ -  \frac{8 i }{(4\pi)^{\frac{d}{2}}} \Gamma \Big(1-\frac{d}{2}\Big) \frac{\Gamma(\frac{d}{2})\Gamma(\frac{d}{2})}{\Gamma(d)} \right] \eta_{mn} (k^2)^{(\frac{d}{2}-1)}.
\eea
Substituting this to (\ref{gauge_self}), we finally obtain the matter-dependent correction to the gauge propagator:
\bea
i\Delta_{mn} (k) =\left[ -8i \frac{\Gamma(1-\frac{d}{2})}{(4\pi)^{\frac{d}{2}}} \frac{\big(\Gamma(\frac{d}{2}) \big)^2}{\Gamma(d-1)}\right] \frac{1}{(k^2)^{3-\frac{d}{2}}}(\eta_{mn} k^2-k_m k_n) \label{gauge_matter}
\eea
Note that, while the gauge fields are allowed only parity-odd propagation at tree level, they acquire parity-even propagation at one-loop level.

It is useful to Fourier transform to the position space.  %The second term in (\ref{gauge_matter}) does not contribute.
%\color{red} IS THIS TRUE? \color{black}
%Hence, corrected gauge propagator in position space is given
In the convention of (\ref{bessel}), the parity-even propagation of the gauge fields takes the form
\bea
i\Delta_{mn}= \left[ \frac{i}{2 \pi^d} \frac{\Gamma(1-\frac{d}{2}) \Gamma(\frac{d}{2}) \Gamma(\frac{d}{2})}{\Gamma(d-1)} \frac{\Gamma(d-2)}{\Gamma(2-\frac{d}{2})} \right] \frac{\eta_{mn}}{(x^2)^{d-2}}
\eea
At $d=3$, the gauge field propagation falls off $\sim 1/x^2$. This is exactly the same behavior as the propagation of the four-dimensional gauge fields. We can understand this from physical considerations. The ABJM theory is conformally invariant, so it must be that the gauge field propagates conformally invariantly. The parity-odd propagation governed by the Chern-Simons term is trivially conformally invariant. The party-even propagation is in general generated, but then it must propagate in $(3+1)$-dimensional space time as this is the only dimension the gauge field propagation is conformally invariant. In fact, our argument is not specific to the ABJM theory; the same argument applies to any conformally invariant theory in any spacetime dimensions.

%%%%%
$\bullet$ self-energy corrections from gauge and ghost fields

The gauge and Fadeev-Popov fields also contribute to the parity-even self-energy corrections. Again, they parallel to the one-loop renormalization in $(3+1)$-dimensional gauge theories except that the cubic interactions are governed by the Chern-Simons term.

From the Chern-Simons cubic interactions of the gauge fields, we have
\bea
i\Pi_{mn}^{\rm gauge}&=&
3\cdot 3 \Big(i \frac{k}{4\pi} \frac{2}{3} \Big)^2 \Big( \frac{2\pi}{k} \Big)^2\int \frac{\rmd^dl}{(2\pi)^d}  \frac{\epsilon_{a b p} l^b}{l^2} \epsilon_{m p q}  \frac{\epsilon_{c d q}(l+k)^d}{(l+k)^2} \epsilon_{a c n} \nonumber \\
&=& -(-)^0 \int \frac{\rmd^dl}{(2\pi)^d} \frac{l^m (l+k)^n + (l+k)^m l^n}{l^2(l+k)^2}.
\eea
Two factors of 3 are from the cubic vertex interaction combinatorics and the minus sign signifies the $(2+1)$-dimensional counterpart of the color anti-screening.

For the Faddeev-Popov ghost loop, we get
\bea
i\Pi_{mn}^{\rm ghost} = - (-1)^{1} \int \frac{d^dl}{(2\pi)^d} \frac{(l+k)_m l_n+(l+k)_n l_m}{l^2(l+k)^2}.
\eea
We see that the gauge and the ghost contributions precisely cancel each other. This should be contrasted with the incomplete cancellation of anti-screening in $(3+1)$-dimensional Yang-Mills theory. Of course, the difference comes about because the cubic interaction in the ABJM theory is governed by the parity-odd Chern-Simons term.

%%%%%%%%%%%%%%%%%%%%%%%%%%%%%%%%%%%%%%%%%%%%%%%%%%%%%%%%%%%%%%%%%%%%%%%%%%%%%%%%%%%%%%%
\section{Mellin-Barnes transformation}
The Mellin-Barnes transformation is frequently used tool for Feynman diagram calculation. Basic transformation rule is given by the integral formula:
\begin{equation}
\frac{1}{(X+Y)^\lambda}=\frac{1}{\Gamma(\lambda)}\int_{-i\infty}^{i\infty} \frac{\rmd z}{2\pi i} \Gamma(-z) \Gamma(\lambda+z) \frac{Y^z}{X^{(\lambda+z)}}
\end{equation}
This is the simplest Mellin-Barnes transformation that disentangles the two terms $X, Y$ in the denominator. The integral is two-fold and one-dimensional because there are two terms $X, Y$ in the denominator. More generally, as the number of independent terms in the denomator increases, its Mellin-Barnes transformation generate higer-fold, higher-dimensional integrals. For instance, for three term denominator, the Mellin-Barnes transformatoin reads
\bea
\frac{1}{(X+Y+Z)^\lambda}=\frac{1}{\Gamma(\lambda)}\int_{-i\infty}^{i\infty} \frac{\rmd z_1}{2\pi i} \int_{-i\infty}^{i\infty} \frac{\rmd z_2}{2\pi i} \Gamma(-z_1) \Gamma(-z_2) \Gamma(\lambda+z_1+z_2) \frac{Z^{z_1} Y^{z_2}}{X^{(\lambda+z_1+z_2)}}.
\eea
The integral contour is chosen in the complex plane that poles from the Gamma functions $\Gamma(\cdots+z)$ lie on left side of the contour and poles from Gamma functions $\Gamma(\cdots-z)$ lie on right side of the contour. Often, the complex integration is more convenient to evaluate.

There is a useful identity called the Barnes lemma. This lemma enable us to reduce higher-dimensional complex integrals to lower-dimensional ones. The first Barnes lemma reads
\bea
\int_{-i\infty}^{i\infty} \frac{\rmd z}{2\pi i} \Gamma(\lambda_1+z) \Gamma(\lambda_2+z) \Gamma(\lambda_3-z) \Gamma(\lambda_4-z) = \frac{\Gamma(\lambda_1+\lambda_3)\Gamma(\lambda_1
+\lambda_4)\Gamma(\lambda_2
+\lambda_3)\Gamma(\lambda_2+\lambda_4)}{\Gamma(\lambda_1+\lambda_2+\lambda_3+\lambda_4)}.
\eea
The second Barnes lemma is:
\bea
&& \int_{-i\infty}^{i\infty} \frac{\rmd z}{2\pi i} \frac{\Gamma(\lambda_1+z)\Gamma(\lambda_2+z)
\Gamma(\lambda_3+z)\Gamma(\lambda_4 -z)
\Gamma(\lambda_5-z)}{\Gamma(\lambda_1+\lambda_2+\lambda_3+\lambda_4+
\lambda_5+z)} \nonumber \\
&& \hskip2cm =  \frac{\Gamma(\lambda_1+\lambda_4)
\Gamma(\lambda_1+\lambda_5)
\Gamma(\lambda_2+\lambda_4)
\Gamma(\lambda_2+\lambda_5)
\Gamma(\lambda_3+\lambda_4)
\Gamma(\lambda_3+\lambda_5)}{\Gamma(\lambda_1+\lambda_2+\lambda_4+\lambda_5)
\Gamma(\lambda_1+\lambda_3+\lambda_4+\lambda_5)
\Gamma(\lambda_2+\lambda_3+\lambda_4+\lambda_5)}.
\eea
In numerical evaluations, these lemmas can be automatized by the Mathematica package {\ttfamily{barnesroutines}}.

%%%%%%%%%%%%%%%%%%%%%%%%%%%%%%%%%%%%%%%%%%%
\section{Ladder Diagrams}
For every ladder diagram, we converted them to multi-dimensional integrals with domains $[0,1]$ for each variable. This form is most useful for applying the Mellin-Barnes transformations and performing numerical evaluation of resulting multi-dimensional integrals.
\hfill\break
$\circledcirc$ $I_{\rm ladder}^{\{4,4,1,1\}}$ \hfill\break
This diagram is the one with $(i,j,k,l) = (4,4,1,1)$ in (\ref{Ladder}).
\bea
I_{\rm ladder}^{\{4,4,1,1\}}=  \int_0^1 d\tau_l \int_0^{\tau_l} d\tau_k \int_0^1 \rmd\tau_j \int_0^{\tau_j} \rmd\tau_i  \ \frac{\epsilon(y_1,y_4,z_1-z_4)}{[(z_1-z_4)^2]^{\frac{d}{2}}} \frac{\epsilon(y_1,y_4,z_1-z_4)}{[(z_1-z_4)^2]^{\frac{d}{2}}}
\eea
This integral can be converted to
\begin{equation}
I_{\rm ladder}^{\{4,4,1,1\}}=\frac{1}{4}\int_0^{1} \rmd\tau_i \int_0^1 \rmd\tau_j \int_0^{1} \rmd\tau_k \int_0^1 \rmd\tau_l \ \frac{\tau_l \tau_j N_{4411}}{\big({\Delta_1^{\{4,4,1,1\}}}\big)^{\frac{d}{2}} \big({\Delta_2^{\{4,4,1,1\}}}\big)^{\frac{d}{2}}}
\end{equation}
where bar notation $\bar{\tau}$ means $1-\tau$ and,
%\color{red}
%DEFINITION OF BARRED VARIABLE
%\color{black}
\bea
N_{4411}&=& (x_{14}^2-x_{15}^2-x_{24}^2+x_{25}^2)(-x_{15}^2x_{24}^2+x_{14}^2x_{25}^2) \nonumber \\
\Delta_1^{\{4,4,1,1\}}&=& x_{14}^2 \bar{\tau_i} \bar{\tau_l}+x_{14}^2 \tau_i \bar{\tau_j} \bar{\tau_l}+x_{15}^2\bar{\tau_i} \tau_l + x_{15}^2 \tau_i\bar{\tau_j} \tau_l + x_{24}^2 \tau_i \tau_j \bar{\tau_l}+ x_{25}^2 \tau_i \tau_j \tau_l\nonumber \\
\Delta_2^{\{4,4,1,1\}}&=& x_{14}^2 \bar{\tau_k}\bar{\tau_j}+x_{14}^2 \tau_k \bar{\tau_l} \bar{\tau_j}+x_{15}^2 \tau_k \tau_l \bar{\tau_j}+x_{24}^2 \tau_j \bar{\tau_k}+x_{24}^2 \tau_j \tau_k \bar{\tau_l}+x_{25}^2 \tau_j \tau_k \tau_l.
\eea
\hfill\break
$\circledcirc$ $I_{\rm ladder}^{\{5,4,1,1\}}$ \hfill\break
This diagram is the one with $(i,j,k,l) = (5,4,1,1)$ in (\ref{Ladder}).
\bea
I_{\rm ladder}^{\{5,4,1,1\}}=\int_0^1 d\tau_l \int_0^1 d\tau_k\int_0^1 d\tau_j \int_0^{\tau_j} \rmd\tau_i    \frac{\epsilon(y_1,y_5,z_1-z_5)}{[(z_1-z_5)^2]^{\frac{d}{2}}} \frac{\epsilon(y_1,y_4,z_1-z_4)}{[(z_1-z_4)^2]^{\frac{d}{2}}}
\eea
This integral can be converted to
\bea
I_{\rm ladder}^{\{5,4,1,1\}}=-\frac{1}{8} \int_0^{1} \rmd\tau_i \int_0^1 \rmd\tau_j \int_0^1 \rmd\tau_k \int_0^1 \rmd\tau_l  \frac{\tau_j N_{5411}}{\big({\Delta_1^{\{5,4,1,1\}}}\big) \big({\Delta_2^{\{5,4,1,1\}}}\big)}
\eea
where
\bea
N_{5411}&=& (x_{15}^2)^2(x_{24}^2+x_{26}^2-x_{46}^2)
+ x_{25}^2(x_{14}^2(x_{25}^2-x_{26}^2)-x_{25}^2x_{46}^2)
\nonumber \\
&-&x_{15}^2(x_{24}^2x_{25}^2-2x_{24}^2x_{26}^2+x_{25}^2x_{26}^2
+x_{14}^2(x_{25}^2+x_{26}^2)-2x_{25}^2x_{46}^2)
 \nonumber \\
\Delta_1^{\{5,4,1,1\}}&=& x_{15}^2 \bar{\tau_i} \bar{\tau_l}+x_{15}^2 \tau_i \bar{\tau_j} \bar{\tau_l}+x_{25}^2 \tau_i \tau_j \bar{\tau_l}+x_{26}^2 \tau_i \tau_j \tau_l \nonumber \\
\Delta_2^{\{5,4,1,1\}}&=& x_{14}^2 \bar{\tau_j} \bar{\tau_k}+x_{15}^2 \bar{\tau_j} \tau_k + x_{24}^2 \tau_j \bar{\tau_k}+x_{25}^2 \tau_j \tau_k.
\eea
\hfill\break
$\circledcirc$ $I_{\rm ladder}^{\{4,3,1,1\}}$ \hfill\break
This diagram is the one with $(i,j,k,l) = (4,3,1,1)$ in (\ref{Ladder}).
\begin{equation}
I_{\rm ladder}^{\{4,3,1,1\}}= \int_0^1 \rmd\tau_l \int_0^1 \rmd\tau_k\int_0^1 \rmd\tau_j \int_0^{\tau_j} \rmd\tau_i
  \ \frac{\epsilon(y_1,y_4,z_1-z_4)}{[(z_1-z_4)^2]^{\frac{d}{2}}} \frac{\epsilon(y_1,y_3,z_1-z_3)}{[(z_1-z_3)^2]^{\frac{d}{2}}}
\end{equation}
The integral can be converted to
\begin{equation}
I_{\rm ladder}^{\{4,3,1,1\}}=-\frac{1}{8}\int_0^{1} \rmd\tau_i \int_0^1 \rmd\tau_j \int_0^1 \rmd\tau_k \int_0^1 \rmd\tau_l \frac{\tau_j \ N_{4311}}{\big(\Delta_1^{\{4,3,1,1\}}\big)^{\frac{d}{2}} \big(\Delta_2^{\{4,3,1,1\}} \big)^{\frac{d}{2}}}
\end{equation}
where
\bea
N_{4311}&=& x_{13}^2x_{24}^2(-x_{14}^2+2x_{15}^2+x_{24}^2)-x_{13}^2(x_{14}^2+x_{24}^2)x_{25}^2 \nonumber \\
&+& (x_{14}^2-x_{24}^2)(-x_{15}^2x_{24}^2+x_{14}^2x_{25}^2-x_{14}^2x_{35}^2+x_{24}^2x_{35}^2) \nonumber \\
\Delta_1^{\{4,3,1,1\}}&=& x_{14}^2 \bar{\tau_i} \bar{\tau_l}+x_{14}^2 \tau_i \bar{\tau_j} \bar{\tau_l}+x_{15}^2 \bar{\tau_i} \tau_l+ x_{15}^2 \tau_i \bar{\tau_j} \tau_l+ x_{24}^2 \tau_i \tau_j \bar{\tau_l}+x_{25}^2 \tau_i \tau_j \tau_l \nonumber \\
\Delta_2^{\{4,3,1,1\}}&=& x_{13}^2 \bar{\tau_j} \bar{\tau_k}+x_{14}^2 \bar{\tau_j} \tau_k+ x_{24}^2 \tau_j \tau_k
\eea
\hfill\break
$\circledcirc$ $I_{\rm ladder}^{\{5,3,1,1\}}$ \hfill\break
This diagram is the one with $(i,j,k,l)_ = (5,3,1,1)$ in (\ref{Ladder}).
\begin{equation}
I_{\rm ladder}^{\{5,3,1,1\}}= \int_0^1 \rmd\tau_l \int_0^1 \rmd\tau_k \int_0^1 \rmd\tau_j \int_0^{\tau_j} \rmd\tau_i \frac{\epsilon(y_1,y_5,z_1-z_5)}{[(z_1-z_5)^2]^{\frac{d}{2}}} \frac{\epsilon(y_1,y_3,z_1-z_3)}{[(z_1-z_3)^2]^{\frac{d}{2}}}
\end{equation}
The integral can be converted to
\begin{equation}
I_{\rm ladder}^{\{5,3,1,1\}} = -\frac{1}{8} \int_0^{1} \rmd\tau_i \int_0^1 \rmd\tau_j \int_0^1 \rmd\tau_k \int_0^1 \rmd\tau_l \frac{\tau_j \ N_{5311}}{\big(\Delta_1^{\{5,3,1,1\}} \big)^{\frac{d}{2}} \big( \Delta_2^{\{5,3,1,1\}} \big)^{\frac{d}{2}}}
\end{equation}
where
\bea
N_{5311}&=& (x_{14}^2-x_{24}^2)(-x_{26}^2x_{35}^2+x_{15}^2(x_{26}^2-x_{36}^2)+x_{25}^2x_{36}^2) \nonumber \\
&-& x_{13}^2(-x_{24}^2x_{25}^2+x_{24}^2x_{26}^2+x_{15}^2(x_{24}^2+x_{26}^2-x_{46}^2)+x_{25}^2x_{46}^2) \nonumber \\
\Delta_1^{\{5,3,1,1\}}&=& x_{15}^2 \bar{\tau_i} \bar{\tau_l}+x_{15}^2 \tau_i \bar{\tau_j} \bar{\tau_l}+x_{25}^2 \tau_i \tau_j \bar{\tau_l}+x_{26}^2 \tau_i \tau_j \tau_l \nonumber \\
\Delta_2^{\{5,3,1,1\}}&=& x_{13}^2 \bar{\tau_j} \bar{\tau_k}+x_{14}^2 \bar{\tau_j} \tau_k+x_{24}^2 \tau_j \tau_k.
\eea
\hfill\break
$\circledcirc$ $I_{\rm ladder}^{\{3,3,1,1\}}$ \hfill\break
This diagram is the one with $(i,j,k,l) = (3,3,1,1)$ in (\ref{Ladder}).
\bea
I_{\rm ladder}^{\{3,3,1,1\}}
= \int_0^1 d\tau_l \int_0^{\tau_l} d\tau_k\int_0^1 \rmd\tau_j \int_0^{\tau_j} \rmd\tau_i \frac{\epsilon(y_1,y_3,z_1-z_3)}{[(z_1-z_3)^2]^{\frac{d}{2}}} \frac{\epsilon(y_1,y_3,z_1-z_3)}{[(z_1-z_3)^2]^{\frac{d}{2}}}
\eea
The integral can be converted to
\begin{equation}
I_{\rm ladder}^{\{3,3,1,1\}}=\frac{1}{4}\int_0^{1} \rmd\tau_i \int_0^1 \rmd\tau_j \int_0^1 \rmd\tau_k \int_0^1 \rmd\tau_l \ \frac{\tau_l \tau_j N_{3311}}{\big(\Delta_1^{\{3,3,1,1\}} \big)^{\frac{d}{2}} \big(\Delta_2^{\{3,3,1,1\}} \big)^{\frac{d}{2}}}
\end{equation}
where
\bea
N_{3311}&=& x^2_{13} x^2_{24} (x^2_{13} - x^2_{14} + x^2_{24}) \nonumber \\
\Delta_1^{\{3,3,1,1\}}&=&  x_{13}^2 \bar{\tau_i} \bar{\tau_l}+x_{13}^2 \tau_i \bar{\tau_j} \bar{\tau_l} +x_{14}^2 \bar{\tau_i} \tau_l+ x_{14}^2 \tau_i \bar{\tau_j} \tau_l+x_{24}^2 \tau_i \tau_j \tau_l\nonumber \\
\Delta_2^{\{3,3,1,1\}}&=& x_{13}^2 \bar{\tau_j} \bar{\tau_k} +x_{13}^2 \bar{\tau_j} \tau_k \bar{\tau_l}+x_{14}^2 \bar{\tau_j} \tau_k \tau_l+x_{24}^2 \tau_j \tau_k \tau_l
\eea
\hfill\break
$\circledcirc$ $I_{\rm ladder}^{\{5,4,2,1\}}$ \hfill\break
This diagram can be obtained by just inserting $(5,4,2,1)$ to $(i,j,k,l)$ in (\ref{Ladder}).
\begin{equation}
I_{\rm ladder}^{\{5,4,2,1\}}=\int_0^1 \rmd\tau_i \int_0^1 \rmd\tau_j \int_0^1 \rmd\tau_k \int_0^1 \rmd \tau_l \ \frac{\epsilon(y_1,y_5,z_1-z_5)}{[(z_1-z_5)^2]^{\frac{d}{2}}} \frac{\epsilon(y_2,y_4,z_2-z_4)}{[(z_2-z_4)^2]^{\frac{d}{2}}}
\end{equation}
The integral can be converted to
\begin{equation}
-\frac{1}{8} \int_0^1 \rmd\tau_i \int_0^1 \rmd\tau_j \int_0^1 \rmd\tau_k \int_0^1 \rmd\tau_l \ \frac{N_{5421}}{[x_{15}^2 \bar{\tau_i} \bar{\tau_l}+x_{25}^2 \tau_i \bar{\tau_l}+x_{26}^2 \tau_i \tau_l]^{\frac{d}{2}}[x_{24}^2 \bar{\tau_j} \bar{\tau_k}+x_{25}^2 \bar{\tau_j}\tau_k+x_{35}^2 \tau_j \tau_k]^{\frac{d}{2}}}
\end{equation}
where
\bea
N_{5421}&=&
-x_{14}^2(x_{25}^2)^2-x_{13}^2(x_{24}^2-x_{25}^2)(x_{25}^2-x_{26}^2)+x_{14}^2x_{25}^2x_{26}^2+x_{14}^2x_{25}^2x_{35}^2+x_{14}^2x_{26}^2x_{35}^2 \nonumber \\
&& -x_{24}^2x_{26}^2x_{35}^2-2x_{14}^2x_{25}^2x_{36}^2+x_{24}^2x_{25}^2x_{36}^2-(x_{25}^2)^2x_{36}^2+2x_{13}^2x_{25}^2x_{46}^2+x_{25}^2(x_{25}^2-x_{35}^2)x_{46}^2\nonumber \\
&&+x_{15}^2(-x_{26}^2x_{35}^2+x_{25}^2x_{36}^2+x_{24}^2(x_{25}^2-x_{26}^2-x_{35}^2+x_{36}^2)-x_{25}^2x_{46}^2+x_{35}^2x_{46}^2)
\eea
%

%%%%%%%%%%%%%%%%%%%%%%%%%%%%%%%%%%%%%
\section{Dimensional Redection Scheme}
When a field theory involves tensorial interactions, such as the $(2+1)$-dimensional ABJM theory, a subtle issue arises on the choice of the regularization scheme. \footnote{This issue was also discussed in the ABJM theory context in \cite{Bianchi:2013a}.} In the ABJM theory,  one of the tensorial interactions involves the Levi-Civita symbol. In the dimensional regularization, vectors such as momentum, position, etc. are extended to $d=3-2\epsilon$ dimension to regulate the UV divergences. For the Levi-Civita symbol, there are seemingly two possible choices of prescription:
\bea
\epsilon^{mpq} \epsilon_{mrs} &=& (\delta_{pr} \delta_{qs} - \delta_{ps} \delta_{qr}) \Gamma(d-1) \hskip2.2cm\mbox{($d$-dimension prescription)} \nonumber \\
\epsilon^{mpq} \epsilon_{mrs} &=& (\delta_{pr} \delta_{qs} - \delta_{ps} \delta_{qr}) \hskip3.5cm \mbox{(3-dimension prescription)}
\eea
The first choice is not consistent with the Slavnov-Taylor identity at two loops, thus violating gauge invariance. Therefore, we must treat spinorial quantities (including gamma function) and the Levi-Civita symbols as objects in 3-dimensions while all other tensors as objects in $d$-dimensions. This so-called dimensional reduction scheme (DRED) is also known consistent with the supersymmetry. For instance, the free energy and the ${1 \over 2}$-BPS circular Wilson loop expectation value in the ABJM theory was computed in DRED scheme, and the result matched perfectly with other exact computations such as the the supersymmetry localization technique.

Another of the tensorial interactions in the ABJM theory involves the metric tensor.
For instance, vertex diagrams consist of tensor integrals. Standard strategy for these tensor integrals is to view as derivatives of some scalar integrals with respect to external momenta. In such manipulations, there appears a $d$-dimensional metric tensor $\tilde{\eta}_{mn}$.
%\color{red} WHAT IS TILDE AND HAT METRIC HERE??\color{black}
% which perfectly isolated from epsilon tensor contraction. The problem is this :
How this tensor should be treated is a subtle matter. Is it ${\eta}_{mn} \tilde{\eta}^{mn} = 3$ or ${\eta}_{mn} \tilde{\eta}^{mn} = 3-2\epsilon$? One cannot decide by just based on gauge invariance and the consequent Slavnov-Taylor identity. Even if we distinguish $\tilde{\eta}^{mn}$ from  ${\eta}^{mn}$, the two-loop  renormalization factors are not modified in pure Chern-Simons part : it still gives 1 as far as the Levi-Civita symbol is defined in 3-dimensions. Here, we argue that ${\eta}_{mn} \tilde{\eta}^{mn} = 3-2\epsilon$ is physically more natural.

To see this, we decompose the $3$-dimensional metric $\eta_{mn}$ to $3-2 \epsilon$ dimensions and $ 2 \epsilon$ dimensions. That is,
\begin{equation}
\eta_{mn} = \hat{\eta}_{mn} + \hat{\hat{\eta}}_{mn}
\end{equation}
where $\hat{\eta}_{mn}$ is the $3-2\epsilon$ dimensional metric tensor while $\hat{\hat{\eta}}_{mn}$ is the $2\epsilon$ dimensional metric tensor.
%block diagonalization for continuous dimension. That is,
With such decomposition, we can treat $\eta^{mn} p_{n}$ as
\begin{displaymath}
\left(\begin{array}{ccc|c}
 &  &  & \\
 &  \hat{\eta}_{\hat{a}\hat{b}}&  & \\
 &  &  & \\
\hline
 &  &  & \hat{\hat{\eta}}_{\hat{\hat{\alpha}} \hat{\hat{\beta}}}
\end{array}\right)
\left(\begin{array}{c}
\\
p^{\hat{b}}\\
\\
\hline
p^{\hat{\hat{b}}}
\end{array}\right)
\end{displaymath}

The decomposition puts the respective spacetime mutually orthogonal, so contractions among themselves go as follows:
\begin{align}
\hat{\eta}_{\hat{m} \hat{n}} \hat{\eta}^{\hat{m} \hat{n}} = 3-2\epsilon, &\quad \quad \hat{\hat{\eta}}_{\hat{\hat{m}} \hat{\hat{n}}} \hat{\hat{\eta}}^{\hat{\hat{m}} \hat{\hat{n}}}= 2\epsilon, \quad \quad \hat{\eta}_{\hat{m} \hat{n}} \hat{\hat{\eta}}^{\hat{\hat{m}} \hat{\hat{n}}} = 0 \nonumber \\
\hat{\eta}_{\hat{m} \hat{n}} p^{\hat{n}} = p_{\hat{m}}, &\quad \quad \hat{\hat{\eta}}_{\hat{\hat{m}} \hat{\hat{n}}} p^{\hat{n}} = 0
\end{align}
From now on, our $p^{m}$ means $p^{\hat{m}}$ since there is no appearance of $p^{\hat{\hat{m}}}$ in our calculation.  \\
\\
When $\eta_{mn}$ or $\hat{\eta}_{\hat{m} \hat{n}}$ acts on $y^{n}$, they may be considered as equivalent since $\eta_{mn} y^{n}=(\hat{\eta}_{\hat{m} \hat{n}}+\hat{\hat{\eta}}_{\hat{\hat{m}} \hat{\hat{n}}}) y^{n}=(\hat{\eta}_{\hat{m} \hat{n}}) y^{n}$. Similarly, $\eta^{mn} \hat{\eta}_{\hat{n} \hat{p}}=(\hat{\eta}^{\hat{m} \hat{n}}+ \hat{\hat{\eta}}^{\hat{\hat{m}} \hat{\hat{n}}}) \hat{\eta}_{\hat{n} \hat{p}}=\hat{\eta}^{\hat{m}}_{\hat{p}}$. However, $\eta_{mn} \eta^{mn}$ and $\eta_{mn} \hat{\eta}^{\hat{m} \hat{n}}$ yield manifestly different results : 3 and $3-2\epsilon$, respectively.  For instance, let us examine this to the following term that appears in the $I_{321}$ integral.
\begin{equation}
\epsilon^{abc} \epsilon^{mnp} \hat{\eta}_{\hat{c} \hat{p}} y_{1,a} y_{2,b} y_{3,m} y_{2,n}
\end{equation}
The metric $\eta^{cp}$ will show up from contracting the Levi-Civita tensors, and this will be further contracted with $\hat{\eta}_{\hat{c} \hat{p}}$. This produces $3-2\epsilon$, not $3$.

%%%%%%%%%%%%%%%%%%%%%%%%%%%%%%%%%%%%%%%%%%%%%%
\section{Triple-Vertex Diagrams}
$\circledcirc I_{\rm vertex}^{\{3,2,1\}}$ \hfill\break
This is the integral for $\{i,j,k\} = {\{3,2,1\}}$ in (\ref{Vertex}).
%In this subsection, we used absolute value notation for convenience. That is, $|x-y|$ means $\big((x-y)^2 \big)^{\frac{1}{2}}$.
\begin{equation}
I_{\rm vertex}^{\{3,2,1\}}=
\int \rmd z_3^m \rmd z_2^n dz_1^p
\epsilon^{abc} \epsilon_{m a r} \epsilon_{n b s} \epsilon_{p c t}
\int \rmd^dw \frac{(w-z_3)^r (w-z_2)^s (w-z_1)^{t}}{|w-z_3|^d|w-z_2|^d|w-z_1|^d}
\end{equation}
Using the parametric expression of the position vectors $z_1,z_2, z_3$ on the edges, this integral equals to
\begin{equation}
\int \prod_{i=1}^3 \rmd \tau_i \ y_3^m y_2^n y_1^p
(-\epsilon_{nm r} \epsilon_{p s t}+\epsilon_{n sr} \epsilon_{p m t})
\int \rmd^dw \frac{(w-z_{32})^s (w)^r (w-z_{12})^t}{|w-z_{32}|^d|w|^d|w-z_{12}|^d}
\end{equation}
Here, we transformed $w-z_2$ to $w$. The numerator can be simplified further.
\begin{equation}
I_{\rm vertex}^{\{3,2,1\}}=\int \prod_{i=1}^3 \rmd\tau_i \int \rmd^dw \frac{-\epsilon(y_1,y_2,w)\epsilon(y_3,y_2,w)}{|w|^d|w-z_{32}|^d|w-z_{12}|^d}
\end{equation}
To proceed further, we replace $w$ by differential operator.
\begin{equation}
I_{\rm vertex}^{\{3,2,1\}}=\frac{1}{(d-2)^2}\int \prod_{i=1}^3 \rmd \tau_i \int \rmd^d w \frac{-\epsilon(y_1,y_2,\partial_{z_1})\epsilon(y_3,y_2,\partial_{z_3})}{|w|^d|w-z_{12}|^{d-2}|w-z_{32}|^{d-2}}
\end{equation}
Introducing the Feynman parameters $\beta_1, \beta_2, \beta_3$,
\bea
I_{\rm vertex}^{\{3,2,1\}}
&=&\frac{-1}{(d-2)^2} \frac{\Gamma(\frac{3d}{2}-2)}{\Gamma(\frac{d}{2})\Gamma(\frac{d}{2}-1)^2}\int \prod_{i=1}^3\rmd \tau_i
 \int \prod_{a=1}^3 \rmd\beta_a
\delta(\sum \beta_a-1) \nonumber \\
&& \hskip3cm \times
\int \rmd^d w \frac{\epsilon(y_1,y_2,\partial_{z_1})\epsilon(y_3,y_2,\partial_{z_3})(\beta_1 \beta_2 \beta_3)^{\frac{d-2}{2}-1} \beta_2}{[\beta_1(w-z_{12})^2+\beta_2 w^2+\beta_3(w-z_{32})^2]^{\frac{3d-4}{2}}} \nonumber \\
&=&\frac{-1}{(d-2)^2} \frac{\Gamma(\frac{3d}{2}-2)}{\Gamma(\frac{d}{2})\Gamma(\frac{d}{2}-1)^2} \int \prod_{i=1}^3 \rmd\tau_i \int \prod_{a=1}^3 \rmd \beta_a \delta(\sum \beta_a-1) \nonumber \\
&& \hskip3cm \times \int \rmd^d l \frac{\epsilon(y_1,y_2,\partial_{z_1})\epsilon(y_3,y_2,\partial_{z_3})(\beta_1 \beta_2 \beta_3)^{\frac{d}{2}-2}\beta_2}{[l^2+\Delta]^{\frac{3d-4}{2}}} \nonumber \\
&=& \frac{-i\pi^{\frac{d}{2}}}{(d-2)^2} \frac{\Gamma(d-2)}{\Gamma(\frac{d}{2})\Gamma(\frac{d}{2}-1)^2}
\int \prod_{i=1}^3 \rmd\tau_i \int \prod_{a=1}^3 \rmd\beta_a\delta(\sum \beta_a-1) \nonumber \\
&& \hskip3cm \times
(\beta_1 \beta_2 \beta_3)^{\frac{d}{2}-2}\beta_2  \epsilon(y_1,y_2,\partial_{z_1})\epsilon(y_3,y_2,\partial_{z_3}) \Delta^{2-d} \nonumber \\
&=& -\frac{i\pi^{\frac{d}{2}}\Gamma(d-2)}{4\Gamma(\frac{d}{2})^3}
\int \prod_{i=1}^3 \rmd\tau_i \int \prod_{a=1}^3 \rmd\beta_a \delta(\sum \beta_a-1) \nonumber \\
&& \hskip3cm \times (\beta_1 \beta_2 \beta_3)^{\frac{d}{2}-2}\beta_2  \epsilon(y_1,y_2,\partial_{z_1})\epsilon(y_3,y_2,\partial_{z_3}) \Delta^{2-d}
\eea
Here, $\Delta$ and $\beta_1, \beta_2, \beta_3$ denote
\bea
\Delta &=& \beta_1 \bar{\beta_1} z_{12}^2+\beta_3 \bar{\beta_3} z_{32}^2-2\beta_1 \beta_3 z_{12} \cdot z_{32} \nonumber \\
&=& \bar{\tau_1}xys(\bar{\tau_3}\bar{y}+\tau_2\bar{x}y)
+\tau_3y\bar{y}t(\bar{\tau_2}\bar{x}+\tau_1x)+\bar{\tau_1}\tau_3uxy\bar{y}
\eea
\bea
\beta_1= x y \quad \beta_2=(1-x)y \quad \beta_3=(1-y) \quad x_{13}^2=s \quad x_{14}^2=u \quad x_{24}^2=t.
\eea
It is the standard trick that the numerator can be expressed by differentiation of the $\Delta$\rq{}s:
\bea
\frac{\partial}{\partial z_1^\gamma} \Delta^{2-d}&=& (2-d) \Delta^{1-d} [2\beta_1 \bar{\beta_1} z_{12,\gamma}-2\beta_1 \beta_3 z_{32,\gamma}] \nonumber \\
\frac{\partial}{\partial z_3^\rho} \frac{\partial}{\partial z_1^\gamma} \Delta^{2-d}&=& (2-d)(1-d)\Delta^{-d}[2\beta_3 \bar{\beta_3}z_{32,\rho}-2\beta_1 \beta_3 z_{12,\rho}][2\beta_1 \bar{\beta_1} z_{12,\gamma}-2\beta_1 \beta_3 z_{32,\gamma}] \nonumber \\
&& -(2-d)\Delta^{1-d} 2\beta_1 \beta_3 \hat{\delta}_{\rho \gamma},
\eea
etc.
Treating the DRED scheme carefully,  the result reads
\bea
I_{\rm vertex}^{\{3,2,1\}}&=& \frac{i\pi^{\frac{d}{2}}\Gamma(d-2)}{8\Gamma(\frac{d}{2})^3}(2-d) \int \prod_{i=1}^3\rmd\tau_i
\int \rmd x \rmd y y(x\bar{x}y^2 \bar{y})^{\frac{d}{2}-2} \bar{x}y
\nonumber \\
&& \hskip2.9cm \times \Big[\frac{2(1-d)}{\Delta^d}(\bar{\tau_1}\tau_3 st(s-u+t)x^2 y^2 \bar{y}^2)-(d-2)\frac{xy\bar{y}}{\Delta^{d-1}}st \Big] \nonumber \\
&=& %\kappa
\kappa st \int \prod_{i=1}^3 \rmd\tau_i \int \rmd x \rmd y (x\bar{x}\bar{y})^{\frac{d}{2}-1} \Big[\frac{(d-2)}{\Delta_y^{d-1}}+\frac{2(d-1)}{\Delta_y^d} x\bar{y} \bar{\tau_1} \tau_3 (s-u+t) \Big].
\nonumber \eea
Here, the coefficient $\kappa$ is given by
\begin{equation}
\kappa=\frac{i\pi^{\frac{d}{2}}\Gamma(d-1)}{8\Gamma(\frac{d}{2})^3}
\end{equation}
and $\Delta_y$ abbreviates division of $\Delta$ by $y$:
\begin{equation}
\Delta_y = \bar{\tau_1}xs(\bar{\tau_3}\bar{y}+\tau_2\bar{x}y)+\tau_3 \bar{y}t(\bar{\tau_2}\bar{x}+\tau_1x)+\bar{\tau_1}\tau_3 ux\bar{y} .\label{321_delta}
\end{equation}
To prevent spurious poles, we only consider the case that Mandelstam invariants are defined in the Euclidean region. In other words, All Mandelstam invariants are positive definite.

\subsection{$I_{\rm vertex}^{\{4,2,1\}}$}
The integral is given by $(i,,j,k) = (4,2,1)$ in (\ref{Vertex}).
\bea
I_{\rm vertex}^{\{4,2,1\}}=\int \rmd z_4^m \rmd z_2^n \rmd  z_1^p \
\epsilon^{abc} \epsilon_{m a r} \epsilon_{n b s} \epsilon_{p c t}
\int \rmd^d w \frac{(w-z_4)^r (w-z_2)^s (w-z_1)^{t}}{|w-z_4|^d|w-z_2|^d|w-z_1|^d}
\eea
The integral can be converted to
\begin{equation}
\int \rmd s_4 \rmd s_2 \rmd s_1 y_4^m y_2^n y_1^p (-\epsilon_{nm r} \epsilon_{p s t}+\epsilon_{n s r} \epsilon_{p m t})
\int \rmd^dw \frac{(w-z_{42})^\sigma (w)^\lambda (w-z_{12})^\tau}{|w-z_{42}|^d|w|^d|w-z_{12}|^d}
\end{equation}
After some algebra, this integral is reduced to
\bea
&&\tilde{\kappa} \int \prod_{i=1}^3 \rmd \beta_i \ \beta_1%^{\frac{d}{2}-1}
(\beta_1\beta_2 \beta_3)^{\frac{d}{2}-2}%\beta_3^{\frac{d}{2}-2}
\delta(\sum\beta_i-1) \Delta^{-d}
\Big[\epsilon(y_4,y_2,\beta_2\beta_3 z_{12}-\beta_3 \bar{\beta_3} z_{42}) \epsilon(y_1,y_2,\beta_2 \beta_3 z_{42}-\beta_2 \bar{\beta_2}z_{12}) \nonumber \\
&& \hskip6cm +\epsilon(y_4,y_2,\beta_2\beta_3 z_{12}-\beta_3 \bar{\beta_3} z_{42}) \epsilon(y_1,y_3,\beta_2 \beta_3 z_{42}-\beta_2 \bar{\beta_2}z_{12})\nonumber \\
&& \hskip6cm -\epsilon(y_3,y_2,\beta_2 \beta_3 z_{12}-\beta_3 \bar{\beta_3} z_{42}) \epsilon(y_1,y_4,\beta_2 \beta_3 z_{42}-\beta_2 \bar{\beta_2} z_{12}) \Big] \nonumber \\
&& -\tilde{\kappa} \int \prod_{i=1}^3 \rmd\beta_i \ \beta_1%^{\frac{d}{2}-1}
(\beta_1 \beta_2 \beta_3)^{\frac{d}{2}-2} %\beta_3^{\frac{d}{2}-1}
\delta(\sum\beta_i-1) \Delta^{-d} \Big[\epsilon(y_4,y_2,y_3)\epsilon(y_1,y_2,\beta_2\beta_3  z_{42}-\beta_2\bar{\beta_2} z_{12}) \nonumber \\
&& \hskip6.5cm +\epsilon(y_4,y_2,y_3)\epsilon(y_1,y_3,\beta_2\beta_3 z_{42}-\beta_2\bar{\beta_2} z_{12})  \nonumber \\
&& \hskip6.5cm +s_4 \epsilon(y_4,y_2,y_3)\epsilon(y_1,y_4,\beta_2\beta_3  z_{42}-\beta_2\bar{\beta_2} z_{12}) \Big] \nonumber \\
&& +\frac{\tilde{\kappa}}{2(1-d)} \int \prod_{i=1}^3 \rmd\beta_i (\beta_1 \beta_2 \beta_3)^{\frac{d}{2}-1} \delta(\sum\beta_i-1) \Delta^{1-d} \nonumber \\
&& \hskip5cm \times  \Big[(y_4 \cdot y_2)(y_1 \cdot y_2)-(y_1 \cdot y_4)(y_2 \cdot y_3)  +(y_3 \cdot y_1)(y_2 \cdot y_4) \Big]. \label{421num}
\eea
Here, $\tilde{\kappa}$ denotes
\begin{equation}
\tilde{\kappa}=i\pi^{\frac{d}{2}}(1-d)\frac{\Gamma(d-1)}{\Gamma(\frac{d}{2})^3}
\end{equation}
We further simplify (\ref{421num}) by converting kinematic variables to Mandelstam invariants and by eliminating the delta function with the substitution
\begin{equation}
\beta_1 = xy, \quad \beta_2 = \bar{x}y, \quad \beta_3=\bar{y}
\end{equation}
This results in
\bea
I_{\rm vertex}^{\{4,2,1\}} =
&-& \tilde{\kappa}\frac{(d-2)}{2(d-1)}
\int \rmd s_{1,2,4} \int \rmd x \rmd y (x\bar{x}\bar{y})^{\frac{d}{2}-1} \frac{1}{\Delta_y^{d-1}} F_{A,421}(x_{i,j}^2)
\nonumber \\
&+&\tilde{\kappa}
\int \rmd s_{1,2,4} \int \rmd x \rmd y (x\bar{x}\bar{y})^{\frac{d}{2}-1} \frac{1}{\Delta_y^d} F_{B,421}(x_{i,j}^2).
\eea
%
%\color{red}
%I DO NOT UNDERSTAND THE NOTATIONS. WHAT ARE BETA\rq{}S AND WHAT ARE S\rq{}S? ARE THEY RELATED ?
%\color{black}
Here, $F_{A,421}(x_{i,j}^2)$ and $F_{B,421}(x_{i,j}^2)$ are function of the Mandelstam invariants obtained from expanding (\ref{421num}). The denominator $\Delta_y$ abbreviates $\Delta$ divided by $y$:
\begin{equation}
\Delta_y = x_{13}^2 \tau_2 xy\bar{\tau_1}\bar{x}+x_{14}^2\bar{\tau_1}\bar{\tau_4}\bar{x}\bar{y}+x_{15}^2\bar{\tau_1}\tau_4\bar{x}\bar{y}+x_{24}^2\bar{\tau_4}\bar{y}(\tau_1\bar{x}+\bar{\tau_2}x)+x_{25}^2 \tau_4\bar{y}(\tau_1\bar{x}+\bar{\tau_2}x)+x_{35}^2 \tau_2 \tau_4 x\bar{y}
\label{deltay}
\end{equation}
The integral is IR finite.

\subsection{$I_{\rm vertex}^{\{4,3,1\}}$}

%This evaluation is very parallel to previous vertex diagrams.
This integral is obtained by inserting $(\i, j, k) = (4,3,1)$ to (\ref{Vertex}).
\bea
I_{\rm vertex}^{\{4,3,1\}}=\int \rmd z_4^m \rmd z_3^n \rmd z_1^p
\epsilon^{abc} \epsilon_{m a r} \epsilon_{n b s} \epsilon_{p c t}
\int \rmd^dw \frac{(w-z_4)^r (w-z_3)^s (w-z_1)^{t}}{|w-z_4|^d|w-z_3|^d|w-z_1|^d}
\eea
After some algebra, the integral becomes
\bea
&+&\tilde{\kappa} \int \prod_{i=1}^3 \rmd \beta_i \ \beta_2 (\beta_1 \beta_2 \beta_3)^{\frac{d}{2}-2} \delta(\sum\beta_i-1) \Delta^{-d} \Big[\epsilon(y_3,y_1,\beta_1\beta_3 z_{43}-\beta_1 \bar{\beta_1} z_{13}) \epsilon(y_4,y_2,\beta_1 \beta_3 z_{13}-\beta_3 \bar{\beta_3}z_{43})\nonumber \\
&& \hskip6.5cm + \epsilon(y_3,y_1,\beta_1 \beta_3 z_{43}-\beta_1 \bar{\beta_1} z_{13}) \epsilon(y_4,y_3,\beta_1 \beta_3 z_{13}-\beta_3 \bar{\beta_3}z_{43})\nonumber \\
&& \hskip6.5cm - \epsilon(y_3,y_2,\beta_1 \beta_3 z_{43}-\beta_1 \bar{\beta_1} z_{13}) \epsilon(y_4,y_1,\beta_1 \beta_3 z_{13}-\beta_3 \bar{\beta_3} z_{43}) \Big] \nonumber \\
&-&\tilde{\kappa} \int \prod_{i=1}^3 \rmd\beta_i \ \beta_1 \beta_2 (\beta_1 \beta_2 \beta_3)^{\frac{d}{2}-2}
\delta(\sum\beta_i-1) \Delta^{-d} \Big[\epsilon(y_3,y_1,y_2)\epsilon(y_4,y_2,\beta_1\beta_3  z_{13}-\beta_3\bar{\beta_3} z_{43}) \nonumber \\
&& \hskip6.5cm +\epsilon(y_3,y_1,y_2)\epsilon(y_4,y_3,\beta_1\beta_3 z_{13}-\beta_3\bar{\beta_3} z_{43})  \nonumber \\
&& \hskip6.5cm +\bar{s_1} \epsilon(y_3,y_1,y_2)\epsilon(y_4,y_1,\beta_1 \beta_3  z_{13}-\beta_3 \bar{\beta_3} z_{43}) \Big] \nonumber \\
&+& \tilde{\kappa} \int \prod_{i=1}^3 \rmd\beta_i \ (\beta_1 \beta_2 \beta_3)^{\frac{d}{2}-1} \delta(\sum\beta_i-1) \Delta^{1-d} \nonumber \\
&& \hskip3cm \times \frac{ 1 }{2(1-d)}\Big[(y_3 \cdot y_1)(y_4 \cdot y_3)-(y_1 \cdot y_4)(y_2 \cdot y_3)  +(y_3 \cdot y_1)(y_2 \cdot y_4) \Big]. \label{431num}
\eea
As before, $\tilde{\kappa}$ is the numerical factor
\begin{equation}
\tilde{\kappa}=i\pi^{\frac{d}{2}}(1-d)\frac{\Gamma(d-1)}{\Gamma(\frac{d}{2})^3}.
\end{equation}
When numerically evaluating, we translated (\ref{431num}) as a function of the Mandelstam invariants. As before, this step reults in a long expression for the numerator. We also eliminate the delta function by reparametrizing $\beta_i$ by two parameters $x,y$:
\begin{equation}
\beta_1 = xy, \quad \beta_2 = \bar{x}y, \quad \beta_3=\bar{y}.
\end{equation}
Finally, we obtain
\bea
I_{\rm vertex}^{\{4,3,1\}} = &-&\tilde{\kappa}
\int \rmd s_{1,3,4} \int \rmd x \rmd y (x\bar{x}\bar{y})^{\frac{d}{2}-1} \frac{(d-2)}{2(d-1)}\frac{1}{\Delta_y^{d-1}} F_{A,431}(x_{i,j}^2)
\nonumber \\
&+& \tilde{\kappa} \int \rmd s_{1,3,4}
\int \rmd x \rmd y (x\bar{x}\bar{y})^{\frac{d}{2}-1} \frac{1}{\Delta_y^d} F_{B,431}(x_{i,j}^2)
\eea
Here $F_{A,431}(x_{i,j}^2)$ and $F_{B,431}(x_{i,j}^2)$ are function of Mandelstam variables obtained from expanding (\ref{431num}). Denominator $\Delta_y$ obtained by dividing the corresponding $\Delta$ by $y$.
\begin{equation}
\Delta_y =-x_{13}^2 x \bar{x} y \bar{\tau}_1 \bar{\tau}_3-x_{14}^2 x \bar{\tau}_1 (\bar{x} y \tau_3 + \bar{y} \bar{\tau}_4) -x_{15}^2 x \bar{y} \bar{\tau}_1 \tau_4 -x_{24}^2 x \tau_1(\bar{x}y \tau_3 + \bar{y} \bar{\tau}_4)-x_{25}^2 x \bar{y} \tau_1 \tau_4 - x_{35}^2 \bar{x} \bar{y} \bar{\tau}_3 \tau_4
\end{equation}
We found that numerical value of $I_{421}$ and $I_{431}$ coincides when the hexagon kinematics satisfy the Gram sub-determinant conditions.
%Due to its complexity, we were not able to understand this from the integrals directly.

\subsection{$I_{\rm vertex}^{\{5,3,1\}}$}
This diagram gives rise to the most complicated integral.  Start from (\ref{Vertex}),
\begin{equation}
I_{\rm vertex}^{\{5,3,1\}}=\int \rmd z_5^m \rmd z_3^n \rmd z_1^p \epsilon^{abc} \epsilon_{ma r} \epsilon_{nb s} \epsilon_{pc t} \int d^dw \frac{(w-z_5)^r (w-z_3)^s (w-z_1)^t}{|w-z_5|^d|w-z_3|^d|w-z_1|^d}
\end{equation}
After straightforward algebra, we get
\bea
I_{\rm vertex}^{\{5,3,1\}}&=&4\frac{(d-1)}{(d-2)}\tau \int[\rmd\beta_3]\int \rmd\tau_{5,3,1} \frac{1}{\Delta^d} H_B(x_{i,j}^2) \nonumber \\
&-& 2\frac{1}{d-2}\tau \int[\rmd\beta_3]\int \rmd\tau_{5,3,1} \frac{1}{\Delta^{d-1}} H_A(x_{i,j}^2) \nonumber \\
&+& 2\frac{(d-1)}{(d-2)}\tilde{\tau} \int[\rmd\tilde{\beta_3}]\int \rmd\tau_{5,3,1} \frac{1}{\Delta^d}  H_C(x_{i,j}^2). \label{vertex_531}
\eea
Here,
\bea
\tau= -i\pi^{\frac{d}{2}}\frac{\Gamma(d-2)}{\Gamma(\frac{3d}{2}-2)},\quad \quad \tilde{\tau}=-i\pi^{\frac{d}{2}}\frac{\Gamma(d-1)}{\Gamma(\frac{3d}{2}-1)}
\eea
and
\bea
\int[\rmd\beta_3]&=& \int_0^1 \prod_{i=1}^3 \rmd\beta_i  \ \beta_2  (\beta_1 \beta_2 \beta_3)^{\frac{d}{2}-2} \ \delta(\sum_i \beta_i-1) \ \frac{\Gamma(\frac{3d}{2}-2)}{\Gamma(\frac{d}{2})\Gamma(\frac{d}{2}-1)^2} \nonumber \\
\int[\rmd\tilde{\beta_3}]&=& \int_0^1 \prod_{i=1}^3 \rmd\beta_i \ {1 \over \beta_3} (\beta_1 \beta_2 \beta_3)^{\frac{d}{2}-1}  \ \delta(\sum_i \beta_i-1) \ \frac{\Gamma(\frac{3d}{2}-1)}{\Gamma(\frac{d}{2})^2 \Gamma(\frac{d}{2}-1)}
\eea
and denominator $\Delta_y$ is given by
\bea
\Delta_y &=& x_{13}^2 x \bar{x} y \bar{\tau_1} \bar{\tau_3}+x_{14}^2 x \bar{x} y \bar{\tau_1} \tau_3 + x_{24}^2 x \bar{x} y \tau_1
+ x_{15}^2 x \bar{y} \bar{\tau_1} \bar{\tau_5}+x_{25}^2 x \bar{y} \tau_1 \bar{\tau_5} \nonumber \\
&+& x_{26}^2 x \bar{y} \tau_1 \tau_5+ x_{35}^2 \bar{x} \bar{y} \bar{\tau_3} \bar{\tau_5}+x_{36}^2 \bar{x} \bar{y} \bar{\tau_3} \tau_5+x_{46}^2 \bar{x} \bar{y}\tau_3 \tau_5.
\eea
Again, $\Delta_y$ abbreviates $\Delta$ divided by $y$.

The functions $H_A(x_{i,j}^2)$, $H_B(x_{i,j}^2)$ and $H_C(x_{i,j}^2)$ are  rather complicated functions of Mandelstam invariants.
We combined the first and the third terms in (\ref{vertex_531}), managed to the following compact expression
\bea
I_{\rm vertex}^{\{5,3,1\}}&=& -i\pi^{\frac{d}{2}}{(d-2)\over 2}\frac{\Gamma(d-1)}{\Gamma(\frac{d}{2})^3} \int_0^1 \rmd x \rmd y \int_0^1 \rmd \tau_{5,3,1} (x \bar{x} \bar{y})^{\frac{d}{2}-1} \frac{1}{\Delta_y^{d-1}} F_{A,531}(x_{i,j}^2)
\nonumber \\
& + & i  \pi^{\frac{d}{2}}\frac{\Gamma(d)}{\Gamma(\frac{d}{2})^3}\int_0^1 \rmd x \rmd y \int_0^1 \rmd \tau_{5,3,1}  (x\bar{x}\bar{y})^{\frac{d}{2}-1} \frac{1}{\Delta_y^d} F_{B,531}(x_{i,j}^2).
\eea
%
%where
%\begin{equation}
%\alpha_1=-i\pi^{\frac{d}{2}} (1-d) \frac{\Gamma(d-1)}{\Gamma(\frac{d}{2})^3}
%\end{equation}
%which gives a value of $16i$ for $d=3$.
%
%For second term in (\ref{vertex_531}),
%\begin{equation}
%\frac{\alpha_1}{2(1-d)} \int_0^1 \rmd x \rmd y \int_0^1 \rmd \tau_{5,3,1} (x \bar{x} \bar{y})^{\frac{d}{2}-1} F_{A,531}(x_{i,j}^2) \frac{(d-2)}{\Delta_y^{d-1}}
%\end{equation}
Since it is a matter of calculation, we do not provide the functions $F_{A,ijk}$ and $F_{B,ijk}$ here.

%%%%%%%%%%%%%%%%%%%%%%%%%%%%%%
\section{$I_{521}$ and $I_{541}$ Integrals}
When extracting the antenna function from the Wilson loop expectation values beyond the hexagion ($n>6$), we encountered extra  triple-vertex diagrams,  $I_{521}$ and $I_{541}$, beyond the hexagon. Recall that, in hexagon, $x_{16}^2$ vanishes identically, so  these diagrams could be obtained from permutating $I_{421}$ or $I_{431}$ therein. However, these diagrams begin to be distinct beyond the octagon $n > 6$. Here, we provide these integrals for $I_{521}$ and $I_{541}$. They are the same for all $n > 6$ because of the relation $2y_i \cdot y_j=x_{i,j+1}^2+x_{i+1,j}^2-x_{i,j}^2-x_{i+1,j+1}^2$ is the same.

$\circledcirc$ Scalar integration $I_{541}$
\hfill\break
Start from
\bea
I_{\rm vertex}^{\{5,4,1\}}=\int \rmd z_5^m \rmd z_4^n \rmd z_1^p \epsilon^{abc} \epsilon_{mar} \epsilon_{nbs} \epsilon_{pct}
\int \rmd^dw \frac{(w-z_5)^r (w-z_4)^s (w-z_1)^t }{|w-z_5|^d|w-z_4|^d|w-z_1|^d}.
\eea
This expression is equivalent to
\bea
\frac{1}{(d-2)^2}\int \rmd^dw \left[
\frac{\epsilon(y_4,y_5,\partial_{z_5})%\epsilon(p_1,p_2,\partial_{z_1})
}{|w|^d|w-z_{54}|^{d-2}|w-z_{14}|^{d-2}} \left(\epsilon(y_1,y_2,\partial_{z_1})+\epsilon(y_1,y_3,\partial_{z_1})
+\epsilon(y_1,y_4,\partial_{z_1})\right)
%+ \frac{\epsilon(p_4,p_5,\partial_{z_5})\epsilon(p_1,p_3,\partial_{z_1})}{|w|^d|w-z_{54}|^{d-2}|w-z_{14}|^{d-2}}
%+ \frac{\epsilon(p_4,p_5,\partial_{z_5})\epsilon(p_1,p_4,\partial_{z_1})}{|w|^d|w-z_{54}|^{d-2}|w-z_{14}|^{d-2}}
\right].
%
%+\frac{1}{(d-2)^2}\int d^dw \frac{\epsilon(p_4,p_5,\partial_{z_5})\epsilon(p_1,p_3,\partial_{z_1})}{|w|^d|w-z_{54}|^{d-2}|w-z_{14}|^{d-2}} \nonumber \\
%&+&\frac{1}{(d-2)^2}\int \rmd^dw \frac{\epsilon(p_4,p_5,\partial_{z_5})\epsilon(p_1,p_4,\partial_{z_1})}{|w|^d|w-z_{54}|^{d-2}|w-z_{14}|^{d-2}}
\eea
Introducing Feynman parameter $\beta_1, \beta_2, \beta_3$ and $\Delta$ as
\bea
\beta_1= xy ,\quad \beta_2=\overline{x}y, \quad \beta_3=\overline{y} \nonumber
\eea
%\bea
%\int[\rmd\beta_3]=\int_0^1 \prod_{i=1}^3 \rmd\beta_i (\beta_1 \beta_2 \beta_3)^{\frac{d}{2}-2} \beta_1 \delta(\sum_i \beta_i-1) \frac{\Gamma(\frac{3d}{2}-2)}{\Gamma(\frac{d}{2})\Gamma(\frac{d}{2}-1)^2} \nonumber
%\eea
\bea
\Delta=-2\beta_2 \beta_3  z_{14}  z_{54}+\beta_2 \overline{\beta}_2 z_{14}^2+ \beta_3 \overline{\beta}_3 z_{54}^2, \nonumber
\eea
we get
\bea
&& I_{\rm vertex}^{\{5,4,1\}} \nonumber \\
& = & -\tilde{\tau} \int \prod_{i=1}^3 d\beta_i \ (\beta_1 \beta_2 \beta_3)^{\frac{d}{2}-1} \delta(\sum\beta_i-1) \Delta^{-d} \Big[ \epsilon(y_4,y_5,\beta_2\beta_3 z_{14}-\beta_3 \bar{\beta_3} z_{54}) \epsilon(y_1,y_2,\beta_2 \beta_3 z_{54}-\beta_2 \bar{\beta_2}z_{14})\nonumber \\
&& \hskip4cm +\epsilon(y_4,y_5,\beta_2\beta_3 z_{14}-\beta_3 \bar{\beta_3} z_{54}) \epsilon(y_1,y_3,\beta_2 \beta_3 z_{54}-\beta_2 \bar{\beta_2}z_{14})\nonumber \\
&& \hskip4cm +\epsilon(y_4,y_5,\beta_2\beta_3 z_{14}-\beta_3 \bar{\beta_3} z_{54}) \epsilon(y_1,y_4,\beta_2 \beta_3 z_{54}-\beta_2 \bar{\beta_2}z_{14}) \Big] \nonumber \\
&& +\frac{1}{2(1-d)} \tilde{\tau} \int \prod_{i=1}^3 \rmd\beta_i (\beta_1 \beta_2 \beta_3)^{\frac{d}{2}-1} \delta(\sum\beta_i-1) \Delta^{1-d} \Big[(y_4 \cdot y_1)(y_5 \cdot y_2)-(y_4 \cdot y_2)(y_5 \cdot y_1) \nonumber \\
&& \hskip5cm +(y_4 \cdot y_1)(y_5 \cdot y_3)-(y_4 \cdot y_3)(y_5 \cdot y_1)+(y_4 \cdot y_1)(y_5 \cdot y_4) \Big]
\eea
where
\begin{equation}
\tilde{\tau}=i\pi^{\frac{d}{2}}(1-d)\frac{\Gamma(d-1)}{\Gamma(\frac{d}{2})^3}.
\end{equation}

$\circledcirc$ Scalar integration $I_{521}$
\hfill\break
This integral can be computed following the same route as $I_{541}$ integral. However, expressions are more complicated. Here we just briefly summarize result. The $I_{521}$ integral reads
\bea
I_{521}&=& \frac{1}{(d-2)^2}\int \rmd^dw
\frac{\epsilon(y_2,y_5,\partial_{z_5})\epsilon(y_1,y_2,\partial_{z_1})}{|w|^d|w-z_{52}|^{d-2}|w-z_{12}|^{d-2}}
+\frac{1}{(d-2)^2}\int \rmd^dw \frac{\epsilon(y_2,y_5,\partial_{z_5})\epsilon(y_1,y_3,\partial_{z_1})}{|w|^d|w-z_{52}|^{d-2}|w-z_{12}|^{d-2}} \nonumber \\
&+& \frac{1}{(d-2)^2}\int \rmd^dw \frac{\epsilon(y_2,y_5,\partial_{z_5})\epsilon(y_1,y_4,\partial_{z_1})}{|w|^d|w-z_{52}|^{d-2}|w-z_{12}|^{d-2}}
-\frac{1}{(d-2)^2}\int \rmd^dw \frac{\epsilon(y_2,y_3,\partial_{z_5})\epsilon(y_1,y_5,\partial_{z_1})}{|w|^d|w-z_{52}|^{d-2}|w-z_{12}|^{d-2}}\nonumber \\
&-&\frac{1}{(d-2)^2}\int \rmd^dw \frac{\epsilon(y_2,y_4,\partial_{z_5})\epsilon(y_1,y_5,\partial_{z_1})}{|w|^d|w-z_{52}|^{d-2}|w-z_{12}|^{d-2}}
-\frac{1}{d-2}\int \rmd^dw  \frac{\epsilon(y_2,y_5,y_3)\epsilon(y_1,y_2,\partial_{z_1})}{|w|^d|w-z_{12}|^{d-2}|w-z_{52}|^{d}} \nonumber \\
&-& \frac{1}{d-2}\int \rmd^dw \frac{\epsilon(y_2,y_5,y_4)\epsilon(y_1,y_2,\partial_{z_1})}{|w|^d|w-z_{12}|^{d-2}|w-z_{52}|^{d}}
-\frac{1}{d-2}\int \rmd^dw  \frac{ \epsilon(y_2,y_5,y_3)\epsilon(y_1,y_3,\partial_{z_1})}{|w|^d|w-z_{12}|^{d-2}|w-z_{52}|^{d}} \nonumber \\
&-& \frac{1}{d-2}\int \rmd^dw  \frac{\epsilon(y_2,y_5,y_4)\epsilon(y_1,y_3,\partial_{z_1})}{|w|^d|w-z_{12}|^{d-2}|w-z_{52}|^{d}}
-\frac{1}{d-2}\int \rmd^dw  \frac{ \epsilon(y_2,y_5,y_3)\epsilon(y_1,y_4,\partial_{z_1})}{|w|^d|w-z_{12}|^{d-2}|w-z_{52}|^{d}} \nonumber \\
&-& \frac{1}{d-2}\int \rmd^dw  \frac{\epsilon(y_2,y_5,y_4)\epsilon(y_1,y_4,\partial_{z_1})}{|w|^d|w-z_{12}|^{d-2}|w-z_{52}|^{d}}
+\frac{1}{d-2}\int \rmd^dw  \frac{s_5 \epsilon(y_2,y_3,y_5)\epsilon(y_1,y_5,\partial_{z_1})}{|w|^d|w-z_{12}|^{d-2}|w-z_{52}|^{d}} \nonumber \\
&+& \frac{1}{d-2}\int \rmd^dw  \frac{s_5 \epsilon(y_2,y_4,y_5)\epsilon(y_1,y_5,\partial_{z_1})}{|w|^d|w-z_{12}|^{d-2}|w-z_{52}|^{d}}
\eea
We do not provide explicit expression of the numerator in terms of the Mandelstam invariants. Though need for computation using the package  {\tt FIESTA2}, they can be obtained straightforwardly from the following expressions.

Note that the latter 8 terms (proportional to $\frac{1}{(d-2)}$) are negligible in the triple collinear limit. This is because the Levi-Civita tensor contains two of $y_2,y_3,y_4$ and they all become parallel in this limit. Although we considered their contribution
in the computation, here we just provide the expressions for the dominant part. The Feynman parameters $\beta_1, \beta_2, \beta_3$ are defined as before, but $\Delta$ is slightly different.
\begin{align}
\beta_1&=xy ,\quad \beta_2=\overline{x}y, \quad \beta_3=\overline{y} \nonumber \\
%\int[d\beta_3]=\int_0^1 d\beta_1 &d\beta_2 d\beta_3 (\beta_1 \beta_2 \beta_3)^{\frac{d}{2}-2} \beta_1 \delta(\sum_i \beta_i-1) \frac{\Gamma(\frac{3d}{2}-2)}{\Gamma(\frac{d}{2})\Gamma(\frac{d}{2}-1)^2} \nonumber \\
\Delta=-2&\beta_2 \beta_3  z_{12} \cdot z_{52}+\beta_2 \overline{\beta}_2 z_{12}^2+ \beta_3 \overline{\beta}_3 z_{52}^2.
\end{align}
The dominant 5-terms are given by
\begin{align}
&-\tilde{\tau} \int \prod_{i=1}^3 \rmd\beta_i \ \beta_1 (\beta_1 \beta_2 \beta_3)^{\frac{d}{2}-2} \delta(\sum\beta_i-1) \Delta^{-d} \Big[ \epsilon(y_2,y_5,\beta_2\beta_3 z_{12}-\beta_3 \bar{\beta_3} z_{52}) \epsilon(y_1,y_2,\beta_2 \beta_3 z_{52}-\beta_2 \bar{\beta_2}z_{12})\nonumber \\
&\hskip6cm +\epsilon(y_2,y_5,\beta_2\beta_3 z_{12}-\beta_3 \bar{\beta_3} z_{52}) \epsilon(y_1,y_3,\beta_2 \beta_3 z_{52}-\beta_2 \bar{\beta_2}z_{12})\nonumber \\
&\hskip6cm +\epsilon(y_2,y_5,\beta_2\beta_3 z_{12}-\beta_3 \bar{\beta_3} z_{52}) \epsilon(y_1,y_4,\beta_2 \beta_3 z_{52}-\beta_2 \bar{\beta_2}z_{12})\nonumber \\
&\hskip6cm -\epsilon(y_2,y_3,\beta_2\beta_3 z_{12}-\beta_3 \bar{\beta_3} z_{52}) \epsilon(y_1,y_5,\beta_2 \beta_3 z_{52}-\beta_2 \bar{\beta_2}z_{12})\nonumber \\
&\hskip6cm -\epsilon(y_2,y_4,\beta_2\beta_3 z_{12}-\beta_3 \bar{\beta_3} z_{52}) \epsilon(y_1,y_5,\beta_2 \beta_3 z_{52}-\beta_2 \bar{\beta_2}z_{12}) \Big] \nonumber \\
&+\frac{\tilde{\tau}}{2(1-d)}  \int \prod_{i=1}^3 \rmd\beta_i (\beta_1 \beta_2 \beta_3)^{\frac{d}{2}-1} \delta(\sum\beta_i-1) \Delta^{1-d} \Big[(y_1 \cdot y_2)(y_2 \cdot y_5)-(y_2 \cdot y_3)(y_1 \cdot y_5) \nonumber \\
& \hskip1.7cm +(y_1 \cdot y_2)(y_4 \cdot y_5) -(y_2 \cdot y_4)(y_1 \cdot y_5)+(y_1 \cdot y_3)(y_2 \cdot y_5) -(y_1 \cdot y_2)(y_4 \cdot y_5)+(y_1 \cdot y_4)(y_2 \cdot y_5) \Big]
\end{align}
where
%\color{red}
%CHECK THE FIRST TERM INSIDE THE BRACKET IN THE LAST LINE. INDICES SEEM NOT RIGHT, AND I TRIED TO CORRECT IT. \color{blue} It was corrected.
\color{black}
\begin{equation}
\tilde{\tau}=i\pi^{\frac{d}{2}}(1-d)\frac{\Gamma(d-1)}{\Gamma(\frac{d}{2})^3}.
\end{equation}
We used this expression for the numerical computations.

\end{document}